\documentclass{article}
\usepackage{amssymb}

\input{tcilatex}

\begin{document}

\begin{center}
{\LARGE GRASSMANN PHASE SPACE METHODS}$\bigskip $

{\LARGE \ FOR FERMIONS.}$\bigskip $

{\LARGE II. FIELD\ THEORY}

\bigskip

B. J Dalton$^{1}$, J. Jeffers$^{2}\smallskip $

\& S. M. Barnett$^{3}\bigskip $

1. Centre for Quantum and Optical Science

Swinburne University of Technology, Melbourne, Victoria 3122,
Australia\smallskip

2. Dept of Physics, University of Strathclyde, Glasgow G4ONG, UK\smallskip

3. School of Physics and Astronomy, University of Glasgow, Glasgow G12 8QQ,
UK

\bigskip
\end{center}

* Corresponding Author:

Tel: +61 0 3 9214 8187; fax: +61 0 3 9214 5160

\textit{Email address: }bdalton@swin.edu.au\pagebreak

\subsection{Abstract}

In both quantum optics and cold atom physics, the behaviour of bosonic
photons and atoms is often treated using phase space methods, where mode
annihilation and creation operators are represented by c-number phase space
variables, with the density operator equivalent to a distribution function
of these variables. The anti-commutation rules for fermion annihilation,
creation operators suggests the possibility of using anti-commuting
Grassmann variables to represent these operators. However, in spite of the
seminal work by Cahill and Glauber and a few applications, the use of
Grassmann phase space methods in quantum - atom optics to treat fermionic
systems is rather rare, though fermion coherent states using Grassmann
variables are widely used in particle physics.

This paper presents a phase space theory for fermion systems based on
distribution functionals, which replace the density operator and involve
Grassmann fields representing anti-commuting fermion field annihilation,
creation operators. It is an extension of a previous phase space theory
paper\ for fermions (Paper I) based on separate modes, in which the density
operator is replaced by a distribution function depending on Grassmann phase
space variables which represent the mode annihilation and creation
operators. This further development of the theory is important for the
situation when large numbers of fermions are involved, resulting in too many
modes to treat separately. Here Grassmann fields, distribution functionals,
functional Fokker-Planck equations and Ito stochastic field equations are
involved. Typical applications to a trapped Fermi gas of interacting spin $%
1/2$ fermionic atoms and to multi-component Fermi gases with non-zero range
interactions are presented, showing that the Ito stochastic field equations
are local in these cases. For the spin $1/2$ case we also show how simple
solutions can be obtained both for the untrapped case and for an optical
lattice trapping potential.\pagebreak

\section{Introduction}

\label{Section - Introduction copy(1)}

This paper treats many-body fermion systems via a phase space theory. It
extends a previous paper (Paper I, \cite{Dalton16b}) on this topic - in
which the modes (or single particle states) that the fermions may occupy
were treated separately, to the situation where fermion field creation and
annihilation operators are involved rather than operators for the separate
modes. This approach is more suitable when large numbers of fermions are
involved since the Pauli exclusion principle guarantees that large numbers
of modes are needed to accommodate them, resulting in too many modes to
treat separately. In the previous paper we have presented phase space theory
as being one of a range of methods (see Ref. \cite{Dalton16b}) used to treat
many-body systems \cite{March67a} in non-relativistic quantum physics.
Briefly, for phase space theories based on separate modes the density
operator describing the quantum state is represented by a distribution
function of phase space variables, which are associated with the mode
annihilation and creation operators. Measureable quantities such as quantum
correlation functions and Fock state probabilities and coherences are
expressed as phase space integrals involving the distribution function and
specific functions of the phase space variables describing the measureable
quantity. The evolution equation for the density operator is replaced by a
Fokker-Planck equation \cite{Risken89a} for the distribution function. In
turn, the phase space variables are replaced by time dependent stochastic
variables, and Ito stochastic equations \cite{Gardiner83a} for these
variables are used to give the same results for measured quantities as would
be obtained by solving the Fokker-Planck equation. Phase space theory for
bosons is now a standard approach, described in many text-books (see for
example \cite{Barnett97a}, \cite{Gardiner91a}). For bosonic systems where
the mode annihilation and creation operators satisfy commutation relations,
the phase space variables are c-numbers. However, in fermionic systems these
operators satisfy anti-commutation rules, and one option for developing a
phase space theory for fermions is for the phase space variables to be
Grassmann variables - since these anti-commute in multiplication. For
fermions c-number phase space theories also exist, but these typically
involve linking pairs of mode operators with the phase space variables (see
Ref. \cite{Dalton16b} for details).

Phase space theory for fermions based on Grassmann phase space variables was
pioneered by Cahill and Glauber \cite{Cahill99a}, although this paper did
not treat dynamical evolution problems -- either time evolution under a
Liouville-von Neumann or master equation or temperature evolution under a
Matsubara equation. Evolution was treated in an important later contribution
by Plimak, Collett and Olsen \cite{Plimak01a}. Applications of Grassman
phase space theory to specific problems (such as in Refs. \cite%
{Scherbakov93a}, \cite{Anastopoulos00a}, \cite{Schresta05a}, \cite{Plimak09a}%
, \cite{Dalton13a}, \cite{Ghosh15a}) have also been made. The theory
presented in Ref. \cite{Dalton16b} is similar to that in Ref. \cite%
{Plimak01a} in using an un-normalised $B$ distribution function, for which
the drift term in the Fokker-Planck equation depends linearly on the phase
space variables. However, the Ito stochastic equations for stochastic
Grassmann variables obtained in Ref. \cite{Dalton16b} were based on a
different approach (developed from a treatment by Gardiner (see pp 95-96 in 
\cite{Gardiner83a}) for bosons)\textbf{\ }in which the phase space average
at any time of an arbitary function and stochastic average of same function
are required to coincide, and for this to occur the terms in the Ito
stochastic equations were related to the drift and diffusion terms in the
Fokker-Planck equation for distribution function. The stochastic equations
obtained in Ref. \cite{Plimak01a} are different, and are based on an ansatz
requiring the distribution function itself at a final (imaginary) time to be
obtained by taking the distribution function at the initial time, replacing
the phase variables by time dependent stochastic Grassmann variables related
via a stochastic c-number linear transformation to the original (time
independent) Grassmann phase variables, and then taking a stochastic average
of the stochastic distribution function multiplied by the determinent
associated with the stochastic c-number linear transformation. The
differences between the approaches in \cite{Plimak01a} and \cite{Dalton16b}
are fully discussed in the latter paper, along with the important issue of
showing how numerical calculations of Fock state populations etc can still
be carried out based on Grassmann phase space theory - without there being a
need to represent Grassmann variables on the computer.

In phase space theories based on fields rather than separate modes, the
density operator describing the quantum state is represented by a
distribution functional (functionals and their calculus are described in
many textbooks, for example Ref. \cite{Dalton15a})\ involving phase space
field functions that are associated with the field annihilation and creation
operators. This extension is important even for bosonic cases where although
most particles may occupy one or two condensate modes, there are sometimes
large numbers of modes that have non-zero occupancy, such as for spatial
squeezing in quantum optics \cite{Gatti97a}. Measureable quantities such as
quantum correlation functions and Fock state probabilities and coherences
are expressed as phase space functional integrals involving the distribution
functional and other functions of the phase space fields specific to the
measureable quantity. The evolution equation for the density operator is
replaced by a functional Fokker-Planck equation for the distribution
functional. In turn, the phase space fields are replaced by time dependent
stochastic fields, and Ito stochastic field equations for these stochastic
fields are used to give the same results for measured quantities as would be
obtained by solving the functional Fokker-Planck equation. Such phase space
distribution functional theories have been widely used in the bosonic case -
such as for photons \cite{Gatti97a}, where the field functions are c-number
fields. The paper by Steel et al \cite{Steele98a} sets out the theory for
bosonic gases, and there have been many further developments such as
including gauge fields \cite{Deuar02a}, hybrid theories \cite{Dalton07b}, 
\cite{Hoffmann08a}, \cite{Dalton11a}, \cite{Dalton12b} involving $P$ and $W$
distributions and double phase spaces, the last allowing for time dependent
mode functions. For fermionic systems c-number field based phase space
theories have also been formulated, involving c-number field functions
either associated with atomic spin field operators \cite{Graham70a} or pairs
of fermion field operators.

However, in developing a field based phase space theory for fermionic
systems, the present paper involves associating the fermion field
annihilation and creation operators directly with field functions - for
analogous reasons to the separate mode case. The theory involves two
independent field functions $\psi $\ and $\psi ^{+}$\ which are associated
with the field annihilation and creation operators respectively. Again, the
anti-commutation rules satisfied by the fermion field operators suggest
using field functions with similar anti-commutation properties, and hence
the present theory employs Grassmann fields rather than c-number fields.
Grassmann fields can be expanded in terms of c-number mode functions, but
now the expansion coefficients are Grassmann variables. It is important to
note however, that the field and mode approaches are interchangeable, as can
be shown via via mode expansions for the Grassmann fields. A Grassmann field
approach was also adopted for fermion problems by Plimak, Collett and Olsen 
\cite{Plimak01a}.

After associating the field annihilation and creation operators with
Grassmann field functions, we\ introduce the Grassmann distribution
functionals that replace the density operator, starting from the
distribution functions based on modal Grassmann phase space variables. The
distribution function for modal phase space variables is equivalent to the
distribution functional for phase space fields. Both $P$ and $B$
distribution functionals are considered. We then derive the expressions for
measurable quantities such as quantum correlation functions, Fock state
probabilities and coherences as Grassmann phase space functional integrals
involving the distribution functional and field functions specific to the
measureable quantity. A careful derivation is given of the functional
Fokker-Planck equation for the distribution functional starting from the
Fokker-Planck equation, with expressions for the drift and diffusion terms
in the functional Fokker-Planck equations related to the corresponding terms
in the original Fokker-Planck equation. The correspondence rules for the
field annihilation and creation operators are also established via mode
expansions. However, they then can be used to derive functional
Fokker-Planck equations directly from the evolution equation for the density
operator when the Hamiltonian is expressed in terms of field operators.
After defining the time dependent stochastic Grassmann fields that replace
the field functions, we then derive the Ito stochastic field equations,
giving explicit expressions for the classical and noise fields in terms of
the related quantities in the Ito stochastic equations based on modes. An
important result establishes the general relationship between the classical
and noise field terms in the Ito stochastic field equations and the drift
and diffusion terms in the functional Fokker-Planck equation. This
relationship is then used to determine the classical and noise terms in the
Ito stochastic field equations directly from the functional Fokker-Planck
equation without requiring mode expansions. For the $B$ distribution case -
also occuring in Plimak et al \cite{Plimak01a} - we show that the stochastic
Grassmann fields at later times are related linearly to these fields at an
earlier time, and where the linear relationships only involve (stochastic)
c-numbers and not Grassmann fields. This allows numerical calculations based
on stochastic Grassmann fields to be carried out. Important issues involved
in carrying out numerical calculations for Grassmann fields need to be
considered - for example the position dependence of the stochastic Grassmann
fields must be represented as points on a spatial grid, and in some cases
non-local Ito stochastic field equations must be solved.

Typical applications, first to a trapped Fermi gas of spin $1/2$ fermionic
atoms with zero range interactions, and second to multi-component Fermi
gases with non-zero range interactions are presented, showing that the Ito
stochastic field equations are local in these cases. For the spin $1/2$ case
we also show how simple solutions can be obtained both for the untrapped
case and for an optical lattice trapping potential.

The plan of this paper is as follows. In Section \ref{Section - Systems of
Identical Fermions} we introduce the field annihilation and creation
operators and a typical Hamiltonian for interacting fermions expressed in
terms of field operators. Fock states for fermion positions are described,
as well as Grassmann fields. In Section \ref{Section - Functional Phase
Space Theory} functional phase space theory for fermions is presented.
Section \ref{Section - Functional Fokker-Planck Equations} covers functional
Fokker-Planck equations and Section \ref{Section - Ito Stochastic Field
Equations} sets out the Ito Stochastic field equations, with a SubSection
dealing with numerical issues. The final Section \ref{Section - Application
to Fermion Field Model} presents applications, the first SubSection treating
a trapped Fermi gas of spin $1/2$ fermionic atoms with zero range
interactions and the second SubSection focuses on multi-component Fermi
gases with non-zero range interactions. The Appendices set out detailed
material too lengthy for the main body of the paper. Appendix \ref{Appendix
- Fermion Position States} describes fermion position states, Appendix \ref%
{Appendix - Grassmann Functional Calculus} sets out Grassmann functional
calculus, Appendix \ref{Appendix - Functional P Distribution} describes the
functional $P$ distribution. Derivations of the Ito stochastic field
equations and the functional Fokker-Planck equation for the application to
spin $1/2$ fermionic atoms are set out in Appendices \ref{Appendix - Ito
Stochastic Field Equations} and \ref{Appendix - Fermi Gas FFPE}. The
Appendices for the previous paper (Ref. \cite{Dalton16b}) also contain
useful background material - including a description of Grassmann algebra
and calculus (Appendix A).

A far more detailed account of Grassmann phase space theory for fermions,
covering both the separate mode and the field cases and the relationship to
c-number phase space theory for bosons is set out in the recently published
book by the present authors entitled \emph{Phase Space Methods for
Degenerate Quantum Gases} \cite{Dalton15a}. Readers are refered to this
textbook for a more complete treatment. Here we focus on presenting the most
essential results. \pagebreak

\section{Systems of Identical Fermions - Key Concepts and Notation}

\label{Section - Systems of Identical Fermions}

In this Section we set out the basic concepts and notation used in this
paper. More details on these items are set out in Ref. \cite{Dalton16b}.

\subsection{Modes, Annihilation and Creation Operators, Fock States.}

In this paper we consider system of identical fermions, which in
quantum-atom optics could be fermionic atoms or molecules. The fermions will
be associated with position and momentum operators describing their centre
of mass motion, but also may exist in different internal states such as
those with differing hyperfine quantum numbers. The approach used here is
that of second quantisation (see for example \cite{March67a}, \cite%
{Schweber61a}), which is based on the existence of \emph{modes} (or single
particle states) $\left\vert \phi _{i}\right\rangle $ - listed in a
prescribed order $1,2,..,i,..,n$ - which the identical fermions may occupy.
Modes associated with different internal states $\left\vert \,\alpha
\right\rangle $ may be designated $\left\vert \phi _{\alpha i}\right\rangle
=\left\vert \phi _{i}^{(\alpha )}\right\rangle \left\vert \,\alpha
\right\rangle $ in which $\left\vert \phi _{i}^{(\alpha )}\right\rangle $
describes the centre of mass motion, but for the moment we will ignore the
internal state since experiments can be carried out with all fermions in the
same internal state. The position representation of the mode $\left\vert
\phi _{i}\right\rangle $ is $\phi _{i}(\mathbf{r})=\left\langle \mathbf{%
r\,|\,}\phi _{i}\right\rangle $ are is referred to as the mode function. The
modes are chosen to be orthonormal and complete. 
\begin{eqnarray}
\int d\mathbf{r\,}\phi _{i}^{\ast }(\mathbf{r})\phi _{j}(\mathbf{r})
&=&\delta _{i;j}\qquad \left\langle \phi _{i}\mathbf{\,|\,}\phi
_{j}\right\rangle =\delta _{i;j}  \label{Eq.Orthonorm} \\
\sum_{i}\phi _{i}^{\ast }(\mathbf{r})\phi _{i}(\mathbf{r}^{\prime }) &=&%
\mathbf{\delta (r-r}^{\prime })\qquad \sum_{i}\left\vert \phi
_{i}\right\rangle \left\langle \phi _{i}\right\vert =\widehat{1}
\label{Eq.Completeness}
\end{eqnarray}

The Pauli exclusion principle precludes the existence of multi-fermion
states where more than one fermion occupies any particular mode.
Multi-fermion \emph{Fock states} (which are represented in first
quantisation as anti-symmetrised products of the occupied single particle
state) can be specified by just stating the number of fermions $\nu _{i}=0,1$
that occupy each mode $\left\vert \phi _{i}\right\rangle $. Fock states can
be written in second quantisation form by introducing fermion \emph{creation}
and \emph{annihilation operators}. These are denoted $\hat{c}_{i}^{\dagger }$%
, $\hat{c}_{i}$ and satisfy the anticommutation rules 
\begin{eqnarray}
&&\{\hat{c}_{i},\hat{c}_{j}^{\dagger }\}\equiv \hat{c}_{i}\hat{c}%
_{j}^{\dagger }+\hat{c}_{j}^{\dagger }\hat{c}_{i}=\delta _{ij},  \nonumber \\
&&\{\hat{c}_{i},\hat{c}\}=0=\{\hat{c}_{i}^{\dagger },\hat{c}_{j}^{\dagger
}\}.  \label{Eq.AntiCommRuleFermionOprsA}
\end{eqnarray}%
We note that $\hat{c}_{i}^{2}=(\hat{c}_{i}^{\dag })^{2}=\widehat{0}$ as may
also be seen from the anti-commutation rules. This similarity to the
Grassmann variables is the reason why Grassmann phase space variables are
chosen for fermions. The Fock states in second quantisation form are 
\begin{equation}
|\nu \rangle =(\hat{c}_{\phi _{1}}^{\dag })^{\nu _{1}}(\hat{c}_{\phi
_{2}}^{\dag })^{\nu _{2}}\cdots (\hat{c}_{\phi _{n}}^{\dag })^{\nu
_{n}}|0\rangle =\left\vert \Phi _{\{\nu \}}\right\rangle
\label{Eq.FermionBasisSet2}
\end{equation}%
where the occupancy notation $\nu \equiv \{\nu _{1},\nu _{2},..,\nu
_{i},..\nu _{n}\}$ includes all the modes. The vacuum state $|0\rangle $ is
an extra state introduced in second quantisation and represents the
situation in which no fermions are present. The convention in (\ref%
{Eq.FermionBasisSet2}) is that the $\hat{c}_{\phi _{i}}^{\dag }$ are
arranged in order with $\hat{c}_{\phi _{1}}^{\dag }$ on the left and $\hat{c}%
_{\phi _{n}}^{\dag }$ on the right, and where $(\hat{c}_{\phi _{i}}^{\dag
})^{0}=\hat{1}$, $(\hat{c}_{\phi _{i}}^{\dag })^{1}=\hat{c}_{\phi
_{i}}^{\dag }$. The modes are listed in the prescribed order to avoid
duplication of the Fock states. The Fock states are orthogonal and
normalised 
\begin{equation}
\langle \nu |\xi \rangle =\delta _{\nu _{1}\xi _{1}}\cdots \delta _{\nu
_{n}\xi _{n}}  \label{Eq.OrthonormBasisStates33}
\end{equation}%
\ and are eigenstates of the fermion \emph{number operator} $\widehat{N}%
=\dsum\limits_{i}\widehat{n}_{i}$, where $\widehat{n}_{i}=\hat{c}%
_{i}^{\dagger }\hat{c}_{i}$ is the number operator for mode $\left\vert \phi
_{i}\right\rangle $. It is easy to show from the anticommutation rules that
the eigenvalues for the fermionic $\widehat{n}_{i}$ are restricted to $0,1$.
We have 
\begin{eqnarray}
\widehat{N}\,|\nu \rangle &=&N\,|\nu \rangle  \label{Eq.NumberOpr} \\
N &=&\dsum\limits_{i}\nu _{i}  \label{Eq.FermionNumber}
\end{eqnarray}%
The Fock states form a basis for a Hilbert space, as we will see.

From the anticommutation rules we can then see that 
\begin{eqnarray}
\hat{c}_{i}|\nu _{1};\cdots ;0;\cdots ,\nu _{n}\rangle &=&0  \nonumber \\
\hat{c}_{i}|\nu _{1};\cdots ;1;\cdots ,\nu _{n}\rangle &=&(-1)^{\eta
_{i}}|\nu _{1};\cdots ;0;\cdots ,\nu _{n}\rangle  \label{Eq.FermionAnnihOprs}
\end{eqnarray}%
where $(-1)^{\eta _{i}}=+1$ or $-1$ according to whether there are an even
or odd number of modes listed preceding the mode $\left\vert \phi
_{i}\right\rangle $ which are occupied $(\eta _{i}=\sum_{j<i}\nu _{j})$.

\subsection{Quantum States and Super-selection Rules}

The general physical \emph{pure state} $|\Psi \rangle $ for $N$ identical
fermions can be written as a quantum superposition of the basis Fock states. 
\begin{eqnarray}
|\Phi \rangle _{N} &=&\sum_{\nu _{1}\cdots \nu _{n}}B_{N}(\nu )|\nu \rangle
\label{Eq.GenPhysPureState} \\
\sum_{\nu }\left\vert B_{N}(\nu )\right\vert ^{2} &=&1,  \label{Eq.Norm2}
\end{eqnarray}%
where the sum runs over the occupancy numbers $\nu \equiv \{\nu _{1}\cdots
\nu _{n}\}$, the $B_{N}(\nu )$ are complex coefficients and where $%
\sum_{i=1}^{n}\nu _{i}=N$. Note that the sum is only over occupation
numbers, subject to the constraint that the total occupancy is $N$. The last
equation ensures that the state is normalised. If we consider all the states 
$|\nu \rangle $ arranged in order of total occupancy $\sum_{i}\nu
_{i}=0,1,2,3\cdots $ a heirarchy of basis states for identical particle
systems with various total particle numbers $N=0,1,2\cdots $ can be listed.
It is then convenient to mathematically define a Hilbert space (\emph{Fock
space}) describing identical fermion systems with all possible total
particle numbers $N$ containing all superpositions of the form 
\begin{equation}
|\Phi \rangle =\sum_{N}C_{N}|\Phi \rangle _{N}=\sum_{\nu _{1}\cdots \nu
_{n}}B(\nu )|\nu \rangle  \label{Eq.NonPhysState}
\end{equation}%
where the $C_{N}$ are complex coefficients. In the last expression the
restriction $\sum_{i=1}^{n}\nu _{i}=N$ does not apply to the $B(\nu )$. Such
a state $|\Phi \rangle $ is not physical, as \emph{super-selection rules}
(SSR) \cite{Wick52a} do not allow superpositions of states with differing
total particle numbers (see \cite{Dalton14a}, \cite{Dalton16c} and
references therein for recent discussions on SSR). However states of this
form are useful mathematically, even though the pure physical states are
restricted to subspaces of this Hilbert space. An example of non-physical
states are the \emph{fermion coherent states} (see Ref. \cite{Dalton16b} for
their definition and properties), where not only are there a linear
combinations of Fock states with differing fermion number, but also the
expansion coefficients are Grassmann numbers instead of c-numbers.

However, the physical states for identical particle systems are not
restricted to pure states of the form (\ref{Eq.GenPhysPureState}). Even when
the number of particles $N$ is prescribed, for \emph{closed systems} there
are \emph{mixed states} described by a quantum density operator $\hat{\rho}%
_{N}$ of the form 
\begin{equation}
\hat{\rho}_{N}=\sum_{\nu }\sum_{\xi }\rho _{N}(\nu ,\xi )|\nu \rangle
\langle \xi |.  \label{Eq.NParticleDensityOpr}
\end{equation}%
rather than a state vector. The complex coefficients $\rho _{N}(\nu ,\xi )$
are the density matrix elements. The requirements that the density operator
is Hermitian, $\hat{\rho}_{N}=\hat{\rho}_{N}^{\dag }$, has unit trace $%
\mbox{Tr}(\hat{\rho}_{N})=1$, and for a mixed state satisfies the condition $%
\mbox{Tr}(\hat{\rho}_{N}^{2})<1$, lead to well-known constraints on the
density matrix elements. For pure states we can write $\hat{\rho}_{N}=|\Phi
\rangle _{N}\langle \Phi |_{N}$ and $\mbox{Tr}\hat{\rho}_{N}^{2}=1$ and in
this case the density matrix elements are $\rho _{N}(\nu ,\xi )=B_{N}(\nu
)B_{N}^{\ast }(\xi ).$

For \emph{open systems} (which particles can enter or leave), physical
states can be prepared in which the number of identical fermions is not
prescribed. A generalisation of the system state is required to incorporate
such cases where the system composition is indefinite. Such states are mixed
and also described by a density operator $\hat{\rho}$. However,
super-selection rules require that the density operator can only be of the
form 
\begin{equation}
\hat{\rho}=\sum_{N}f_{N}\hat{\rho}_{N}  \label{Eq.GeneralDensityOpr}
\end{equation}%
where $\sum_{N}f_{N}=1,f_{N}\geq 0$, which only involves component density
operators $\hat{\rho}_{N}$ for systems with specified total fermion numbers $%
N$, each weighted by real, positive $f_{N}$. The density operator satisfies
the earlier requirements for mixed states. Pure states have only one $%
f_{N}=1 $ and all others vanish. More general density operators such as 
\begin{equation}
\hat{\rho}=\sum_{N}\sum_{M}\rho _{N,M}|\Phi \rangle _{N}\langle \Phi |_{M}
\label{Eq.NonPhysicalDensityOpr}
\end{equation}%
with non-zero $\rho _{N,M}$ do not represent physical states, even though
they are mathematical operators in the general mathematical Hilbert space.
The physically allowed density operators both for pure or mixed states all
commute with the number operator%
\begin{equation}
\lbrack \widehat{N},\hat{\rho}]=0  \label{Eq.PhysicalDensOpr}
\end{equation}

As pointed out in Ref. \cite{Dalton16b}, the super-selection rule leads to
restrictions on the phase space distribution function, and a brief
discussion of its origin is presented in Ref. \cite{Dalton16b}.

\subsection{Field Operators and Hamiltonian}

In systems which contain a large number of modes, the distribution function
treatment becomes unwieldy and a switch to a treatment avoiding a
consideration of separate modes is highly desirable. The system is then
described in terms of \emph{field} \emph{operators} $\hat{\Psi}^{\dag }(%
\mathbf{r}),\hat{\Psi}(\mathbf{r})$, where $\mathbf{r}$ is the particle
position. The field operators may be defined via mode expansions with the
mode annihilation and creation operators being the expansion coefficients 
\begin{equation}
\hat{\Psi}(\mathbf{r)}=\sum_{i}\hat{c}_{_{i}}\phi _{i}(\mathbf{r}),\hspace{%
0.3cm}\hat{\Psi}^{\dag }(\mathbf{r})=\sum_{i}\hat{c}_{i}^{\dag }\phi
_{i}^{\ast }(\mathbf{r}).  \label{Eq.FermionFieldOprs2}
\end{equation}%
These operators are associated with the creation, destruction of fermionic
particles at particular positions. For the field operators our fundamental
discrete mode anticommutation relations are replaced by 
\begin{eqnarray}
\{\hat{\Psi}(\mathbf{r}),\hat{\Psi}^{\dag }(\mathbf{r}^{\prime })\}
&=&\delta (\mathbf{r}-\mathbf{r}^{\prime })  \nonumber \\
\{\hat{\Psi}(\mathbf{r}),\hat{\Psi}(\mathbf{r}^{\prime })\} &=&0=\{\hat{\Psi}%
^{\dag }(\mathbf{r}),\hat{\Psi}^{\dag }(\mathbf{r}^{\prime })\}
\label{Eq.FermiFieldCommRulesA}
\end{eqnarray}%
which follow from Eq. (\ref{Eq.AntiCommRuleFermionOprsA}) and the mode
completeness result Eq. (\ref{Eq.Completeness}).

The field creation operators create a particle at a particular position in
the given internal state. For a set of distinct particle positions $\mathbf{r%
}_{1},\mathbf{r}_{2}\cdots \mathbf{r}_{N}$, where to avoid duplication an
ordering convention is invoked, so that $\mathbf{r}_{1}<\mathbf{r}%
_{2}<\cdots <\mathbf{r}_{N}$. The Fock state $|\mathbf{r}_{1}\cdots \mathbf{r%
}_{N}\rangle $ which has one fermion at each of the positions $\mathbf{r}%
_{1},\mathbf{r}_{2}\cdots \mathbf{r}_{N}$ is given by 
\begin{equation}
|\mathbf{r}_{1}\cdots \mathbf{r}_{N}\rangle =\hat{\Psi}^{\dag }(\mathbf{r}%
_{1})\cdots \hat{\Psi}^{\dag }(\mathbf{r}_{N})|0\rangle
\label{Eq.MultiFermionPositionState}
\end{equation}%
where the normalisation and orthogonality conditions for the position states
involve delta functions 
\begin{equation}
\left\langle \mathbf{r}_{1}\cdots \mathbf{r}_{N}|\,\mathbf{s}_{1}\cdots 
\mathbf{s}_{N}\right\rangle =\delta (\mathbf{r}_{1}-\mathbf{s}_{1})\cdots
\delta (\mathbf{r}_{N}-\mathbf{s}_{N}).
\end{equation}%
The proof of these results is given in Appendix \ref{Appendix - Fermion
Position States}.

The vacuum state $\left\vert 0\right\rangle $, which contains no particles
in any mode, is associated with the \emph{vacuum projector} $\left\vert
0\right\rangle \left\langle 0\right\vert $. This projector can be written
using \emph{normally ordered} forms of products of exponential operators
based on number operators. The definition of normal ordering is set out in
Ref. \cite{Dalton16b}. We have\textbf{\ } 
\begin{eqnarray}
\left\vert 0\right\rangle \left\langle 0\right\vert &=&\mathcal{N}\left(
\prod\limits_{i}\exp (-\hat{c}_{i}^{\dag }\hat{c}_{i})\right)  \nonumber \\
&=&\mathcal{N}\left[ \exp \left( -\sum\limits_{i}\hat{c}_{i}^{\dag }\hat{c}%
_{i}\right) \right] =\mathcal{N}\left[ \exp \left( -\dint d\mathbf{r}\,\hat{%
\Psi}^{\dag }(\mathbf{r})\hat{\Psi}(\mathbf{r})\right) \right]  \nonumber \\
&&
\end{eqnarray}%
in terms of mode or field operators .

Field annihilation and creation operators do not represent physical
quantities and have the effect of changing $N$ particle physical states into 
$N\pm 1$ particle physical states. Physical quantities can be constructed
from the field operators, however, which always have the same number of
creation and annihilation operators. For example, in terms of field
operators a typical \emph{Hamiltonian }for a fermion system with spin $\frac{%
1}{2}$ fermions of mass $m$ is 
\begin{eqnarray}
\hat{H}_{f}\mbox{\rule{-0.5mm}{0mm}} &=&\mbox{\rule{-1mm}{0mm}}\int %
\mbox{\rule{-1mm}{0mm}}d\mathbf{r}\left( \sum_{\alpha }\frac{\hbar ^{2}}{2m}%
\nabla \hat{\Psi}_{\alpha }(\mathbf{r})^{\dag }\mbox{\rule{-0.5mm}{0mm}}%
\cdot \mbox{\rule{-0.5mm}{0mm}}\nabla \hat{\Psi}_{\alpha }(\mathbf{\ r}%
)+\sum_{\alpha }\hat{\Psi}_{\alpha }(\mathbf{r})^{\dag }V_{\alpha }\hat{\Psi}%
_{\alpha }(\mathbf{r})\right.  \nonumber \\
&&\hspace{2cm}\left. +\frac{g_{f}}{2}\sum_{\alpha }\hat{\Psi}_{\alpha }(%
\mathbf{r})^{\dag }\hat{\Psi}_{-\alpha }(\mathbf{r})^{\dag }\hat{\Psi}%
_{-\alpha }(\mathbf{r})\hat{\Psi}_{\alpha }(\mathbf{r})\right)
\label{Eq.HamFermionFields}
\end{eqnarray}%
where $\alpha =-\frac{1}{2},+\frac{1}{2}$ lists two internal states with $%
M_{f}=-\frac{1}{2},+\frac{1}{2}$ and the zero range approximation is used
for interactions between the particles and $V_{f}$ is the trapping
potential. The field operators are now generalised to allow for internal
states. The fermion-fermion interaction is required to be between fermions
of opposite spin because $\hat{\Psi}_{\alpha }^{2}(\mathbf{r})=0$. For
bosons strong $s$ wave interactions occur, but for fermions only weak $p$
wave collisions occur unless the fermions have opposite spin, because the
Pauli principle prevents two fermions with the same spin from being at the
same position. Typical Hamiltonians for interacting fermions can also be
expressed in terms of mode annihilation and creation operators (see Eqs.
(37), (38) and (39) in Ref. \cite{Dalton16b}).

\subsection{Grassmann Fields}

In the case of bosons, the field operators $\hat{\Psi}(r\mathbf{),}\hat{\Psi}%
^{\dag }(r\mathbf{)}$ are associated with \emph{c-number fields }$\psi
(r),\psi ^{+}(r)$. Here $r$ refers to spatial position - which may be in 1D,
2D or 3D. Such fields may be expanded in terms of orthonormal mode functions 
$\phi _{i}(r)$ in the form 
\begin{eqnarray}
\psi (r) &=&\sum\limits_{i}\alpha _{i}\,\phi _{i}(r)
\label{Eq.BosonFieldFn2} \\
\psi ^{+}(r) &=&\sum\limits_{i}\phi _{i}^{\ast }(r)\alpha _{i}^{+}.
\label{Eq.BosonFieldFn3}
\end{eqnarray}%
where $\alpha _{i}$, $\alpha _{i}^{+}$ are c-number phase space variables.
The mode functions satisfy orthonormality and completeness conditions
analogous to (\ref{Eq.Orthonorm}), (\ref{Eq.Completeness})

For fermions, the field operators $\hat{\Psi}(r\mathbf{),}\hat{\Psi}^{\dag
}(r\mathbf{)}$ are associated with \emph{Grassmann fields }$\psi (r),\psi
^{+}(r)$. Grassmann fields can also be expanded in terms of a suitable
orthonormal set of mode functions but now with Grassmann phase space
variables $g_{k}$ or $g_{k}^{+}$ as expansion coefficients. Thus we have for
Grassmann functions $\psi (r),\psi ^{+}(r)$ associated with field
annihilation and creation operators 
\begin{eqnarray}
\psi (r) &=&\sum\limits_{i}g_{i}\,\phi _{i}(r)  \label{Eq.GrassmannFieldFn2}
\\
\psi ^{+}(r) &=&\sum\limits_{i}\phi _{i}^{\ast }(r)g_{i}^{+}.
\label{Eq.GrassmannFunction3}
\end{eqnarray}%
Again, if $g_{i}^{+}\,=g_{i}^{\ast }\,$\ then $\psi ^{+}(r)=\psi ^{\ast }(r)$%
, the complex conjugate Grassmann field.

The Grassmann fields are odd \emph{Grassmann functions} of the first order.
The following results for the expansion coefficients can easily be obtained 
\begin{eqnarray}
g_{k}\, &=&\int dr\,\phi _{k}^{\ast }(r)\psi (r)
\label{Eq.GrassmannExpnCoefts1} \\
g_{k}^{+}\, &=&\int dr\,\phi _{k}(r)\psi ^{+}(r)
\label{Eq.GrassmannExpnCoefts2}
\end{eqnarray}%
which has the same form as for c-number fields.

Grassmann fields differ from bosonic fields in that they anti-commute and
their square and higher powers are zero. Thus with $\eta
(r)=\sum\limits_{l}h_{l}\,\phi _{l}(r)$ a second Grassmann field%
\begin{eqnarray}
\psi (r)\eta (s)+\eta (s)\psi (r) &=&0 \\
(\psi (r))^{2} &=&(\psi (r))^{3}=\cdots =0.
\end{eqnarray}%
These rules follow from the anti-commutation properties of Grassman numbers.
\pagebreak

\section{Functional Phase Space Theory}

\label{Section - Functional Phase Space Theory}

In systems with a large number of fermions the Pauli exclusion principle
implies there must also be a large number of mode involved. In this case the
distribution function treatment based on treating mode separately becomes
unwieldy and a switch to a treatment avoiding a consideration of separate
modes is highly desirable. Field annihilation and creation \ operators now
replace the separate mode operators. In \emph{functional phase space theory}
the density operator is now represented by a \emph{distribution functional}
rather than a distribution function, and the field operators are represented
by \emph{Grassmann fields}, rather than mode operators being represented by
Grassmann phase space variables. It should be noted though that the
distribution functional is entirely \emph{equivalent} to the distribution
function and the two approaches are inter-changeable. If the Grassmann
fields are expanded in terms of spatial mode functions with the Grassmann
phase space variables as coefficients, then the distribution functional is
equivalent to a Grassmann function of the phase space variables. We simply%
\emph{\ choose }the distribution functional to be such that the equivalent
distribution function is the \emph{same} as the previous distribution
function. As previously, distribution functionals of the $P$ type and the
un-normalised $B$ type both occur, though here we focus on the latter
because of its greater usefulness in numerical work. When using functional
phase space theory \emph{functional Fokker-Planck} equations now replace
Fokker-Planck equations and can be \emph{derived} either from the
Fokker-Planck equations or more simply by establishing \emph{correspondence
rules} involving the Grassmann fields and functional derivatives that are
equivalent to the original correspondence rules that involved Grassmann
phase space variables and Grassmann derivatives. Similarly \emph{Ito
stochastic field} equations for \emph{Grassman stochastic fields} replace
Ito stochastic equations for stochastic Grassmann phase variables and can be 
\emph{derived} from these. The Ito stochastic field equations can also be
obtained \emph{more directly} from the drift and diffusion terms in the
functional Fokker-Planck equation \ As functional calculus involving
Grassmann fields is not widely known, for convenience its key features are
set out in Appendix \ref{Appendix - Grassmann Functional Calculus}.
\smallskip

\subsection{Grassmann Functionals - Basic Idea}

The concept of a functional $F[\psi (r)]$ is well-established in the case of
c-number fields. Essentially, a functional $F[\psi (r)]$ maps a c-number
field $\psi (r)$ onto a c-number, which depends on all the values of $\psi
(x)$ over its entire range of $r$. There has been a widespread use of
functionals both in classical and quantum physics (see Ref. \cite{Dalton16b}
for background references), so their properties and calculus involving
functional derivatives and integrals will not be presented here. A recent
textbook Ref. \cite{Dalton15a} outlines these key aspects of c-number
functionals.

It is important to note that as the value of the function at any point in
the range for $x$ is determined uniquely by the expansion coefficients $%
\{\alpha _{k}\}$, then the functional $F[\psi (r)]$ must therefore also just
depend on the c-number expansion coefficients, and hence may also be viewed
as a c-number function $f(\alpha _{1},\cdots ,\alpha _{k},\cdots \alpha
_{n}) $ of the expansion coefficients. This equivalence is useful when
functional differentiation and integration are considered.%
\begin{equation}
F[\psi (r)]\equiv f(\alpha _{1},\cdots ,\alpha _{k},\cdots \alpha _{n})
\end{equation}%
This feature also applies to functionals of the form $F[\psi (r),\psi
^{+}(r)]$, which again are equivalent to a c-number function of both sets of
expansion coefficients.

\begin{equation}
F[\psi (r),\psi ^{+}(r)]\equiv f(\alpha _{1},\cdots ,\alpha _{k},\cdots
\alpha _{n},\alpha _{1}^{+},\cdots ,\alpha _{k}^{+},\cdots \alpha _{n}^{+})
\end{equation}

The idea of a functional $F[\psi (r)]$ can be extended to cases where $\psi
(r)$ is a Grassmann field rather than a c-number field. Analogous to the
situation for c-number functionals, a \emph{Grassmann functional} $F[\psi
(r)]$ maps the g-number function $\psi (r)$ onto a Grassmann function that
depends on \emph{all} the values of $\psi (r)$ over its entire range.
Examples of Grassmann functionals are given in Appendix \ref{Appendix -
Grassmann Functional Calculus}. Grassmann \emph{functional} \emph{derivatives%
} and Grassmann \emph{functional integrals} can be defined for Grassmann
functionals, following analogous definitions for c-number functionals. These
definitions and associated rules for these processes are outlined in
Appendix \ref{Appendix - Grassmann Functional Calculus}.

As for c-number fields, the expansion coefficients $\{g_{k}\}$ determine the
Grassmann field at any point in the range for $x$, so the functional $F[\psi
(r)]$ must therefore also just depend on the g-number expansion
coefficients. Hence the Grassmann functional $F[\psi (r)]$ is equivalent to
a Grassmann function $f(g_{1},\cdots ,g_{k},\cdots g_{n})$ of the expansion
coefficients. This equivalence is used to relate Grassmann functional
differentiation and integration to ordinary Grassmann differentiation and
Grassmann integration with respect to the Grassmann phase space variables.%
\begin{equation}
F[\psi (r)]\equiv f(g_{1},\cdots ,g_{k},\cdots g_{n})
\label{Eq.GrassmannEquivFunction}
\end{equation}%
This feature also applies to functionals of the form $F[\psi (r),\psi
^{+}(r)]$, which are also equivalent to a Grassmann function of all the
expansion coefficients.

\begin{equation}
F[\psi (r),\psi ^{+}(r)]\equiv f(g_{1},\cdots ,g_{k},\cdots
g_{n},g_{1}^{+},\cdots ,g_{k}^{+},\cdots g_{n}^{+})
\label{Eq.GrassmannEquivFunction2}
\end{equation}

This result is of central importance in the present paper, where we consider
un-normalised $B$ Grassmann distribution functionals of fields $\psi
(r),\psi ^{+}(r)$. The Grassmann functional $B[\psi (r),\psi ^{+}(r)]$ is 
\emph{equivalent} to the previous un-normalised $B$ Grassmann distribution
function $B(g_{k},g_{k}^{+})$ of the Grassmann phase space variables $%
g_{k},g_{k}^{+}$ treated in Ref. \cite{Dalton16b} 
\begin{equation}
B[\psi (r),\psi ^{+}(r)]\equiv B(g_{k},g_{k}^{+})
\label{Eq.BDistnFnalBDistFnEquiv}
\end{equation}%
The same symbol will be used for both, but note the $[..]$ for functionals, $%
(..)$ for functions. There is a similar equivalence for normalised $P$
Grassmann distribution functionals and functions. Essentially, the
distribution functional $B[\psi (r),\psi ^{+}(r)]$ and the distribution
function $B(g_{k},g_{k}^{+})$ are just two different ways of representing
the same density operator. This equivalence enables us to use results
previously obtained for the separate mode theory in Paper I (Ref. \cite%
{Dalton16b}) to derive equivalent results for the field theory approach
outlined in the present paper. \smallskip

\subsection{Field Operators - New Notation}

It is now useful to change the notation to both allow for the presence of
differing internal states $\left\vert \alpha \right\rangle $ for the
fermions and to treat field annihilation and creation operators and their
mode expansions via unified expressions. The field annihilation, creation
operators $\widehat{\psi }_{\alpha }(r),\widehat{\psi }_{\alpha }^{\dag }(r)$
for internal state $\left\vert \alpha \right\rangle $ will now be denoted $%
\widehat{\psi }_{\alpha A}(r)$, with $A=1,2$ distinguishing annihilation and
creation operators and $r$ denoting the spatial position The expansion in
terms of mode annihilation, creation operators $\widehat{c}_{\alpha i},%
\widehat{{\small c}}_{\alpha i}^{\dag }$ (denoted $\widehat{c}_{\alpha
i}^{A} $ with $A=1,2$ for annihilation and creation operators) and
orthnormal mode functions and their conjugates $\phi _{\alpha i}(r),\phi
_{\alpha i}^{\ast }(r)$ now denoted $\xi _{\alpha i}^{A}(r)$ with $A=1,2$
will now be written as%
\begin{equation}
\widehat{\psi }_{\alpha A}(r)=\dsum\limits_{i}\widehat{c}_{\alpha
i}^{A}\;\xi _{\alpha i}^{A}(r)  \label{Eq.FermionFieldOprExpn}
\end{equation}%
\smallskip

\subsection{Distribution Functionals - $B$ Distribution}

To formulate functional phase space theory we first associate the field
annihilation and creation operators $\widehat{\psi }_{\alpha }(r),\widehat{%
\psi }_{\alpha }^{\dag }(r)$ for each internal state $\left\vert \alpha
\right\rangle $ associated with \emph{Grassmann field functions} $\psi
_{\alpha }(r),\psi _{\alpha }^{+}(r)$, which in accord with the new notation
will be denoted $\psi _{\alpha A}(r)$ with $A=1,2$. These can be expanded in
terms of mode functions $\xi _{\alpha i}^{A}(r)$ with Grassmann phase
variables $g_{\alpha i},g_{\alpha i}^{+}$ (denoted $g_{\alpha i}^{A}$ with $%
A=1,2$ and $i=1,2,..,n_{\alpha }$) for each of the $n_{\alpha }$ spatial
modes associated with this internal state acting as the expansion
coefficients%
\begin{equation}
\psi _{\alpha A}(r)=\dsum\limits_{i}g_{\alpha i}^{A}\;\xi _{\alpha i}^{A}(r)
\label{Eq.FermionFieldFnExpn}
\end{equation}%
Clearly these are linear and odd Grassmann fuctions, and the Grassmann
fields are equivalent to the Grassmann phase variables.

If we now denote $\psi (r)\equiv \{\psi _{\alpha }(r),\psi _{\alpha
}^{+}(r)\}$ and $g\equiv \{g_{\alpha i},g_{\alpha i}^{+}\}$ we then
represent the density operator $\widehat{\rho }$ by a $B$ distribution
functional $B[\psi (r)]$ of the Grassmann fields. As explained above, the
distribution functional $B[\psi (r)]$ is equivalent to the previous
distribution function $B(g)$. 
\begin{equation}
B[\psi (r)]\equiv B(g)  \label{Eq.BDistnFnalBDistFnEquiv2}
\end{equation}

Similarly, a $P$ distribution functional $P[\psi (r)]$ is equivalent to the
previous $P$ distribution function $P(g)$. The two distribution functionals
are related via 
\begin{equation}
P[\psi (r)]=B[\psi (r)]\,\exp (+\dint dx\,\psi (r)\psi ^{+}(r))
\label{Eq.BandPDistnFnalRelnA}
\end{equation}%
as can easily be established via mode expansions. \smallskip

\subsection{Quantum Correlation Functions}

The previous normally ordered quantum correlation functions based on
separate modes (see Eq. (27) in Ref. \cite{Dalton16b}) now involve the field
creation and annihilation operators $\hat{\Psi}^{\dag }(\mathbf{r}),\hat{\Psi%
}(\mathbf{r})$ and are of the form 
\begin{eqnarray}
&&G^{(p,q)}(\mathbf{r}_{1}\mathbf{,r}_{2}\cdots \mathbf{r}_{p};\mathbf{s}%
_{q}\cdots \mathbf{s}_{2},\mathbf{s}_{1})  \nonumber \\
&=&\langle \hat{\Psi}(\mathbf{r}_{1})^{\dag }\cdots \hat{\Psi}(\mathbf{r}%
_{p})^{\dag }\hat{\Psi}(\mathbf{s}_{q})\cdots \hat{\Psi}(\mathbf{s}%
_{1})\rangle  \nonumber \\
&=&\mbox{Tr}(\hat{\Psi}(\mathbf{s}_{q})\cdots \hat{\Psi}(\mathbf{s}_{1})\,%
\hat{\rho}\,\hat{\Psi}(\mathbf{r}_{1})^{\dag }\cdots \hat{\Psi}(\mathbf{r}%
_{p})^{\dag }).\quad  \label{Eq.QuantumCorrelnFnGeneralA}
\end{eqnarray}%
where for an $N$ particle system we require $p,q\leq N$ to give a non-zero
result. For the fermion case we also require $p-q=0,\pm 2,\pm 4,\cdots $ due
to the SSR. As we will see, this result can be written in terms of phase
space functional integrals using the general result relating Grassmann
functional integrals and Grassmann phase space integrals in Eq. (\ref%
{Eq.GrassmannPhaseSpaceIntegral4}) in Appendix \ref{Appendix - Grassmann
Functional Calculus}. 
\begin{eqnarray}
&&\int D\psi ^{+}\,D\psi \,F[\psi (r),\psi ^{+}(r)]  \nonumber \\
&=&\lim_{n\rightarrow \infty }\lim_{\epsilon \rightarrow 0}\int \cdots \int
dg_{n}^{+}\cdots dg_{k}^{+}\cdots dg_{1}^{+}dg_{n}\cdots dg_{k}\cdots
dg_{1}\,  \nonumber \\
&\times &f(g_{1},\cdots ,g_{k},\cdots g_{n},g_{1}^{+},\cdots
,g_{k}^{+},\cdots g_{n}^{+})  \label{Eq.GrassmannPhaseSpaceIntegral4B}
\end{eqnarray}%
where the functional $F[\psi (r),\psi ^{+}(r)]$ and the Grassmann function $%
f(g,g^{+})$ are equivalent.

If we substitute mode expansions (\ref{Eq.FermionFieldOprs2}) for the field
operators then using (\ref{Eq.QuantumCorrelnFnGeneralA}) the quantum
correlation functions can be expressed in terms of the $P$ distribution
functional as follows%
\begin{eqnarray}
&&G^{(p,q)}(\mathbf{r}_{1}\mathbf{,r}_{2}\cdots \mathbf{r}_{p};\mathbf{s}%
_{q}\cdots \mathbf{s}_{2},\mathbf{s}_{1})  \nonumber \\
&=&\sum\limits_{m_{q}\cdots m_{1}}\sum\limits_{l_{1}\cdots l_{p}}\phi
_{m_{q}}(\mathbf{s}_{q})\cdots \phi _{m_{1}}(\mathbf{s}_{1})\phi
_{l_{1}}^{\ast }(\mathbf{r}_{1})\cdots \phi _{l_{p}}^{\ast }(\mathbf{r}_{p})%
\mbox{Tr}(\hat{c}_{m_{q}}\cdots \hat{c}_{m_{1}}\hat{\rho}(t)\hat{c}%
_{l_{1}}^{\dag }\cdots \hat{c}_{l_{p}}^{\dag }).  \nonumber \\
&=&\sum\limits_{m_{q}\cdots m_{1}}\sum\limits_{l_{1}\cdots l_{p}}\phi
_{m_{q}}(\mathbf{s}_{q})\cdots \phi _{m_{1}}(\mathbf{s}_{1})\phi
_{l_{1}}^{\ast }(\mathbf{r}_{1})\cdots \phi _{l_{p}}^{\ast }(\mathbf{r}_{p})
\nonumber \\
&&\times \int
\prod\limits_{i}dg_{i}^{+}\prod\limits_{i}dg_{i}\,(g_{m_{q}}\cdots
g_{m_{2}}g_{m_{1}})P(g,g^{+})(g_{l_{1}}^{+}g_{l_{2}}^{+}\cdots g_{l_{p}}^{+})
\nonumber \\
&=&\int \prod\limits_{i}dg_{i}^{+}\prod\limits_{i}dg_{i}\,(\psi (\mathbf{s}%
_{q})\cdots \psi (\mathbf{s}_{1}))\,P(g,g^{+})(\psi ^{+}(\mathbf{r}%
_{1})\cdots \psi ^{+}(\mathbf{r}_{p}))  \nonumber \\
&=&\int \int D\psi ^{+}D\psi \,\psi (\mathbf{s}_{q})\cdots \psi (\mathbf{\ s}%
_{1})P[\psi (r),\psi ^{+}(r)]\psi ^{+}(\mathbf{r}_{1})\cdots \psi ^{+}(%
\mathbf{r}_{p}))  \nonumber \\
&&  \label{Eq.QCorrFnFermiFunctionalIntegral}
\end{eqnarray}%
where in the last step we have replaced the Grassmann phase space integral
by the equivalent Grassmann functional integral via (\ref%
{Eq.GrassmannPhaseSpaceIntegral4B}). Hence we see that the quantum
correlation functions involving fermion field operators are given as
Grassmann functional integrals in which the Grassmann fields have replaced
the field operators. \smallskip

\subsection{Fock State Populations and Coherences}

We consider fermion position eigenstates as in Eq.(\ref%
{Eq.MultiFermionPositionState}) 
\begin{eqnarray}
\left\vert \Phi \{\mathbf{r}\}\right\rangle &=&|\mathbf{r}_{1}\cdots \mathbf{%
r}_{p}\rangle =\hat{\Psi}_{f}^{\dag }(\mathbf{r}_{1})\cdots \hat{\Psi}%
_{f}^{\dag }(\mathbf{r}_{p})|0\rangle  \nonumber \\
\left\vert \Phi \{\mathbf{s}\}\right\rangle &=&|\mathbf{s}_{1}\cdots \mathbf{%
s}_{p}\rangle =\hat{\Psi}_{f}^{\dag }(\mathbf{s}_{1})\cdots \hat{\Psi}%
_{f}^{\dag }(\mathbf{s}_{p})|0\rangle  \label{Eq.FermionPositionEigesntates}
\end{eqnarray}%
The population for state $\left\vert \Phi \{\mathbf{r}\}\right\rangle $ and
the coherence between the state $\left\vert \Phi \{\mathbf{r}\}\right\rangle 
$ and the state $\left\vert \Phi \{\mathbf{s}\}\right\rangle $ are given by%
\textbf{\ } 
\begin{eqnarray}
P(\Phi \{\mathbf{r}\}) &=&Tr(\left\vert \Phi \{\mathbf{r}\}\right\rangle
\left\langle \Phi \{\mathbf{r}\}\right\vert \,\hat{\rho})
\label{Eq.FermiPositionPopn} \\
C(\Phi \{\mathbf{s}\};\Phi \{\mathbf{r}\}) &=&Tr(\left\vert \Phi \{\mathbf{s}%
\}\right\rangle \left\langle \Phi \{\mathbf{r}\}\right\vert \hat{\rho})
\label{Eq.FermionPositionCoherence}
\end{eqnarray}

Substituting for the field operators from Eqs.(\ref{Eq.FermionFieldOprs2})
and using the results (30) and (31) in Ref. \cite{Dalton16b} we see that for
the fermion position probability can be expressed in terms of the $B$
distribution functional as 
\begin{eqnarray}
P(\Phi \{\mathbf{r}\}) &=&\sum\limits_{l_{1},\cdots
l_{p}}\sum\limits_{m_{1},\cdots m_{p}}\phi _{l_{1}}^{\ast }(\mathbf{r}%
_{1})\cdots \phi _{l_{p}}^{\ast }(\mathbf{r}_{p})\,\phi _{m_{p}}(\mathbf{r}%
_{p})\cdots \phi _{m_{1}}(\mathbf{r}_{1})  \nonumber \\
&&\times \int d\mathbf{g}^{+}d\mathbf{g}\,\,\,g_{m_{p}}\cdots
g_{m_{1}}\,B(g,g^{+})\,g_{l_{1}}^{+}\cdots g_{l_{p}}^{+}  \nonumber \\
&=&\int D\psi ^{+}D\psi \,\psi (\mathbf{r}_{p})\cdots \psi (\mathbf{r}%
_{1})\,B[\psi (r),\psi ^{+}(r)]\,\psi ^{+}(\mathbf{r}_{1})\cdots \psi ^{+}(%
\mathbf{r}_{p})  \nonumber \\
&&  \label{Eq.FermionPositionPopnResult}
\end{eqnarray}%
and similarly for the fermion position coherence%
\begin{eqnarray}
C(\Phi \{\mathbf{s}\};\Phi \{\mathbf{r}\}) &=&\int D\psi ^{+}D\psi \,\psi (%
\mathbf{r}_{p})\cdots \psi (\mathbf{r}_{1})B[\psi (r),\psi ^{+}(r)]\,\psi
^{+}(\mathbf{s}_{1})\cdots \psi ^{+}(\mathbf{s}_{p})  \nonumber \\
&&  \label{Eq.FermionPositionCoherenceResult}
\end{eqnarray}%
where the phase space integrals have been converted into functional
integrals involving the fermion $B$ distribution functional. These results
are useful for discussing similtaneous position measurements - note the
probability density factors such as\textbf{\ }$\psi ^{+}(\mathbf{r}_{1})\psi
(\mathbf{r}_{1})$ - and for discussing spatial coherence effects in systems
such as Fermi gases. In both cases the result is given as a functional
average of products of field functions, rather similar to equivalent
classical formulae.\smallskip

\subsection{Characteristic Functional}

For completeness, we also define the characteristic functional $\chi \lbrack
h(r),h^{+}(r)]$ via 
\begin{eqnarray}
&&\chi \lbrack h(r),h^{+}(r)]=\mbox{Tr}(\hat{\Omega}^{+}[h^{+}(r)]\,\,\hat{%
\rho}\,\hat{\Omega}^{-}[h(r)])  \label{Eq.FermiCharFnal} \\
\hat{\Omega}^{+}[h^{+}(r)] &=&\exp i\int dr\,\hat{\Psi}(r)h^{+}(r)=1+i\int
dr\,\hat{\Psi}(r)h^{+}(r)  \nonumber \\
\hat{\Omega}^{-}[h(r)] &=&\exp i\int dx\,h(r)\hat{\Psi}^{\dag }(r)=1+i\int
dr\,h(r)\hat{\Psi}^{\dag }(r)  \label{Eq.FermiCharFnal2}
\end{eqnarray}%
where for the field operators $\hat{\Psi}(r)$ and $\hat{\Psi}^{\dag }(r)$ we
associate a pair of Grassmann fields $h^{+}(r)$ and $h(r)$. Note that $%
h,h^{+}$ are both complex, and are not related to each other. Note that $%
\chi \lbrack h(r),h^{+}(r)]$ only depends on the $h(r),h^{+}(r)$ and not on
their complex conjugates $h^{\ast }(r),h^{+\ast }(r)$. The simplification of
the exponentials follows from second and higher powers of the field
operators being zero.

The characteristic functional $\chi \lbrack h(r),h^{+}(r)]$ is related to
the $P$ distribution functional $P[\psi (r),\psi ^{+}(r)]$ via a phase space
functional integral 
\begin{eqnarray}
&&\chi \lbrack h(r),h^{+}(r)]  \nonumber \\
&=&\int D\psi ^{+}D\psi \,\exp i\left\{ \int dr\,\psi (r)h^{+}(r)\right\}
P[\psi (r),\psi ^{+}(r)]\,\exp i\left\{ \int dr\,h(r)\psi ^{+}(r)\right\} 
\nonumber \\
&&  \label{Eq.CharFnalDistnFnalReln}
\end{eqnarray}%
Note the similarity of this relationship to that in Eq. (34) in Ref. \cite%
{Dalton16b} for the characteristic and distribution functions. The proof is
given in Appendix \ref{Appendix - Functional P Distribution}. A derivation
of the result (\ref{Eq.QCorrFnFermiFunctionalIntegral}) for the quantum
correlation function can also be carried out starting from the
characteristic functional, and involves functional differentiation of
functional integrals.\pagebreak

\section{Functional Fokker- Planck Equations}

\label{Section - Functional Fokker-Planck Equations}

In the case of separate modes, the distribution function satisfies a
Fokker-Planck equation of the form (see Eq. (49)\textbf{\ }in Ref. \cite%
{Dalton16b}) 
\begin{equation}
\frac{\partial }{\partial t}B(g,g^{+})=\,-\tsum%
\limits_{p=1}^{2n}(A_{p}B(g,g^{+}))\frac{\overleftarrow{\partial }}{\partial
g_{p}}+\frac{1}{2}\tsum\limits_{p,q=1}^{2n}(D_{pq}B(g,g^{+}))\frac{%
\overleftarrow{\partial }}{\partial g_{q}}\frac{\overleftarrow{\partial }}{%
\partial g_{p}}  \label{Eq.FermionFPEUnNormBDist2}
\end{equation}%
where $A_{p}$ is the drift vector and $D_{pq}$ is the diffusion matrix.
Explicit expressions for the drift vector and diffusion matrix are set out
in Ref. \cite{Dalton16b} (see Eqs. (50), (51) and (52) therein) for the case
where the density operator satisfies a Markovian master equation. However,
as Grassmann derivatives can be related to functional derivatives - as in
Eqs.(\ref{Eq.GrassRightModeDerivC}) and (\ref{Eq.GrassRightModeDerivD}) in
Appendix \ref{Appendix - Grassmann Functional Calculus} 
\begin{equation}
\frac{\overleftarrow{\partial }}{\partial g_{k}}=\int dr\phi _{k}(r)\left( 
\frac{\overleftarrow{\delta }}{\delta \psi (r)}\right) _{r}\qquad \frac{%
\overleftarrow{\partial }}{\partial g_{k}^{+}}=\int dr\phi _{k}^{\ast
}(r)\left( \frac{\overleftarrow{\delta }}{\delta \psi ^{+}(r)}\right) _{r}.
\label{Eq.GrassRightModeDerivBoth}
\end{equation}%
and as the distribution function and distribution functional are equivalent,
we can convert the Fokker-Planck equation into an equivalent functional
Fokker-Planck equation.

\subsection{Functional Fokker-Planck Equation - B Distribution}

In terms of the new notation with $g_{p}\rightarrow g_{\alpha i}^{A}$, $%
A_{p}\rightarrow A_{\alpha i}^{A}$, $D_{pq}\rightarrow D_{\alpha i\,\beta
j}^{A\;B}$ we find that the \emph{functional Fokker-Planck equation} is 
\begin{eqnarray}
\frac{\partial }{\partial t}B[\psi ] &=&-\tsum\limits_{\alpha A}\dint
dr\,(A_{\alpha A}[r]\,B[\psi ])\frac{\overleftarrow{\delta }}{\delta \psi
_{\alpha A}(r)}  \nonumber \\
&&{\small +}\frac{1}{2}\tsum\limits_{\alpha A,\beta B}\diint
dr\,ds\,(D_{\alpha A\,\beta B}[r,s]\,B[\psi ])\frac{\overleftarrow{\delta }}{%
\delta \psi _{\beta B}(s)}\frac{\overleftarrow{\delta }}{\delta \psi
_{\alpha A}(r)}  \nonumber \\
&&  \label{Eq.FermionFFPE}
\end{eqnarray}

The drift vector and diffusion matrix functionals are given by%
\begin{eqnarray}
A_{\alpha A}[\psi (r),r] &=&\dsum\limits_{i}\xi _{\alpha
\,i}^{A}(r)\,A_{\alpha i}^{A}(g)  \label{Eq.DriftFnal} \\
D_{\alpha A\,\beta B}[\psi (r),r;\psi (s),s] &=&\dsum\limits_{ij}\xi
_{\alpha \,i}^{A}(r)\,D_{\alpha i\,\beta j}^{A\;B}(g)\,\xi _{\beta
\,j}^{B}(s){\small \,}  \label{Eq.DiffnFnal}
\end{eqnarray}%
showing how the Fokker-Planck equation could be turned into its equivalent
functional Fokker-Planck equation. Note the spatial integrals, single for
the drift term, double for the diffusion term. The drift vector and
diffusion matrix can be considered as Grassmann functionals. Fortunately,
the diffusion matrix often only depends on one spatial coordinate such as in
the case where the zero range approximation applies to the
Hamiltonian.\smallskip

\subsection{Correspondence Rules - B Distribution}

The functional Fokker-Planck equation may also be obtained more directly via
establishing the correspondence rules associated with the field operators.
These correspondence rules may be applied directly to the master, Liouville
or Matsubara equation for the evolution of the density operator to give the
functional Fokker-Planck equation, assuming the Hamiltonian operator and the
relaxation operators are expressed in terms of the field operators rather
than the mode operators. This approach is used in Section \ref{Section -
Application to Fermion Field Model}.

The correspondence rules are derived from those in Eq. (45) in Ref. \cite%
{Dalton16b} for the separate modes distribution function. \ Noting the mode
expansion (\ref{Eq.FermionFieldOprExpn})\ for the field operators we see that%
\begin{eqnarray}
\hat{\rho} &\Rightarrow &\sum_{i}\phi _{\alpha i}(r)\hat{c}_{\alpha i}\,\hat{%
\rho}\qquad B(g)\Rightarrow \sum_{i}\phi _{\alpha i}(r)g_{\alpha i}\,B[\psi ]
\label{Eq.GrassCorrA} \\
\hat{\rho} &\Rightarrow &\hat{\rho}\,\sum_{i}\phi _{\alpha i}(r)\hat{c}%
_{\alpha i}\qquad B(g)\Rightarrow B[\psi ]\,\sum_{i}\phi _{\alpha
i}(r)\left( +\frac{\overleftarrow{\partial }}{\partial g_{\alpha i}^{+}}%
\right)  \label{Eq.GrassCorrB} \\
\hat{\rho} &\Rightarrow &\sum_{i}\phi _{\alpha i}^{\ast }(r)\hat{c}_{\alpha
i}^{\dag }\,\hat{\rho}\qquad B(g)\Rightarrow \sum_{i}\phi _{\alpha i}^{\ast
}(r)\left( +\frac{\overrightarrow{\partial }}{\partial g_{\alpha i}}\right)
\,B[\psi ]  \label{Eq.GrassCorrC} \\
\hat{\rho} &\Rightarrow &\hat{\rho}\sum_{i}\phi _{\alpha i}^{\ast }(r)\hat{c}%
_{\alpha i}^{\dag }\qquad B(g)\Rightarrow \,B[\psi ]\,\sum_{i}\phi _{\alpha
i}^{\ast }(r)g_{\alpha i}^{+}.  \label{Eq.GrassCorrD}
\end{eqnarray}%
Then if the distribution function $B(g)$ is replaced by the distribution
functional $B[\psi (r)]$ we see from Eqs. (\ref%
{Eq.GrassmannLeftFnalDerivResult}), (\ref{Eq.GrassmannRightFnalDerivResult}%
), (\ref{Eq.GrassmannLeftFnalDerivResultB}) and (\ref%
{Eq.GrassmannRightFnalDerivResultB}) in Appendix \ref{Appendix - Grassmann
Functional Calculus} that the correspondence rules are 
\begin{eqnarray}
\hat{\rho} &\Rightarrow &\widehat{{\small \psi }}_{\alpha }(r)\,\hat{\rho}%
\quad B[\psi ]\Rightarrow \psi _{\alpha }(r)\,B[\psi ]\qquad \qquad \hat{\rho%
}\Rightarrow \hat{\rho}\,\widehat{{\small \psi }}_{\alpha }(r)\qquad B[\psi
]\Rightarrow B[\psi ]\,(+\frac{\overleftarrow{\delta }}{\delta \psi _{\alpha
}^{+}(r)})  \nonumber \\
\hat{\rho} &\Rightarrow &\widehat{{\small \psi }}_{\alpha }^{\dag }(r)\,\hat{%
\rho}\quad B[\psi ]\Rightarrow (+\frac{\overrightarrow{\delta }}{\delta \psi
_{\alpha }(r)})\,B[\psi ]\qquad \hat{\rho}\Rightarrow \hat{\rho}\,\widehat{%
{\small \psi }}_{\alpha }^{\dag }(r)\quad B[\psi ]\Rightarrow \,B[\psi
]\,\psi _{\alpha }^{+}(r)  \label{Eq.FermionFieldOprsCorresRules}
\end{eqnarray}

Analogous correspondence rules apply to the $P$ distribution functional. For
completeness these are set out in Appendix \ref{Appendix - Functional P
Distribution}\smallskip

\subsection{Case of $P$ Distribution}

In the case of the $P$ distribution functional the functional Fokker-Planck
has the same general features as that for in (\ref{Eq.FermionFFPE}) for the $%
B$ distribution. The drift vector and diffusion matrix functionals will of
course differ from those for the $B$ distribution, since the correspondence
rules (see Appendix \ref{Appendix - Functional P Distribution}) are not the
same. The relationship between the drift vector and diffusion matrix
functionals and the drift vector and diffusion matrix functions for the $P$
distribution is the same as in (\ref{Eq.DriftFnal}) and (\ref{Eq.DriftFnal}).

\pagebreak

\section{Ito Stochastic Field Equations}

\label{Section - Ito Stochastic Field Equations}

In Paper I (Ref. \cite{Dalton16b}) we established Ito stochastic equations
that were equivalent to the Fokker-Planck equation by requiring that the
phase space average at any time $t$\ of an arbitrary function $F(g,g^{+})$\
and stochastic average of same function $F(\widetilde{g}(t),\widetilde{g}%
^{+}(t))$ coincide (see Eq. (53) in Ref. \cite{Dalton16b}) when the phase
space variables $g_{p}$ are replaced by stochastic variables $\widetilde{g}%
_{p}(t)$, and for this to occur the Ito stochastic equations for $\widetilde{%
g}_{p}(t)$\ must be suitably related to Fokker-Planck equation for
distribution function $B(g,g^{+},t)$.

\begin{equation}
\left\langle {\small F(}g,g^{+}{\small )}\right\rangle _{t}=\,\overline{F%
{\small (}\widetilde{g}{\small (t),}\widetilde{g}^{+}{\small (t))}}
\label{Eq.PhaseSpaceStochAverEquiv2}
\end{equation}%
This approach is based on a treatment by Gardiner (see pp 95-96 in \cite%
{Gardiner83a}) which established the relationship between Fokker-Planck and
Ito stochastic equations for bosons.

In the case of separate modes, the stochastic phase variables satisfy Ito
stochastic equations given by Eq. (60) in Ref. \cite{Dalton16b}).

\begin{equation}
\frac{d}{dt}\widetilde{g}_{p}(t)=C^{p}(\widetilde{g}(t))+\tsum%
\limits_{a}B_{a}^{p}(\widetilde{g}(t))\;\Gamma _{a}(t_{+})
\label{Eq.ItoSDEGrassmann2}
\end{equation}%
where $C^{p}(\widetilde{g}(t))$ and $B_{a}^{p}(\widetilde{g}(t))$ are odd
Grassmann functions. The quantities $\Gamma _{a}(t_{+})$ are c-number
Gaussian-Markoff random noise terms \cite{Gardiner91a}. The $C^{p}$ and $%
B_{a}^{p}$ terms were shown (see Eqs. (69) and (70) in Ref. \cite{Dalton16b}%
)\ to be related to the drift vector and diffusion matrix in the
Fokker-Planck equation via 
\begin{eqnarray}
C^{p}(g,g^{+}) &=&-\,A_{p}(g,g^{+})  \label{Eq.FPEItoDrift2} \\
\lbrack B(g,g^{+})B^{{\small T}}(g,g^{+})]_{qp} &=&D_{qp}(g,g^{+})
\label{Eq.FPEItoDiffn2}
\end{eqnarray}

The basic stochastic average properties of the Gaussian-Markoff random noise
terms are

\begin{eqnarray}
\overline{\Gamma _{{\small a}}(t_{1})}\, &=&0  \nonumber \\
\overline{\Gamma _{{\small a}}(t_{1})\Gamma _{b}(t_{2})}\, &=&\delta _{%
{\small ab}}\delta (t_{1}-t_{2})  \nonumber \\
\overline{\Gamma _{a}(t_{1})\Gamma _{b}(t_{2})\Gamma _{c}(t_{3})} &=&0 
\nonumber \\
\overline{\Gamma _{a}(t_{1})\Gamma _{b}(t_{2})\Gamma _{c}(t_{3})\Gamma
_{d}(t_{4})} &=&\overline{\Gamma _{a}(t_{1})\Gamma _{b}(t_{2}})\overline{%
\Gamma _{c}(t_{3})\Gamma _{d}(t_{4}})+\overline{\Gamma _{a}(t_{1})\Gamma
_{c}(t_{3}})\overline{\Gamma _{b}(t_{2})\Gamma _{d}(t_{4}})  \nonumber \\
&&+\overline{\Gamma _{a}(t_{1})\Gamma _{d}(t_{4}})\overline{\Gamma
_{b}(t_{2})\Gamma _{c}(t_{3}})  \nonumber \\
&&  \label{Eq.GaussianMarkoffProps}
\end{eqnarray}%
showing that the stochastic averages of a single $\Gamma $ is zero and the
stochastic average of the product of two $\Gamma $'s is zero if they are
different and delta function correlated in the time difference if they are
the same. In addition, the stochastic averages of products of odd numbers of 
$\Gamma $\ are zero and stochastic averages of products of even numbers of $%
\Gamma $\ are the sums of products of stochastic averages of pairs of $%
\Gamma $. The $\Gamma _{a}$ are listed as $a=1,2,\cdots ,2n^{2}$, where the
number of modes is $n$.

A further important property is that any $F(\widetilde{g}(t))$\ and the
products of any $\Gamma _{a}(t_{+})$\ at later times $t_{+}$ are
uncorrelated 
\begin{eqnarray}
&&\overline{F(\widetilde{g}(t_{1}))\Gamma _{a}(t_{2})\Gamma
_{b}(t_{3})\Gamma _{c}(t_{4})..\Gamma _{k}(t_{l})}  \nonumber \\
&=&\overline{F(\widetilde{g}(t_{1}))}\,\;\overline{\Gamma _{a}(t_{2})\Gamma
_{b}(t_{3})\Gamma _{c}(t_{4})..\Gamma _{k}(t_{l})}\qquad
t_{1}<t_{2},t_{3},..,t_{l}  \label{Eq.UncorrelProperty}
\end{eqnarray}

It was first pointed out in \cite{Plimak01a} (see also Paper I (Ref. \cite%
{Dalton16b}) that for the $B$ distribution the expression for $C^{p}$ in the
Ito stochastic equation can be written as (see Eq. (80) of Ref. \cite%
{Dalton16b}) 
\begin{equation}
C^{p}(g,g^{+})=\tsum\limits_{r=1}^{2n}L_{r}^{p}\,g_{r}
\label{Eq.ClassicalTermResult2}
\end{equation}%
and in Ref. \cite{Dalton16b} the expression for $B_{a}^{p}$ in the Ito
stochastic equation was shown to be given by (see Eq. (75) of Ref. \cite%
{Dalton16b})%
\begin{equation}
B_{a}^{p}(g,g^{+})=\tsum\limits_{r=1}^{2n}K_{r,a}^{p}\,g_{r}
\label{Eq.NoiseTermResult2}
\end{equation}%
where the $L_{r}^{p}$ and $_{r,a}^{p}\,$are c-numbers. Both terms involve a
linear dependence on the Grassmann phase variables, resulting in a linear
Ito stochastic equation for the $\widetilde{g}_{p}(t)$. We note that if we
now introduce Grassmann stochastic fields via the expansions%
\begin{equation}
\widetilde{\psi }(r)=\sum\limits_{i}\widetilde{g}_{i}\,\phi _{i}(r)\qquad 
\widetilde{\psi }^{+}(r)=\sum\limits_{i}\phi _{i}^{\ast }(r)\widetilde{g}%
_{i}^{+}.  \label{Eq.GrassmannStochasticFields}
\end{equation}%
we can easily convert the Ito stochastic equations for $\widetilde{g}_{i}(t),%
\widetilde{g}_{i}^{+}(t)$ into linear Ito stochastic field equations for $%
\widetilde{\psi }(r),\widetilde{\psi }^{+}(r)$.

In view of the key relationship (\ref{Eq.PhaseSpaceStochAverEquiv2}) between
the phase space integral of an arbitrary Grassmann function and the
stochastic average of this function, we can establish a relationship between
the phase space functional integral of products of Grassmann fields (as
occur in expressions for quantum correlation functions, Fock state
populations and coherences) and stochastic averages of the same products of
stochastic fields. This result is the basis for numerical calculations based
on Grassmann stochastic fields for fermions rather than solving the
functional Fokker-Planck equation and calculating functional integrals
involving the non-stochastic Grassmann fields in order to determine quantum
correlation functions etc. . The proof is similar to that for (\ref%
{Eq.QCorrFnFermiFunctionalIntegral}), but in reverse. Using (\ref%
{Eq.GrassmannPhaseSpaceIntegral4B}) with $F[\psi (r),\psi ^{+}(r)]\equiv $ $%
\psi (\mathbf{s}_{q})\cdots \psi (\mathbf{\ s}_{1})P[\psi (r),\psi
^{+}(r)]\psi ^{+}(\mathbf{r}_{1})\cdots \psi ^{+}(\mathbf{r}_{p})$ and Eqs. (%
\ref{Eq.GrassmannFieldFn2}), (\ref{Eq.GrassmannFunction3}) we have 
\begin{eqnarray}
&&\int \int D\psi ^{+}D\psi \,(\psi (\mathbf{s}_{q})\cdots \psi (\mathbf{\ s}%
_{1})P[\psi (r),\psi ^{+}(r)]\psi ^{+}(\mathbf{r}_{1})\cdots \psi ^{+}(%
\mathbf{r}_{p}))  \nonumber \\
&=&\int \prod\limits_{i}dg_{i}^{+}\prod\limits_{i}dg_{i}\,(\psi (\mathbf{s}%
_{q})\cdots \psi (\mathbf{s}_{1}))\,P(g,g^{+})(\psi ^{+}(\mathbf{r}%
_{1})\cdots \psi ^{+}(\mathbf{r}_{p}))  \nonumber \\
&=&\sum\limits_{m_{q}\cdots m_{1}}\sum\limits_{l_{1}\cdots l_{p}}\phi
_{m_{q}}(\mathbf{s}_{q})\cdots \phi _{m_{1}}(\mathbf{s}_{1})\phi
_{l_{1}}^{\ast }(\mathbf{r}_{1})\cdots \phi _{l_{p}}^{\ast }(\mathbf{r}_{p})
\nonumber \\
&&\times \int
\prod\limits_{i}dg_{i}^{+}\prod\limits_{i}dg_{i}\,(g_{m_{q}}\cdots
g_{m_{2}}g_{m_{1}})P(g,g^{+})(g_{l_{1}}^{+}g_{l_{2}}^{+}\cdots g_{l_{p}}^{+})
\nonumber \\
&=&\sum\limits_{m_{q}\cdots m_{1}}\sum\limits_{l_{1}\cdots l_{p}}\phi
_{m_{q}}(\mathbf{s}_{q})\cdots \phi _{m_{1}}(\mathbf{s}_{1})\phi
_{l_{1}}^{\ast }(\mathbf{r}_{1})\cdots \phi _{l_{p}}^{\ast }(\mathbf{r}_{p})
\nonumber \\
&&\times \,\overline{(\widetilde{g}_{m_{q}}(t)\cdots \widetilde{g}_{m_{2}}(t)%
\widetilde{g}_{m_{1}}(t))(\widetilde{g}_{l_{1}}^{+}(t)\widetilde{g}%
_{l_{2}}^{+}(t)\cdots \widetilde{g}_{l_{p}}^{+}(t))}  \nonumber \\
&=&\overline{\widetilde{\psi }(\mathbf{s}_{q},t)\cdots \widetilde{\psi }(%
\mathbf{\ s}_{1},t)\widetilde{\psi }^{+}(\mathbf{r}_{1},t)\cdots \widetilde{%
\psi }^{+}(\mathbf{r}_{p},t)}  \label{Eq.FunctionalIntegralAverStochAverReln}
\end{eqnarray}%
The step from line two to line three follows from Eq. (\ref%
{Eq.PhaseSpaceStochAverEquiv2}).

\subsection{Ito Stochastic Field Equations - B Distribution}

In terms of the general notation, the fermionic stochastic fields $%
\widetilde{\psi }_{\alpha }(r,t),\widetilde{\psi }_{\alpha }^{+}(r,t)$
(denoted $\widetilde{\psi }_{\alpha A}(r,t),A=1,2$) are defined by the same
mode expansion (\ref{Eq.FermionFieldFnExpn}) as for the time independent
Grassmann fields $\psi _{\alpha }(r),\psi _{\alpha }^{+}(r)$ but now with
the Grassmann phase space variables $g_{\alpha i},g_{\alpha i}^{+}$ replaced
by time dependent Grassman stochastic variables $\widetilde{g}_{\alpha i},%
\widetilde{g}_{\alpha i}^{+}$ (denoted $\widetilde{g}_{\alpha i}^{A},A=1,2$)
for each mode.

\begin{equation}
\widetilde{\psi }_{\alpha A}(r)=\dsum\limits_{i}\widetilde{g}_{\alpha
i}^{A}\;\xi _{\alpha i}^{A}(r)  \label{Eq.ItoStochFieldExpn}
\end{equation}%
The $t$ dependence will be left understood for simplicity of notation.

In terms of the new notation the Ito stochastic equations for the stochastic
phase variables are 
\begin{equation}
\frac{d}{dt}\widetilde{g}_{\alpha i}^{A}(t)=C_{\alpha i}^{A}(\widetilde{g}%
)+\tsum\limits_{a}B_{a}^{\alpha i\,A}(\widetilde{g}(t))\;\Gamma _{a}(t_{+})
\label{Eq.ItoStochEqns2}
\end{equation}

We have shown in Paper I (Ref. \cite{Dalton16b}) that for separate modes the
Ito stochastic equations for the $\widetilde{g}_{p}$ in Eq. (\ref%
{Eq.ItoSDEGrassmann2}) (see Eqs.(60) or (63) in Ref. \cite{Dalton16b}) are
equivalent to the Fokker-Planck equation in Eq.(\ref%
{Eq.FermionFPEUnNormBDist2}) for the case of the $B$ distribution function
(see Eq. (49) in Ref. \cite{Dalton16b}), and as this Fokker-Planck equation
is equivalent to the functional Fokker-Planck equation in Eq.(\ref%
{Eq.FermionFFPE}) for the $B$ distribution functional, we can derive the
equivalent Ito stochastic field equation for the $\widetilde{\psi }_{\alpha
A}(r)$ by starting from the Ito stochastic equations (\ref{Eq.ItoStochEqns2}%
)\ for the $\widetilde{g}_{\alpha i}^{A}$, multiplying by the mode function $%
\xi _{\alpha i}^{A}(r)$ and summing over the modes for the same internal
component $\alpha $.

The Ito stochastic field equations obtained are%
\begin{eqnarray}
\frac{\partial }{\partial t}\widetilde{\psi }_{\alpha A}(r) &=&C_{\alpha A}[%
\widetilde{\psi }(r),r]+\tsum\limits_{a}B_{a}^{\alpha A}[\widetilde{\psi }%
(r),r]\;\Gamma _{a}(t_{+})  \label{Eq.ItoStochFieldEqn} \\
&=&\left( \frac{\partial }{\partial t}\widetilde{\psi }_{\alpha A}({\small r}%
)\right) _{class}+\left( \frac{\partial }{\partial t}\widetilde{\psi }%
_{\alpha A}({\small r})\right) _{noise}  \label{Eq.ClassicaNoiseTerms}
\end{eqnarray}%
The right side is the sum of a classical field term and a noise field term.
The classical and noise field terms involve Grassmann functionals $C_{\alpha
A}[\widetilde{\psi }(r),r]$ and $B_{a}^{\alpha A}[\widetilde{\psi }(r),r]$
defined by\ 
\begin{eqnarray}
C_{\alpha A}[\widetilde{\psi }(r),r] &=&\dsum\limits_{i}C_{\alpha i}^{A}(%
\widetilde{g})\,\xi _{\alpha \,i}^{A}(r)\,  \label{Eq.FFPEItoFldEqnDrift} \\
B_{a}^{\alpha A}[\widetilde{\psi }(r),r] &=&\dsum\limits_{i}B_{a}^{\alpha
i\,A}(\widetilde{g})\,\xi _{\alpha i}^{A}(r)
\label{Eq.FFPEItoStochFldEqnDiffn}
\end{eqnarray}%
Note that the classical field term is also stochastic because it involves a
functional of the stochastic fields $\widetilde{\psi }_{\alpha A}(r)$. The
noise field term is stochastic, not only for this reason but also because it
involves the Gaussian-Markoff noise terms. The derivation of the Ito
stochastic field equations is set out in Appendix \ref{Appendix - Ito
Stochastic Field Equations}.

We can then show that the relationships between these quantities and the
drift (\ref{Eq.DriftFnal}) and diffusion (\ref{Eq.DiffnFnal}) terms in the
functional Fokker-Planck equation are

\begin{eqnarray}
C_{\alpha A}[\widetilde{\psi }(r),r] &=&-A_{\alpha A}[\widetilde{\psi }(r),r]
\label{Eq.ClassicalFieldResult} \\
(BB^{T})_{\alpha A\,\beta B} &=&\dsum\limits_{a}B_{a}^{\alpha A}[\widetilde{%
\psi }(r),r]\,B_{a}^{\beta B}[\widetilde{\psi }(s),s]  \nonumber \\
&=&D_{\alpha A\,\beta B}[\widetilde{\psi }(r),r;\widetilde{\psi }(s),s]
\label{Eq.NoiseFieldResult}
\end{eqnarray}%
The first relationship follows from (\ref{Eq.FFPEItoFldEqnDrift}) and (\ref%
{Eq.FPEItoDrift2}). The second result follows from (\ref%
{Eq.FFPEItoStochFldEqnDiffn}) and (\ref{Eq.FPEItoDiffn2}), the last equation
being%
\begin{equation}
\tsum\limits_{a}B_{a}^{\alpha i\,A}(g)\,B_{a}^{\beta j\,B}(g)=D_{\alpha
i\,\beta j}^{A\;B}(g)\,  \label{Eq.NoiseTermProp2}
\end{equation}%
in the new notation.

In practice mode expansions for fields not needed to derive the Ito
stochastic field equations. Having derived the functional Fokker-Planck
equations using the correspondence rules for the field operators, we can
then simply use the general results we have established in Eqs. (\ref%
{Eq.ClassicalFieldResult}) and (\ref{Eq.NoiseFieldResult}) to obtain the
classical and noise field quantities $C_{\alpha A}[\widetilde{\psi }(r),r]$
and $B_{a}^{\alpha A}[\widetilde{\psi }(r),r].$ The classical field quantity
is simply $C=-A$ from the drift functional, whilst the noise field quantity
requires a factorisation of the diffusion functional $D$ as in $BB^{T}=D$.
\smallskip

\subsection{Form of Drift and Diffusion Functionals}

Having obtained the mode based expressions in (\ref{Eq.FFPEItoFldEqnDrift})
and (\ref{Eq.FFPEItoStochFldEqnDiffn}) for the terms in the Ito stochastic
field equations (\ref{Eq.ItoStochFieldEqn}) we can develop these expressions
further to express $C_{\alpha A}[\widetilde{\psi }(r),r]$ and $B_{a}^{\alpha
A}[\widetilde{\psi }(r),r]$ in terms of the stochastic fields $\widetilde{%
\psi }_{\beta A}(s)$. As we will see, the relationship is \emph{linear} but
in general will be \emph{non-local}. The linearlity feature follows from the
drift vector, diffusion matrix in the Fokker-Planck equation for the $B(g)$
distribution function depending linearly, bilinearly on the Grassmann phase
space variables.

In the case of the $B$ distribution the drift terms $C_{\alpha i}^{A}(%
\widetilde{g})$ are linear functions of the stochastic phase variables $%
\widetilde{g}_{\alpha i}^{A}$, based on Eq. (\ref{Eq.ClassicalTermResult2})
(see Eq. (80) in Ref. \cite{Dalton16b}) 
\begin{equation}
C_{\alpha i}^{A}(\widetilde{g})=\tsum\limits_{\beta \,j}L_{\beta
j}^{A\,\alpha i}\,\widetilde{g}_{\beta j}^{A}
\end{equation}%
so that from (\ref{Eq.FFPEItoFldEqnDrift}) and (\ref{Eq.ItoStochFieldExpn})%
\begin{equation}
C_{\alpha A}[\widetilde{\psi }(r),r]=\tsum\limits_{i\beta \,j}L_{\beta
j}^{A\,\alpha i}\dint ds\,\xi _{\alpha \,i}^{A}(r)\,\xi _{\beta
\,j}^{A}(s)^{\ast }\,\widetilde{\psi }_{\beta A}(s)\,\,\,
\label{Eq.DriftFnal2}
\end{equation}%
which indicates that $C_{\alpha A}[\widetilde{\psi }(r),r]$ is a linear
functional of the stochastic fields.

The terms $B_{a}^{\alpha i\,A}(\widetilde{g})$ related to the diffusion
matrix are obtained from Eq. (\ref{Eq.NoiseTermResult2}) (see Eq. (75) in
Ref. \cite{Dalton16b}) as 
\begin{equation}
B_{a}^{\alpha i\,A}(\widetilde{g})=\tsum\limits_{B\,\beta j}K_{B\,\beta
j,\,a}^{A\,\alpha i}\,\widetilde{g}_{\beta j}^{B}
\end{equation}%
so that from (\ref{Eq.FFPEItoStochFldEqnDiffn}) and (\ref%
{Eq.ItoStochFieldExpn})%
\begin{equation}
B_{a}^{\alpha A}[\widetilde{\psi }(r),r]=\dsum\limits_{iB\beta j}K_{B\,\beta
j,\,a}^{A\,\alpha i}\dint ds\,\xi _{\alpha \,i}^{A}(r)\,\xi _{\beta
\,j}^{B}(s)^{\ast }\,\widetilde{\psi }_{\beta B}(s)\,\,\,
\label{Eq.DiffnFnal2}
\end{equation}%
which indicates that $B_{\alpha }^{\alpha A}[\widetilde{\psi }(r),r]$ is a
linear functional of the stochastic fields.

As noted above both (\ref{Eq.DriftFnal2}) and (\ref{Eq.DiffnFnal2}) give the
drift and diffusion field terms at spatial position $r$ in the Ito
stochastic field equations as spatial integrals of the stochastic fields at
other positions $s$. Thus in general the ito stochastic field equations are 
\emph{non-local} - which increases the complexity for numerical
calculations. However, as we will see in Section \ref{Section - Application
to Fermion Field Model} there are important cases where the stochastic field
equations remain\emph{\ local}.

Thus the classical and noise field terms are given by 
\begin{eqnarray}
\left( \frac{\partial }{\partial t}\widetilde{\psi }_{\alpha A}({\small r}%
)\right) _{class} &=&\tsum\limits_{i\beta \,j}L_{\beta j}^{A\,\alpha i}\dint
ds\,\xi _{\alpha \,i}^{A}(r)\,\xi _{\beta \,j}^{A}(s)^{\ast }\,\widetilde{%
\psi }_{\beta A}(s)\,\,  \label{Eq.ClassField2} \\
\left( \frac{\partial }{\partial t}\widetilde{\psi }_{\alpha A}({\small r}%
)\right) _{noise} &=&\tsum\limits_{a}\dsum\limits_{iB\beta j}K_{B\,\beta
j,\,a}^{A\,\alpha i}\dint ds\,\xi _{\alpha \,i}^{A}(r)\,\xi _{\beta
\,j}^{B}(s)^{\ast }\,\widetilde{\psi }_{\beta B}(s)\;\Gamma _{a}(t_{+})\, 
\nonumber \\
&&  \label{Eq.NoiseFld2}
\end{eqnarray}%
This shows the key result in the stochastic field case that the Ito
stochastic field equations are still linear in the stochastic fields, and
and thus the fields at time $t+\delta t$ are related linearly to the fields
at time $t$ and also the linear transformation (though stochastic due to the 
$\Gamma _{a}$) only involves c-numbers. This feature still applies even if
the equations are non-local. \smallskip

\subsection{Stochastic Averages - Classical and Noise Fields}

In the case of the classical field 
\begin{equation}
\overline{\left( \frac{\partial }{\partial {\small t}}\widetilde{\psi }%
_{\alpha A}({\small r,t}_{1+})\right) _{class}}=0  \label{Eq.ClassFieldAver}
\end{equation}%
so the stochastic average is zero.

For the noise field - average fluctuations zero, average product two noise $%
\rightarrow $ diffusion matrix and delta correlated in time 
\begin{eqnarray}
&&\overline{\left( \frac{\partial }{\partial t}\widetilde{\psi }_{\alpha A}(%
{\small r,t}_{1+})\right) _{noise}}=0  \nonumber \\
&&\overline{\left( \frac{\partial }{\partial t}\widetilde{\psi }_{\alpha A}(%
{\small r,t}_{1+})\right) _{noise}\left( \frac{\partial }{\partial t}%
\widetilde{\psi }_{\beta B}(s{\small ,t}_{2+})\right) _{noise}}  \nonumber \\
&=&\overline{D_{\alpha A\,\beta B}(\widetilde{\psi }(r,t_{1,2}),r;\widetilde{%
\psi }(s,t_{1,2}),s)}\times \delta (t_{1}-t_{2}).  \label{Eq.NoiseFieldAver}
\end{eqnarray}%
so the stochastic average is zero and the stochastic average of the product
of two noise fields are delta correlated in the time difference and given by
the stochastic average of the relevant diffusion matrix elements. The
stochastic average of the products of odd numbers of noise fields are zero,
and for products of even numbers of noise fields the result is the sums of
products of pairs of noise fields.\smallskip\ 

\subsection{Case of $P$ Distribution}

In the case of the $P$ distribution functional the Ito stochastic field
equations have the same general features as that for in (\ref%
{Eq.ItoStochFieldEqn}) for the $B$ distribution. The classical and noise
fields will of course differ from those for the $B$ distribution, since the
functional Fokker-Planck equations are not the same. The relationship
between the classical and noise field quantities and the drift vector and
diffusion matrix functionals in the functional Fokker-Planck equation for
the $P$ distribution functional is the same as in (\ref%
{Eq.ClassicalFieldResult}) and (\ref{Eq.NoiseFieldResult}). However, the
classical field term is no longer a linear functional of the stochastic
fields.

\subsection{Numerical Issues}

Analogous to the treatment based on separate modes, the Ito stochastic field
equations (\ref{Eq.ItoStochFieldEqn}) can be written in the form

\begin{eqnarray}
&&\widetilde{\psi }_{\alpha A}(r,t+\delta t)  \nonumber \\
&=&\tsum\limits_{\beta B}\tint ds\left( 
\begin{array}{c}
\delta _{\alpha A,\beta B}\,\delta (r-s)+\delta
_{A,B}\tsum\limits_{i\,j}\,L_{\beta j}^{A\,\alpha i}\,\xi _{\alpha
\,i}^{A}(r)\,\xi _{\beta \,j}^{A}(s)^{\ast }\,\delta t \\ 
+\tsum\limits_{a}\dsum\limits_{ij}K_{B\,\beta j,\,a}^{A\,\alpha i}\,\xi
_{\alpha \,i}^{A}(r)\,\xi _{\beta \,j}^{B}(s)^{\ast }\,\delta w_{a}(t_{+})%
\end{array}%
\right) \widetilde{\psi }_{\beta B}(s,t)  \nonumber \\
&&  \label{Eq.LinearRelnStochFieldsSmallTimeInterval} \\
&=&\tsum\limits_{\beta B}\tint ds\;\Theta _{\alpha A,\,\beta B}(r,s,t_{+})\,%
\widetilde{\psi }_{\beta B}(s,t)  \label{Eq.DefnThetaFn2}
\end{eqnarray}%
where we have used the linearity features (\ref{Eq.DriftFnal2}) and (\ref%
{Eq.DiffnFnal2}) for the Ito terms $C_{\alpha A}[\widetilde{\psi }(r),r]$
and $B_{a}^{\alpha A}[\widetilde{\psi }(r),r]$ and where the \emph{Wiener
stochastic variable} $\widetilde{w}_{a}(t)$ and its increment are%
\begin{equation}
\widetilde{w}_{a}(t)=\int_{t_{0}}^{t}dt_{1}\Gamma _{a}(t_{1})\qquad \delta 
\widetilde{w}_{a}(t_{+})=\dint\limits_{t+}^{t+\delta t}dt_{1}\,\Gamma
_{a}(t_{1})  \label{Eq.WienerInc2}
\end{equation}%
An important result for the stochastic average of the product of two Wiener
increments is 
\begin{equation}
\overline{\delta \widetilde{w}_{a}(t_{+})\delta \widetilde{w}_{b}(t_{+})}%
=\delta _{a,b}\,\delta t
\end{equation}%
The properties of Wiener increments are set out in Eq. (85) in Ref. \cite%
{Dalton16b}.

The result in (\ref{Eq.LinearRelnStochFieldsSmallTimeInterval}) shows that
the stochastic field $\widetilde{\psi }_{\alpha A}(r,t+\delta t)$ at
position $r$ at time $t+\delta t$ is related \emph{linearly} to the
stochastic fields $\widetilde{\psi }_{\beta B}(s,t)$ at various positions $s$
at time $t$, and that the quantities involved in this linear relationship
are all\emph{\ c-numbers} - these involving both stochastic variables $%
\delta w_{a}(t_{+})$ and non-stochastic quantities such as mode functions $%
\xi _{\alpha \,i}^{A}(r),\,\xi _{\beta \,j}^{B}(s)$ and quantities such as $%
L_{\beta j}^{A\,\alpha i}$ and $K_{B\,\beta j,\,a}^{A\,\alpha i}$ obtainable
from the functional Fokker-Planck equation. This outcome is analogous to
that occuring for separate modes and is the basis for possible numerical
treatments. Note however the additional complexity due to spatial
integration and differing internal components. In practice, the spatial
integration process is often not required - the stochastic fields on each
side of (\ref{Eq.LinearRelnStochFieldsSmallTimeInterval}) involving the same
spatial position $r$.

We can then use (\ref{Eq.DefnThetaFn2}) and the uncorrelation property (\ref%
{Eq.UncorrelProperty}) to derive a result for the stochastic average of
products of the $\widetilde{\psi }_{\alpha A}(r,t+\delta t)$ at time $%
t+\delta t$ showing that it involves stochastic average of products of the $%
\widetilde{\psi }_{\beta B}(r,t)$ at time $t$ with the same number of
factors, and that the relationship between the two sets of stochastic
averages is linear - with a transformation matrix that only involves a
stochastic average of c-numbers. These c-numbers are of course stochastic as
they involve Wiener increments as well as quantities obtained from the drift
and diffusion terms in the functional Fokker-Planck equation. We find that 
\begin{eqnarray}
&&\overline{\widetilde{\psi }_{\alpha A}(r_{1},t+\delta t)\widetilde{\psi }%
_{\beta B}(r_{2},t+\delta t)...\widetilde{\psi }_{\sigma S}(r_{p},t+\delta t)%
}  \nonumber \\
&=&\tsum\limits_{\eta R}\tint ds_{1}\;\tsum\limits_{\theta Q}\tint
ds_{2}\;...\tsum\limits_{\omega Z}\tint ds_{p}\;  \nonumber \\
&&\times \left[ \Theta _{\alpha A,\,\eta R}(r_{1},s_{1},t_{+})\,\,\Theta
_{\beta B,\,\theta Q}(r_{2},s_{2},t_{+})\,...\Theta _{\sigma S,\,\omega
Z}(r_{p},s_{p},t_{+})\right] \,_{stoch\,aver}\,  \nonumber \\
&&\times \overline{\widetilde{\psi }_{\eta R}(s_{1},t)\,\widetilde{\psi }%
_{\theta Q}(s_{2},t)\,...\widetilde{\psi }_{\omega Z}(s_{p},t)}
\label{Eq.ResultStochAverProd2}
\end{eqnarray}%
This result is analogous to Eq.\textbf{\ }(88) in Ref. \cite{Dalton16b}. In
principle, this result enables stochastic averages of products of stochastic
Grassmann fields at the end of a small time interval to be determined from
the set of such stochastic averages (of the same number of factors) at the
beginning of the interval. By dividing a finite time interval up into a
number of small intervals the set of stochastic averages of a given number
of factors can be evolved from the set at an initial time with the same
number of factors to that applying at a later time. Then the initial set of
stochastic averages can be obtained from the initial density operator using
the expressions (\ref{Eq.FermionPositionPopnResult}) and (\ref%
{Eq.FermionPositionCoherenceResult}) for initial position Fock state
populations and coherences - the phase space functional integrals being
equivalent to the stochastic averages in accord with (\ref%
{Eq.PhaseSpaceStochAverEquiv2}) and (\ref{Eq.GrassmannPhaseSpaceIntegral4}).
The calculation is of the same order of complexity as for the separate mode
case if the total number of space grid elements is the same as the number of
modes involved.\pagebreak

\section{Application to Fermion Field Models}

\label{Section - Application to Fermion Field Model}

\subsection{A. Two Component Zero Range Model}

As a typical application of fermion field models we consider a trapped Fermi
gas of spin $1/2$ fermionic atoms with spin conserving collisions of zero
range between pairs of atoms .Here there are two distinct internal states
corresponding to spin up and spin down atoms, designated by $\alpha
=u(\uparrow ),d(\downarrow )$, with $-\alpha $ referring to the opposite
spin state. The Fermi gas is isolated from the environment so no relaxation
effects are involved. This model was also considered by Plimak et al \cite%
{Plimak01a}.

\subsubsection{Hamiltonian}

The Hamiltonian is written in terms of the field operators $\hat{\Psi}%
_{\alpha }(r)$, $\hat{\Psi}_{\alpha }(r)^{\dag }$ as

\begin{eqnarray}
\hat{H}_{f}\mbox{\rule{-0.5mm}{0mm}} &=&\mbox{\rule{-1mm}{0mm}}\int %
\mbox{\rule{-1mm}{0mm}}dr\left( \sum_{\alpha }\frac{\hbar ^{2}}{2m}\nabla 
\hat{\Psi}_{\alpha }(r)^{\dag }\mbox{\rule{-0.5mm}{0mm}}\cdot %
\mbox{\rule{-0.5mm}{0mm}}\nabla \hat{\Psi}_{\alpha }(\ r)+\sum_{\alpha }\hat{%
\Psi}_{\alpha }(r)^{\dag }V_{\alpha }\hat{\Psi}_{\alpha }(r)\right. 
\nonumber \\
&&\hspace{2cm}\left. +\frac{g_{f}}{2}\sum_{\alpha }\hat{\Psi}_{\alpha
}(r)^{\dag }\hat{\Psi}_{-\alpha }(r)^{\dag }\hat{\Psi}_{-\alpha }(r)\hat{\Psi%
}_{\alpha }(r)\right)  \label{Eq.HamiltonianFieldModel} \\
&=&\widehat{{\small K}}+\widehat{{\small V}}+\widehat{{\small U}}
\label{Eq.HamiltTerms}
\end{eqnarray}%
and is the sum of kinetic energy, trap potential energy and collision
interaction energy terms.\smallskip\ 

\subsubsection{Functional Fokker-Planck Equation - B Distribution}

The functional Fokker-Planck equation can be obtained via applying the
correspondence rules to the Liouville-non Neumann equation. The
contributions can be written as a sum of terms from $\widehat{{\small K}},%
\widehat{{\small V}},\widehat{{\small U}}$. As the derivation requires some
complex manipulation it is presented in Appendix \ref{Appendix - Fermi Gas
FFPE}. In the case considered the functional Fokker-Planck equations can be
written in a simpler notation in which the field operators $\hat{\Psi}%
_{\alpha }(r)$, $\hat{\Psi}_{\alpha }(r)^{\dag }$\ are represented via
Grassmann fields $\psi _{\alpha }(r)$\ and $\psi _{\alpha }^{+}(r)$\
respectively - so in presenting the final equations we will not use the $%
A=1,2$\ notation to distinguish $\psi _{\alpha }(r)$\ and $\psi _{\alpha
}^{+}(r)$\ as we did for the general theory.

The kinetic energy term is 
\begin{eqnarray}
&&\left( \frac{\partial }{\partial t}B[\psi (r)]\right) _{K}  \nonumber \\
&=&\frac{i}{\hbar }\int ds\,\left[ \left\{ \left( \frac{\hbar ^{2}}{2m}%
\nabla ^{2}{\small \psi }_{u}{\small (s)\,}B[\ \psi (r)]\right) \frac{%
\overleftarrow{\delta }}{\delta \psi _{u}(s)}\right\} \right.  \nonumber \\
&&\left. +\left\{ \left( \frac{\hbar ^{2}}{2m}\nabla ^{2}{\small \psi }_{d}%
{\small (s)\,}B[\ \psi (r)]\right) \frac{\overleftarrow{\delta }}{\delta
\psi _{d}(s)}\right\} \right.  \nonumber \\
&&\left. -\left\{ \left( \frac{\hbar ^{2}}{2m}\nabla ^{2}{\small \psi }%
_{u}^{+}{\small (s)\,}B[\ \psi (r)]\right) \frac{\overleftarrow{\delta }}{%
\delta \psi _{u}^{+}(s)}\right\} \right.  \nonumber \\
&&\left. -\left\{ \left( \frac{\hbar ^{2}}{2m}\nabla ^{2}{\small \psi }%
_{d}^{+}{\small (s)\,}B[\ \psi (r)]\right) \frac{\overleftarrow{\delta }}{%
\delta \psi _{d}^{+}(s)}\right\} \right]  \nonumber \\
&&  \label{Eq.FFPEKineticFermiFldModel}
\end{eqnarray}%
and only contributes to the drift term.

The trap potential energy term is 
\begin{eqnarray}
&&\left( \frac{\partial }{\partial t}B[\psi (r)]\right) _{V}  \nonumber \\
&=&\frac{i}{\hbar }\int {\small ds}\,\left[ -\left\{ \left( V_{u}\psi
_{u}(s)\,B[\ \psi (r)]\right) \frac{\overleftarrow{\delta }}{\delta \psi
_{u}(s)}\right\} -\left\{ \left( V_{d}\psi _{d}(s)\,B[\ \psi (r)]\right) 
\frac{\overleftarrow{\delta }}{\delta \psi _{d}(s)}\right\} \right. 
\nonumber \\
&&\left. +\left\{ \left( V_{u}\psi _{u}^{+}(s)\,B[\ \psi (r)]\right) \frac{%
\overleftarrow{\delta }}{\delta \psi _{u}^{+}(s)}\right\} +\left\{ \left(
V_{d}\psi _{d}^{+}(s)\,B[\ \psi (r)]\right) \frac{\overleftarrow{\delta }}{%
\delta \psi _{d}^{+}(s)}\right\} \right]  \nonumber \\
&&  \label{Eq.FFPEPotentialFermiFldModel}
\end{eqnarray}%
and also only contributes to the drift term.

The fermion-fermion interaction term is 
\begin{eqnarray}
&&\left( \frac{\partial }{\partial t}B[\psi (r)]\right) _{U}  \nonumber \\
&{\small =}&\frac{i}{\hbar }\frac{g}{2}\int {\small ds}\,\left[ \left\{ \psi
_{d}(s)\psi _{u}(s)\,B[\psi (r)]\frac{\overleftarrow{\delta }}{\delta \psi
_{d}(s)}\frac{\overleftarrow{\delta }}{\delta \psi _{u}(s)}\right\} \right. 
\nonumber \\
&&\left. +\left\{ \psi _{u}(s)\psi _{d}(s)\,B[\psi (r)]\frac{\overleftarrow{%
\delta }}{\delta \psi _{u}(s)}\frac{\overleftarrow{\delta }}{\delta \psi
_{d}(s)}\right\} \right.  \nonumber \\
&&\left. -\left\{ \psi _{d}^{+}(s)\psi _{u}^{+}(s)\,B[\psi (r)]\frac{%
\overleftarrow{\delta }}{\delta \psi _{d}^{+}(s)}\frac{\overleftarrow{\delta 
}}{\delta \psi _{u}^{+}(s)}\right\} \right.  \nonumber \\
&&\left. -\left\{ \psi _{u}^{+}(s)\psi _{d}^{+}(s)\,B[\psi (r)]\frac{%
\overleftarrow{\delta }}{\delta \psi _{u}^{+}(s)}\frac{\overleftarrow{\delta 
}}{\delta \psi _{d}^{+}(s)}\right\} \right]
\label{Eq.FFPEInteractFermiFldModel}
\end{eqnarray}%
and is the only contribution to the diffusion term.\smallskip

\subsubsection{Ito Stochastic Field Equations}

A straightforward application of the general theory using (\ref%
{Eq.ClassicalFieldResult}) and (\ref{Eq.NoiseFieldResult}) to determine the
classical and noise field terms gives the Ito stochastic field equations in
the form (\ref{Eq.ItoStochFieldEqn}). The diffusion matrix is simple to
factorise into the Takagi form \cite{Takagi25a}. After minor algebra the Ito
field equations can be put in the form analogous to Eq. (81) in Ref. \cite%
{Dalton16b} (see also Eq.(\ref{Eq.ItoStochFldEqnIntegralForm}) in Appendix %
\ref{Appendix - Ito Stochastic Field Equations} for the intermediate step).
We have

\begin{eqnarray}
&&\left[ 
\begin{tabular}{|l|}
\hline
$\tilde{\psi}_{u}(s,t+\delta t)$ \\ \hline
$\tilde{\psi}_{d}(s,t+\delta t)$ \\ \hline
$\tilde{\psi}_{u}^{+}(s,t+\delta t)$ \\ \hline
$\tilde{\psi}_{d}^{+}(s,t+\delta t)$ \\ \hline
\end{tabular}%
\ \right]  \nonumber \\
&=&\left[ 
\begin{tabular}{|l|l|l|l|}
\hline
$\Theta _{u,u}(t^{{\small +}})$ & $\Theta _{u,d}(t^{+})$ & $0$ & $0$ \\ 
\hline
$\Theta _{d,u}(t^{+})$ & $\Theta _{d,d}(t^{+})$ & $0$ & $0$ \\ \hline
$0$ & $0$ & $\Theta _{u,u}^{+}(t^{+})$ & $\Theta _{u,d}^{+}(t^{+})$ \\ \hline
$0$ & $0$ & $\Theta _{d,u}^{+}(t^{+})$ & $\Theta _{d,d}^{+}(t^{+})$ \\ \hline
\end{tabular}%
\ \right] \left[ 
\begin{tabular}{|l|}
\hline
$\tilde{\psi}_{u}(s,t)$ \\ \hline
$\tilde{\psi}_{d}(s,t)$ \\ \hline
$\tilde{\psi}_{u}^{+}(s,t)$ \\ \hline
$\tilde{\psi}_{d}^{+}(s,t)$ \\ \hline
\end{tabular}%
\ \right]  \nonumber \\
&&  \label{Eq.ItoStochFieldEqnsFermiFldModel}
\end{eqnarray}%
where the $\Theta $ quantities are given by%
\begin{eqnarray}
\lbrack \Theta ({\small t}^{{\small +}})] &=&\left[ 
\begin{tabular}{|l|l|}
\hline
$1-\frac{i}{\hbar }\left\{ -\frac{\hbar ^{2}}{2m}{\small \nabla }^{2}\,+%
{\small V}_{u}\right\} \delta t$ & $\sqrt{\frac{ig}{2\hbar }}\{\delta 
\widetilde{w}_{u,d}+i\,\delta \widetilde{w}_{d,u}\}$ \\ \hline
$\sqrt{\frac{ig}{2\hbar }}\{\delta \widetilde{w}_{u,d}-i\,\delta \widetilde{w%
}_{d,u}\}$ & $1-\frac{i}{\hbar }\left\{ -\frac{\hbar ^{2}}{2m}{\small \nabla 
}^{2}{\small \,+V}_{d}\right\} \delta t$ \\ \hline
\end{tabular}%
\ \right]  \nonumber \\
\lbrack \Theta ^{+}({\small t}^{{\small +}})] &=&\left[ 
\begin{tabular}{|l|l|}
\hline
$1+\frac{i}{\hbar }\left\{ -\frac{\hbar ^{2}}{2m}{\small \nabla }^{2}{\small %
\,+V}_{u}\right\} \delta t$ & $\sqrt{\frac{ig}{2\hbar }}\{\delta \widetilde{w%
}_{u+,d+}+i\,\delta \widetilde{w}_{d+,u+}\}$ \\ \hline
$\sqrt{\frac{ig}{2\hbar }}\{-\delta \widetilde{w}_{u+,d+}+i\,\delta 
\widetilde{w}_{d+,u+}\}$ & $1+\frac{i}{\hbar }\left\{ -\frac{\hbar ^{2}}{2m}%
{\small \nabla }^{2}{\small \,+V}_{d}\right\} \delta t$ \\ \hline
\end{tabular}%
\ \right]  \nonumber \\
&&  \label{Eq.ThetaMatrxFermiFldModel}
\end{eqnarray}%
and involve the Laplacian, the trap potentials, Wiener increments (\ref%
{Eq.WienerInc2}) as well as collision and mass parameters from the
Hamiltonian. Note that the Ito stochastic field equations are \emph{local}
in this case.

This form of the Ito stochastic field equations shows that the Grassmann
stochastic fields at time $t+\delta t$\ are linearly related to the
Grassmann stochastic fields at time $t$\ via quantities that only involve
c-numbers. The quantities include stochastic Wiener increments as well as
the Laplacian and spatially dependent potentials, but nevertheless no
Grassmann variables are involved. A similar Grassmann independent feature
applied to the situation when separate modes were treated (see Eq.(81) in
Ref. \cite{Dalton16b}), and is a feature restricted to treatments involving
the $B$ distribution, as Plimak et al first pointed out \cite{Plimak01a}. .

The result in (\ref{Eq.ItoStochFieldEqnsFermiFldModel}) (analogous to Eq.
(81) for separate modes in Ref. \cite{Dalton16b}) shows how Grassmann
stochastic fields at any time $t_{f}$\ can be related to Grassmann
stochastic fields at initial time $t_{0}$\ via c-number stochastic
quantities. The final\ stochastic averages of products of Grassmann
stochastic fields at initial time $t_{0}$ are determined from initial
conditions via c-number calculations, just as in the separate modes
case.\smallskip

\subsubsection{Case of Free Fermi Gas}

In the case of the free Fermi gas where the trap potential is zero, a useful
development of the result (\ref{Eq.ItoStochFieldEqnsFermiFldModel}) can be
obtained via introducing spatial Fourier transforms. In the case of the
stochastic fields the stochastic Fourier fields $\tilde{\phi}_{\alpha
}(k,t), $ $\tilde{\phi}_{\alpha }^{+}(k,t)$ are defined by 
\begin{eqnarray}
\tilde{\psi}_{\alpha }(s,t) &=&\int \frac{{\small dk}}{({\small 2\pi })^{3/2}%
}\,\exp (ik\cdot s)\,\tilde{\phi}_{\alpha }(k,t)  \nonumber \\
\tilde{\psi}_{\alpha }^{+}(s,t) &=&\int \frac{{\small dk}}{{\small (2\pi )}%
^{3/2}}\,\exp (-ik\cdot s)\,\tilde{\phi}_{\alpha }^{+}(k,t)
\label{Eq.FourierStochFields}
\end{eqnarray}%
and the inverse equations. The equations for $\tilde{\psi}_{\alpha }(s,t)$\
and $\tilde{\psi}_{\alpha }^{+}(s,t)$ become stochastic equations for the
Fourier transforms $\,\tilde{\phi}_{\alpha }(k,t)$ and $\,\tilde{\phi}%
_{\alpha }^{+}(k,t)$. We then have 
\begin{eqnarray}
&&\left[ 
\begin{tabular}{|l|}
\hline
$\tilde{\phi}_{u}(k,t+\delta t)$ \\ \hline
$\tilde{\phi}_{d}(k,t+\delta t)$ \\ \hline
$\tilde{\phi}_{u}^{+}(k,t+\delta t)$ \\ \hline
$\tilde{\phi}_{d}^{+}(k,t+\delta t)$ \\ \hline
\end{tabular}%
\ \right]  \nonumber \\
&=&\left[ 
\begin{tabular}{|l|l|l|l|}
\hline
$F_{u,u}(t)$ & $F_{u,d}(t)$ & $0$ & $0$ \\ \hline
$F_{d,u}(t)$ & $F_{d,d}(t)$ & $0$ & $0$ \\ \hline
$0$ & $0$ & $F_{u+,u+}^{+}(t)$ & $F_{u+,d+}^{+}(t)$ \\ \hline
$0$ & $0$ & $F_{d+,u+}^{+}(t)$ & $F_{d+,d+}^{+}(t)$ \\ \hline
\end{tabular}%
\ \right] \left[ 
\begin{tabular}{|l|}
\hline
$\tilde{\phi}_{u}(k\mathbf{,}t)$ \\ \hline
$\tilde{\phi}_{d}(k\mathbf{,}t)$ \\ \hline
$\tilde{\phi}_{u}^{+}(k,t)$ \\ \hline
$\tilde{\phi}_{d}^{+}(k,t)$ \\ \hline
\end{tabular}%
\ \right]  \nonumber \\
&&  \label{Eq.ItoStochFldsFourierFermi}
\end{eqnarray}%
where the sub-matrices are%
\begin{eqnarray}
\lbrack F(t)] &=&\left[ 
\begin{tabular}{|l|l|}
\hline
$1-\frac{i}{\hbar }\left\{ \frac{\hbar ^{2}}{2m}k^{2}\right\} \delta t$ & $%
\sqrt{\frac{ig}{2\hbar }}\{\delta \widetilde{w}_{u,d}+i\delta \widetilde{w}%
_{d,u}\}$ \\ \hline
$\sqrt{\frac{ig}{2\hbar }}\{\delta \widetilde{w}_{u,d}-i\delta \widetilde{w}%
_{d,u}\}$ & $1-\frac{i}{\hbar }\left\{ \frac{\hbar ^{2}}{2m}k^{2}\right\}
\delta t$ \\ \hline
\end{tabular}%
\ \right]  \nonumber \\
\lbrack F^{+}(t)] &=&\left[ 
\begin{tabular}{|l|l|}
\hline
$1+\frac{i}{\hbar }\left\{ \frac{\hbar ^{2}}{2m}k^{2}\right\} \delta t$ & $%
\sqrt{\frac{ig}{2\hbar }}\{\delta \widetilde{w}_{u+,d+}+i\delta \widetilde{w}%
_{d+,u+}\}$ \\ \hline
$\sqrt{\frac{ig}{2\hbar }}\{-\delta \widetilde{w}_{u+,d+}+i\delta \widetilde{%
w}_{d+,u+}\}$ & $1+\frac{i}{\hbar }\left\{ \frac{\hbar ^{2}}{2m}%
k^{2}\right\} \delta t$ \\ \hline
\end{tabular}%
\ \right] .  \nonumber \\
&&  \label{Eq.SubMatricesFermiFourier}
\end{eqnarray}%
which though stochastic no longer involve the Laplacian operator, which is
replaced by the c-number $k^{2}$. The equations can then be solved
numerically.

The field operators $\widehat{\psi }_{\alpha }(r)$ and $\widehat{\psi }%
_{\alpha }^{\dag }(r)$ themselves may be replaced by momentum field
operators $\,\widehat{\phi }_{\alpha }(k)$ and $\,\widehat{\phi }_{\alpha
}^{\dag }(k)$ using equations analogous to (\ref{Eq.FourierStochFields}).
The creation field momentum operators $\,\widehat{\phi }_{\alpha }^{\dag
}(k) $ create a fermionic atom with internal state $\alpha $ and having
momentum $k$, and these are the continuum versions of the mode creation
operators $\widehat{c}_{k\,\alpha }$ discussed in Section 6 of Paper I (\cite%
{Dalton16b}). Allowing for the difference between box normalisation and
delta function normalisation of the free space mode functions, the
application of the approach in the present section to the simple four mode
model in Section 6 of Paper I (\cite{Dalton16b}) yields the same results as
before for the coherence between momentum Fock states. \smallskip

\subsubsection{Case of Optical Lattice}

In the case where the Fermi gas is trapped in an optical lattice the trap
potential is periodic, and a useful development of the result (\ref%
{Eq.ItoStochFieldEqnsFermiFldModel}) can be obtained via introducing
transforms.based on Bloch functions. The stochastic field functions in terms
of Bloch functions $\chi _{\alpha }^{k,a}(\mathbf{r})$, where $k$ would
range over the Brillouin zone and $a$ would list the different bands. Such
functions obey an eigenvalue equation of the form 
\begin{equation}
\left\{ -\frac{\hbar ^{2}}{2m}\nabla ^{2}\,+V_{\alpha }\right\} \chi
_{\alpha }^{k,a}(r)=\hbar \omega _{\alpha }^{k,a}\chi _{\alpha }^{k,a}(r).
\label{Eq.Bloch}
\end{equation}%
This enables the Laplacian and the trap potential to be eliminated in favour
of a c-number, the Bloch energy $\hbar \omega _{\alpha }^{k,a}$. A solution
via Bloch function transforms can be developed, and applied in numerical
work.

\subsection{B. Multi-Component Fields with Non-Zero Range Interactions}

We can also treat the more general case of fermion fields with several
components, where the fields have non-zero range interactions as well as
each being coupled to external potentials. Many situations in cold Fermi
gases can be studied using this model. Here there are several distinct
internal spin states designated by $\alpha $. The Fermi gas is isolated from
the environment so no relaxation effects are involved. All atoms have the
same mass $m$.

\subsubsection{Hamiltonian}

The Hamiltonian is written in terms of the field operators $\hat{\Psi}%
_{\alpha }(r)$, $\hat{\Psi}_{\alpha }(r)^{\dag }$ as%
\begin{eqnarray}
\hat{H}_{f}\mbox{\rule{-0.5mm}{0mm}} &=&\mbox{\rule{-1mm}{0mm}}\int %
\mbox{\rule{-1mm}{0mm}}dr\left( \sum_{\alpha }\frac{\hbar ^{2}}{2m}\nabla 
\hat{\Psi}_{\alpha }(r)^{\dag }\mbox{\rule{-0.5mm}{0mm}}\cdot %
\mbox{\rule{-0.5mm}{0mm}}\nabla \hat{\Psi}_{\alpha }(\mathbf{\ }%
r)+\sum_{\alpha \beta }\hat{\Psi}_{\alpha }(r)^{\dag }\,V_{r;r}^{\alpha
;\beta }\,\hat{\Psi}_{\beta }(r)\right.  \nonumber \\
&&\hspace{2cm}\left. +\frac{1}{2}\int \mbox{\rule{-1mm}{0mm}}ds\sum_{\alpha
\beta ;\gamma \delta }\hat{\Psi}_{\alpha }(r)^{\dag }\hat{\Psi}_{\beta
}(s)^{\dag }\,V_{rs;rs}^{\alpha \beta ;\gamma \delta }\,\hat{\Psi}_{\delta
}(s)\hat{\Psi}_{\gamma }(r)\right)  \label{Eq.MultiCompHam} \\
&=&\widehat{{\small K}}+\widehat{{\small V}}+\widehat{{\small U}}
\label{Eq.TermsMultiCompHam}
\end{eqnarray}%
and is the sum of kinetic energy, one body interaction energy and two body
interaction energy terms. For simplicity the quantities $V_{r;r}^{\alpha
;\beta }\,$\ and $V_{rs;rs}^{\alpha \beta ;\gamma \delta }$ are real. The
one body interaction energy terms describe both the interaction of the
fermions with trapping potentials $(\alpha =\beta )$ as well as interactions
with coupling potentials $(\alpha \neq \beta )$. Thus a fermion at position $%
r$ with internal component $\alpha $ is changed into a fermion at position $%
r $ but with different component $\beta $. The quantities $V_{r;r}^{\alpha
;\beta }$ reflect the position dependence of these trapping potential and
coupling interactions. In the two body interaction energy terms a fermion at
position $r$ and internal component $\gamma $ interacts with a fermion at
position $s$ and internal component $\delta $, resulting in the fermion at
position $r$ changing into internal component $\alpha $ and the fermion at
position $s$ changing into internal component $\beta $. The quantities $%
V_{rs;rs}^{\alpha \beta ;\gamma \delta }$ reflect the position dependences
of these two body interactions, including there being a non-zero range
involved - hence the double spatial integral over both $r$ and $s$.

By interchanging $r$ and $s$ and from the hermitiancy of the one and two
body energy terms several symmetry relations can be established. We find
that 
\begin{eqnarray}
V_{r;r}^{\alpha ;\beta } &=&V_{r;r}^{\beta ;\alpha }  \label{Eq.OneBodySymm}
\\
V_{rs;rs}^{\alpha \beta ;\gamma \delta } &=&V_{sr;sr}^{\beta \alpha ;\delta
\gamma }\qquad \qquad V_{rs;rs}^{\alpha \beta ;\gamma \delta
}=V_{rs;rs}^{\gamma \delta ;\alpha \beta }  \label{Eq.TwoBodySymm}
\end{eqnarray}

\subsubsection{Functional Fokker-Planck Equation - B Distribution}

In the case considered the functional Fokker-Planck equations can be wriiten
in a simpler notation in which the field operators $\hat{\Psi}_{\alpha }(r)$%
, $\hat{\Psi}_{\alpha }(r)^{\dag }$ are represented via Grassmann fields $%
\psi _{\alpha }(r)$ and $\psi _{\alpha }^{+}(r)$ respectively - here we will
not use the $A=1,2$ notation to distinguish $\psi _{\alpha }(r)$ and $\psi
_{\alpha }^{+}(r)$ as we did previously. The functional Fokker-Planck
equations can be obtained via the application of the correspondence rules in
(\ref{Eq.FermionFieldOprsCorresRules}). They are given by 
\begin{eqnarray}
\frac{\partial }{\partial t}B[\psi ] &=&-\tsum\limits_{\alpha }\dint
dr\,(A_{\alpha }[r]\,B[\psi ])\frac{\overleftarrow{\delta }}{\delta \psi
_{\alpha }(r)}-\tsum\limits_{\alpha }\dint dr\,(A_{\alpha }^{+}[r]\,B[\psi ])%
\frac{\overleftarrow{\delta }}{\delta \psi _{\alpha }^{+}(r)}  \nonumber \\
&&{\small +}\frac{1}{2}\tsum\limits_{\alpha ,\beta }\diint
dr\,ds\,(D_{\alpha \,\beta }[r,s]\,B[\psi ])\frac{\overleftarrow{\delta }}{%
\delta \psi _{\beta }(s)}\frac{\overleftarrow{\delta }}{\delta \psi _{\alpha
}(r)}  \nonumber \\
&&{\small +}\frac{1}{2}\tsum\limits_{\alpha ,\beta }\diint
dr\,ds\,(D_{\alpha \,\beta }^{+;+}[r,s]\,B[\psi ])\frac{\overleftarrow{%
\delta }}{\delta \psi _{\beta }^{+}(s)}\frac{\overleftarrow{\delta }}{\delta
\psi _{\alpha }^{+}(r)}  \label{Eq.FFPENonZeroRange}
\end{eqnarray}%
where the drift vector and diffusiom matrix elements are given by%
\begin{eqnarray}
A_{\alpha }[r] &=&+\frac{i}{\hbar }\left( -\frac{\hbar ^{2}}{2m}\nabla
^{2}\psi _{\alpha }(r)+\dsum\limits_{\beta }V_{r;r}^{\alpha ;\beta }\psi
_{\beta }(r)\right)  \nonumber \\
A_{\alpha }^{+}[r] &=&-\frac{i}{\hbar }\left( -\frac{\hbar ^{2}}{2m}\nabla
^{2}\psi _{\alpha }^{+}(r)+\dsum\limits_{\beta }V_{r;r}^{\alpha ;\beta }\psi
_{\beta }^{+}(r)\right)  \label{Eq.NonZeroDriftFFPE}
\end{eqnarray}%
and 
\begin{eqnarray}
D_{\alpha \,\beta }[r,s] &=&-\frac{i}{\hbar }\left( \dsum\limits_{\gamma
\delta }\psi _{\gamma }(r)\,V_{rs;rs}^{\alpha \beta ;\gamma \delta }\,\psi
_{\delta }(s)\right)  \nonumber \\
D_{\alpha \,\beta }^{+;+}[r,s]\, &=&+\frac{i}{\hbar }\left(
\dsum\limits_{\gamma \delta }\psi _{\gamma }^{+}(r)\,V_{rs;rs}^{\alpha \beta
;\gamma \delta }\,\psi _{\delta }^{+}(s)\right)  \label{Eq.NonZeroDiffnFFPE}
\end{eqnarray}%
Note that there are no cross terms $D_{\alpha \,\beta }^{+;-}[r,s]\,$or $%
D_{\alpha \,\beta }^{-;+}[r,s].\,$\ The drift tems depend linearly on the
Grassmann fields, whilst the diffusion terms depend bilinearly on the
Grassmann fields. Furthermore, the Grassmann fields $\psi _{\alpha }(r)$ and 
$\psi _{\alpha }^{+}(r)$ are not coupled to each other, so the Ito
stochastic field equations for $\widetilde{\psi }_{\alpha }(r)$ do not
couple to those for $\widetilde{\psi }_{\alpha }^{+}(r)$ - so this
simplifies the theory. The same situation applied previously to the two
component zero range interaction case treated above.

We also see that the diffusion matrix is \emph{anti-symmetric}. 
\begin{equation}
D_{\alpha \,\beta }[r,s]=-D_{\beta \,\alpha }[s,r]\qquad \qquad D_{\alpha
\,\beta }^{+;+}[r,s]=-D_{\beta \,\alpha }^{+;+}[s,r]
\label{Eq.AntiSymmDiffnNonZero}
\end{equation}%
This result is easily established using the symmetry properties (\ref%
{Eq.TwoBodySymm}) for the $V_{rs;rs}^{\alpha \beta ;\gamma \delta }$
together with the anti-commutation of the Grassmann field functions.

\subsubsection{Ito Stochastic Field Equations}

We first must find matrices $B$ and $B^{+}$ such that $D=BB^{T}$ and $%
D^{+;+}=B^{+}(B^{+})^{T}$. This can be carried out via a generalisation of
the proceedure in Section 5 of Paper I \cite{Dalton16b} for the separate
modes case. We write $D_{\alpha \,\beta }[r,s]$ and $D_{\alpha \,\beta
}^{+;+}[r,s]$ in the forms 
\[
D_{\alpha r;\beta s}=\dsum\limits_{\gamma \delta }Q_{\gamma r;\,\delta
s}^{\alpha r;\,\beta s}\,\psi _{\gamma }(r)\psi _{\delta }(s)\qquad
D_{\alpha r;\beta s}^{+;\,+}=\dsum\limits_{\gamma \delta }(Q^{+})_{\gamma
r;\,\delta s}^{\alpha r;\,\beta s}\,\psi _{\gamma }^{+}(r)\psi _{\delta
}^{+}(s) 
\]%
where 
\[
(Q)_{\gamma r;\,\delta s}^{\alpha r;\,\beta s}=-\frac{i}{\hbar }%
V_{rs;rs}^{\alpha \beta ;\gamma \delta }\qquad (Q^{+})_{\gamma r;\,\delta
s}^{\alpha r;\,\beta s}=+\frac{i}{\hbar }V_{rs;rs}^{\alpha \beta ;\gamma
\delta } 
\]%
are new matrices created from the two body interaction terms $%
V_{rs;rs}^{\alpha \beta ;\gamma \delta }$. These matrices $Q$ and $Q^{+}$
are both \emph{symmetric} - which follows from the antisymmetry of $D$ and $%
D^{+;+}$. The matrices $Q$ and $Q^{+}$ are such that the rows are listed by
the double symbol $\alpha r,\gamma r$ and the columns by the double symbol $%
\beta s$,$\delta s$. The proceedure is analogous to the creation of the
matrix elements $Q_{r;s}^{p;q}$ in Section \ref{Section - Ito Stochastic
Equations} where the rows were listed by the double symbol $p,r$\ and
columns listed as $q,s$. However, in this field theory situation we have
implicitly discretised the spatial positions first.

As in Section 5 of Paper I \cite{Dalton16b} we can then use Takagi
factorisation \cite{Takagi25a} of the symmetric matrices $Q$ and $Q^{+}$ to
write%
\begin{equation}
(Q)_{\gamma r;\,\delta s}^{\alpha r;\,\beta s}=\dsum\limits_{a}K_{\gamma
r;\,a}^{\alpha r}\,K_{\delta s;\,a}^{\beta s}\,\qquad (Q^{+})_{\gamma
r;\,\delta s}^{\alpha r;\,\beta s}=\dsum\limits_{b}(K^{+})_{\gamma
r;\,b}^{\alpha r}\,(K^{+})_{\delta s;\,b}^{\beta s}
\label{Eq.TakagiNonZeroFlds}
\end{equation}%
and then it follows that with 
\begin{equation}
B_{a}^{\alpha r}=\dsum\limits_{\gamma }K_{\gamma r;\,a}^{\alpha r}\,\psi
_{\gamma }(r)\qquad (B^{+})_{b}^{\alpha r}=\dsum\limits_{\gamma
}(K^{+})_{\gamma r;\,b}^{\alpha r}\,\psi _{\gamma }^{+}(r)
\label{Eq.NoiseFactorsNonZero}
\end{equation}%
Note that $K$ and $K^{+}$ are inter-related via $K^{+}=iK$, since $%
KK^{T}=-(K^{+})(K^{+})^{T}$.

Hence using the general results (\ref{Eq.ItoStochFieldEqn}), (\ref%
{Eq.ClassicalFieldResult}), (\ref{Eq.NoiseFieldResult}) and the particular
formulae for the present case (\ref{Eq.NonZeroDriftFFPE}) and (\ref%
{Eq.NoiseFactorsNonZero}) the Ito stochastic field equations are given by%
\begin{eqnarray}
\delta \widetilde{\psi }_{\alpha }(r) &\equiv &\widetilde{\psi }_{\alpha
}(r,t+\delta t)-\widetilde{\psi }_{\alpha }(r,t)  \nonumber \\
&=&-\frac{i}{\hbar }\left( -\frac{\hbar ^{2}}{2m}\nabla ^{2}\widetilde{\psi }%
_{\alpha }(r,t)+\dsum\limits_{\beta }V_{r;r}^{\alpha ;\beta }\widetilde{\psi 
}_{\beta }(r,t)\right) \delta t  \nonumber \\
&&+\dsum\limits_{a}\dsum\limits_{\gamma }K_{\gamma r;\,a}^{\alpha r}\,%
\widetilde{\psi }_{\gamma }(r,t)\;\delta \widetilde{W}_{a}(t_{+})
\label{Eq.ItoSFENonZero1}
\end{eqnarray}%
and 
\begin{eqnarray}
\delta \widetilde{\psi }_{\alpha }^{+}(r) &\equiv &\widetilde{\psi }_{\alpha
}^{+}(r,t+\delta t)-\widetilde{\psi }_{\alpha }^{+}(r,t)  \nonumber \\
&=&+\frac{i}{\hbar }\left( -\frac{\hbar ^{2}}{2m}\nabla ^{2}\widetilde{\psi }%
_{\alpha }^{+}(r,t)+\dsum\limits_{\beta }V_{r;r}^{\alpha ;\beta }\widetilde{%
\psi }_{\beta }^{+}(r,t)\right) \delta t  \nonumber \\
&&+\dsum\limits_{b}\dsum\limits_{\gamma }(K^{+})_{\gamma r;\,b}^{\alpha r}\,%
\widetilde{\psi }_{\gamma }^{+}(r,t)\;\delta \widetilde{W}_{b}(t_{+})
\label{Eq.ItoSFENonZero2}
\end{eqnarray}%
Once again the Ito stochastic field equations are purely local and the Ito
stochastic fields at time $t+\delta t$ \ are related linearly to those at
time $t$ via quantities that only involve c-numbers - those these are
stochastic due to the Wiener noise increments $\delta \widetilde{W}%
_{a,b}(t_{+})$. We also note that the equations for $\widetilde{\psi }%
_{\alpha }(r)$ and $\widetilde{\psi }_{\alpha }^{+}(r)$ are not coupled.
Depending on the number of component fields and on the number of spatial
grid points needed to represent the non-zero range two body terms $%
V_{rs;rs}^{\alpha \beta ;\gamma \delta }$, the numbers of Wiener increments
needed could be very large. However, at least in principle the Ito
stochastic field equations are numerically solvable \pagebreak

\section{Summary and Conclusions}

\label{Section - Summary and Conclusion}

A phase space theory for fermions has been presented for fermion systems
based on distribution functionals, which replace the density operator and
involve Grassmann fields representing anti-commuting fermion field
annihilation, creation operators. This paper is an extension of a previous
paper (Paper I, Ref. \cite{Dalton16b})\ for fermion systems based on
separate modes, in which the density operator is replaced by a distribution
function depending on Grassmann phase space variables which represent the
mode annihilation and creation operators. The distribution functional
involving Grassmann fields and the distribution function involving Grassmann
phase space variables are equivalent - being two alternative ways of
representing the density operator, since the Grassmann fields can be
expanded in terms of orthonormal mode functions with the Grassmann phase
variables as expansion coefficients. This field theory extension is
important in the case when large numbers of fermions are involved, since the
Pauli exclusion principle results in too many modes to treat separately.
Quantum correlation functions, Fock state populations and coherences are
given as phase space Grassmann functional integrals. Functional
Fokker-Planck equations for the distribution functional have been obtained
both starting from the Fokker-Planck equations for the distribution
function, and also based on the correspondence rules for the effect of
fermion field annihilation, creation operators on the density operator. Ito
stochastic field equations are derived, both starting from the Ito
stochastic equations for stochastic phase space variables and also directly
from the functional Fokker-Planck equation. The Ito stochastic field
equations can be non-local, but in many cases they are local. The classical
and noise fields shown to be determined from the drift and diffusion terms
in functional Fokker-Planck equation. For the $B$ distribution functional
case the stochastic Grassmann fields at a later time are related linearly to
those at an earlier time via c-number quantities involving Wiener increments
and other terms in the Ito stochastic field equations. This makes numerical
calculations possible, with the stochastic averages of products Grassmann
fields at a later time being linearly related to such products (of the same
order) at an initial time. These initial time stochastic averages are
obtainable from the initial conditions. Applications of the theory to a
trapped interacting Fermi gas is presented, both for a two component case
with zero range interactions and multi-component cases with non-zero range
interactions. The Ito stochastic field equations are local in both the zero
range and finite range interaction cases treated.

It will be of interest to apply the theory to various topics in degenerate
Fermi gases. The utility of the theory could first be tested on some
well-understood fermion systems that have been treated by other methods,
such determining the size of a Cooper pair or treating Feshbach resonance in
Fermi gases or deriving the two fluid hydrodynamuic equations for Fermi
liquids. Some further development of the formalism could be worthwhile, such
as obtaining formulae for two-time quantum correlation functions, as these
are linked to fermionic excitations. Some work on this already exists \cite%
{Plimak09a}. Degenerate quantum gases involving both fermions and bosons
often occur, so expanding the formalism to include both is desirable, and
again some work has already been done \cite{Dalton13a}, \cite{Dalton15a}.
Also, extending the phase space method to include stochastic gauges could be
worthwhile to facilitate numerical calculations, and here some work on this
for both bosons \cite{Deuar06a} and fermions \cite{Corney06b} using c-number
phase space theory has been carried out. It is clearly desirable to carry
out numerical applications on fermion systems to fully test out the
Grassmann phase space theory. \medskip

\section{Acknowledgements}

This work was supported by the Australian Research Council Centre of
Excellence for Quantum Atom Optics (Grant Number CE0348178), the Royal
Society and the Wolfson foundation. The authors thank J. Corney, P.
Drummond, M. Olsen and L. Plimak for helpful discussions. The authors are
grateful to Dr C Gilson, Dept. of Applied Maths., University of Glasgow for
the first derivation of a solution to the factorisation $D=BB^{T}$ for
anti-symmetric Grassmann matrices $D$ and to a referee for directing our
attention to the locality issue and other aspects involved in numerical
calculations.

\pagebreak

\section{Appendix A - Fermion Position States}

\label{Appendix - Fermion Position States}

To show that the field creation operators do indeed create a particle at a
particular position we note that the single particle state $|\phi
_{i}\rangle $ can be expressed in terms of Dirac delta function normalised 
\emph{position eigenstates} $\left\vert \mathbf{r}\right\rangle $ via 
\begin{eqnarray}
|\phi _{i}\rangle &=&\int d\mathbf{r\,}\left\vert \mathbf{r}\right\rangle
\left\langle \mathbf{r\,|}\phi _{i}\right\rangle =\int d\mathbf{r\,}%
\left\vert \mathbf{r}\right\rangle \phi _{i}(\mathbf{r})  \nonumber \\
\left\langle \mathbf{r\,|\mathbf{r}^{\prime }}\right\rangle &=&\mathbf{%
\delta (r-r}^{\prime })  \label{Eq.PositionRepn}
\end{eqnarray}%
where the mode function $\phi _{i}(\mathbf{r})$ is the position
representation of $|\phi _{i}\rangle $. Using completeness it then follows
that 
\begin{equation}
\left\vert \mathbf{r}\right\rangle =\sum_{i}\phi _{i}^{\ast }(\mathbf{r}%
)|\phi _{i}\rangle  \label{Eq.SingleParticlePositionState}
\end{equation}%
gives the position eigenstates in terms of the single particle states $|\phi
_{i}\rangle $. The sum is over all modes.

We now consider a set of distinct particle positions $\mathbf{r}_{1},\mathbf{%
r}_{2}\cdots \mathbf{r}_{N}$. As in the case of the single particle modes an
ordering convention is invoked to avoid duplication. Thus $\mathbf{r}_{1}<%
\mathbf{r}_{2}<\cdots <\mathbf{r}_{N}$. In first quantisation we introduce
the state $\left\vert \Pi \right\rangle $ obtained from modes $%
l_{1},l_{2},...,l_{N}$ by taking the product of $\phi _{l_{1}}^{\ast }(%
\mathbf{r}_{1})\cdots \phi _{l_{N}}^{\ast }(\mathbf{r}_{N})$ with the first
quantisation form of the Fock state $|l_{1}\cdots l_{N}\rangle $ in which
the modes are in any order, and then summing over all the $%
l_{1},l_{2},...,l_{N}$. Note that terms in which the same mode appears twice
or more will be incuded, but as these are zero due to the Pauli principle
this is not a problem. Substituting for $|l_{1}\cdots l_{N}\rangle $ using
the anti-symmetrising operator $\mathcal{A}$, which involves the sum of all $%
N!$ permutation operators for the $N$ particles, each permuation being
multiplied by $+1,-1$ according to whether is is even,odd, and then using (%
\ref{Eq.SingleParticlePositionState}).we find that 
\begin{eqnarray}
\left\vert \Pi \right\rangle &=&\sum\limits_{l_{1},\cdots ,l_{N}}%
\mbox{\rule{-2mm}{0mm}}\phi _{l_{1}}^{\ast }(\mathbf{r}_{1})\cdots \phi
_{l_{N}}^{\ast }(\mathbf{r}_{N})|l_{1}\cdots l_{N}\rangle  \nonumber \\
&=&\mbox{\rule{-2mm}{0mm}}\sum\limits_{l_{1},\cdots ,l_{N}}%
\mbox{\rule{-2mm}{0mm}}\phi _{l_{1}}^{\ast }(\mathbf{r}_{1})\cdots \phi
_{l_{N}}^{\ast }(\mathbf{r}_{N})(\mathcal{A})|l_{1}(1)\rangle \cdots
|l_{N}(N)\rangle  \nonumber \\
&=&(\mathcal{A})|\mathbf{r}_{1}(1)\rangle \cdots |\mathbf{r}_{N}(N)\rangle 
\nonumber \\
&=&|\mathbf{r}_{1}\cdots \mathbf{r}_{N}\rangle
\end{eqnarray}%
which is a symmetrised state $|\mathbf{r}_{1}\cdots \mathbf{r}_{N}\rangle $
in which there is one particle at $\mathbf{r}_{1}$, a second at $\mathbf{r}%
_{2}$, $\cdots $ , an $N$th at $\mathbf{r}_{N}$. But in second quantisation
the state $\left\vert \Pi \right\rangle $ can be written 
\begin{eqnarray}
\left\vert \Pi \right\rangle &=&\sum\limits_{l_{1},\cdots ,l_{N}}%
\mbox{\rule{-2mm}{0mm}}\phi _{l_{1}}^{\ast }(\mathbf{r}_{1})\cdots \phi
_{l_{N}}^{\ast }(\mathbf{r}_{N})|l_{1}\cdots l_{N}\rangle  \nonumber \\
&=&\mbox{\rule{-2mm}{0mm}}\sum\limits_{l_{1},\cdots ,l_{N}}%
\mbox{\rule{-2mm}{0mm}}\phi _{l_{1}}^{\ast }(\mathbf{r}_{1})\cdots \phi
_{l_{N}}^{\ast }(\mathbf{r}_{N})\hat{c}_{l_{1}}^{\dag }\cdots \hat{c}%
_{l_{P}}^{\dag }|0\rangle  \nonumber \\
&=&\hat{\Psi}^{\dag }(\mathbf{r}_{1})\cdots \hat{\Psi}^{\dag }(\mathbf{r}%
_{N})|0\rangle
\end{eqnarray}%
where again for terms where a given mode appears twice or more, such terms
are zero because they involve repeated mode creation operators and therefore
can be included with no change in $\left\vert \Pi \right\rangle $. Also if
the $l_{1},l_{2},...,l_{N}$ are not in conventional order, the result for
distinct occupied modes $|l_{1}\cdots l_{N}\rangle =\hat{c}_{l_{1}}^{\dag
}\cdots \hat{c}_{l_{P}}^{\dag }|0\rangle $ still applies, since both sides
can be changed to conventional order by multiplying by $+1$ or $-1$. The
last expression for $\left\vert \Pi \right\rangle $ is obtained using the
expression (\ref{Eq.FermionFieldOprs2}) for the field creation operators.
Comparing the two results for $\left\vert \Pi \right\rangle $ we see that
the symmetrized state $|\mathbf{r}_{1}\cdots \mathbf{r}_{N}\rangle $ is
given by 
\begin{equation}
|\mathbf{r}_{1}\cdots \mathbf{r}_{N}\rangle =\hat{\Psi}^{\dag }(\mathbf{r}%
_{1})\cdots \hat{\Psi}^{\dag }(\mathbf{r}_{N})|0\rangle
\label{Eq.MultiFermionPositionStateB}
\end{equation}%
Thus the field creation operators do in fact create particles at particular
positions, ordered as $\mathbf{r}_{1}<\mathbf{r}_{2}<\cdots <\mathbf{r}_{N}$%
. The normalisation condition for the position states are 
\begin{equation}
\left\langle \mathbf{r}_{1}\cdots \mathbf{r}_{N}|\,\mathbf{s}_{1}\cdots 
\mathbf{s}_{N}\right\rangle =\delta (\mathbf{r}_{1}-\mathbf{s}_{1})\cdots
\delta (\mathbf{r}_{N}-\mathbf{s}_{N}).  \label{Eq.FockPositionNorm}
\end{equation}%
can be obtained using the anti-commutation rules.

\pagebreak

\section{Appendix B - Grassmann Functional Calculus}

\label{Appendix - Grassmann Functional Calculus}

In this Appendix the main features of Grassmann functional calculus will be
set out. For simplicity we will only consider functionals of a single
Grassmann field $\psi (x)$ or a pair of Grassmann fields $\psi (x),\psi
^{+}(x)$. The variable $x$ refers to position, which may be in 1D, 2D or 3D.

\subsection{Examples of Grassman-Number Functionals}

A somewhat trivial application of the Grassmann functional concept is to
express a Grassmann field $\psi (y)$ as a functional $F_y[\psi (x)]$ of $%
\psi (x)$ 
\begin{eqnarray}
F_y[\psi (x)] &\equiv &\psi (y). \\
&=&\int dx\,\delta (y-x)\,\psi (x)  \label{Eq.GrassmannFnalExample1}
\end{eqnarray}
Here this specific functional involves the Dirac delta function as a kernel.

Another example involves the spatial derivative $\bigtriangledown _{y}\psi
(y)$ which may also be expressed as a functional $F_{\nabla y}[\psi (x)]$ 
\begin{eqnarray}
F_{\nabla y}[\psi (x)] &\equiv &\bigtriangledown _{y}\psi (y) \\
&=&\int dx\,\delta (y-x)\,\smallskip \bigtriangledown _{x}\psi (x)  \nonumber
\\
&=&-\int dx\,\{\bigtriangledown _{x}\delta (y-x)\}\,\smallskip \psi (x) 
\nonumber \\
&=&\int dx\,\{\bigtriangledown _{y}\delta (y-x)\}\,\smallskip \psi (x)
\label{Eq.GrassmannFnalExample2}
\end{eqnarray}%
Here the functional involves $\bigtriangledown _{y}\delta (y-x)$ as a kernel.

A functional is said to be linear if 
\begin{equation}
F[c_{1}\psi _{1}(x)+c_{2}\psi _{2}(x)]=c_{1}F[\psi _{1}(x)]+c_{2}F[\psi
_{2}(x)]  \label{Eq.GrassmannLinearFnal}
\end{equation}%
where $c_{1},c_{2}$ are constants. Both the Grassmann function $\psi (y)$
and its spatial derivative are linear functionals.

An example of a non-linear Grassmann functional is 
\begin{eqnarray}
F_{\chi (2)} &=&\left( \int dx\,\chi ^{\ast }(x)\,\psi (x)\right) ^{2} 
\nonumber \\
&=&\int \int dx\,dy\,\chi ^{\ast }(x)\,\chi ^{\ast }(y)\,\psi (x)\,\psi (y)
\label{Eq.NonLinGrassFnal}
\end{eqnarray}%
\smallskip

\subsection{Functional Differentiation for Grassmann Fields}

\label{Sec10.7}

The left functional derivative$\frac{\overrightarrow{{\LARGE \delta }}}{%
{\LARGE \delta \psi (x)}}F[\psi (x)]$ is defined by 
\begin{equation}
F[\psi (x)+\delta \psi (x)]\approx F[\psi (x)]+\int dx\,\delta \psi
(x)\,\left( \frac{\overrightarrow{\delta }}{\delta \psi (x)}F[\psi
(x)]\right) _{x}  \label{Eq.GrassmannFuncDeriv1}
\end{equation}%
where $F[\psi (x)+\delta \psi (x)]-F[\psi (x)]$ is evaluated correct to the
first order in a change $\delta \psi (x)$ in the Grassmann field. Note that
the idea of smallness does not apply to the Grassmann field change $\delta
\psi (x)$. In Eq.(\ref{Eq.GrassmannFuncDeriv1}) the left side is a
functional of $\psi (x)+\delta \psi (x)$ and the first term on the right
side is a functional of $\psi (x)$. Both these quantities and the second
term on the right side are Grassmann functions. The latter term is a
functional of the Grassmann field $\delta \psi (x)$ and thus the functional
derivative must be a Grassmann function of $x$, hence the subscript $x$. In
most situations this subscript will be left understood. Note that $\Delta
F=F[\psi (x)+\delta \psi (x)]-F[\psi (x)]$ will in general involve terms
that are non-linear in $\delta \psi (x)$. For the example in Eq.(\ref%
{Eq.NonLinGrassFnal}) $\Delta F=\int \int dx\,dy\,\chi ^{\ast }(x)\,\chi
^{\ast }(y)\,\delta \psi (x)\,\delta \psi (y)$ and here the functional
derivative is zero.

In addition we note that the Grassmann function $\delta \psi (x)$ does not
necessarily commute with the g-function $\left( \frac{\overrightarrow{%
{\Large \delta }}}{{\Large \delta \psi (x})}F[\psi (x)]\right) _{x}$. This
therefore means that there is also a right functional derivative $\left(
F[\psi (x)]\frac{\overleftarrow{{\LARGE \delta }}}{{\LARGE \delta \psi (x)}}%
\right) _{x}$ defined by 
\begin{equation}
F[\psi (x)+\delta \psi (x)]\approx F[\psi (x)]+\int dx\,\left( F[\psi (x)]%
\frac{\overleftarrow{\delta }}{\delta \psi (x)}\right) _{x}\,\delta \psi (x)
\label{Eq.GrassmannRightFuncDeriv1}
\end{equation}%
We emphasise again: the functional derivative of a g-number functional is a
Grassmann function, not a functional. The specific examples below illustrate
this feature.

For functionals of the form $F[\psi (x),\psi ^{+}(x)]$ we have similar
expressions for the left and right functional derivatives but now with
respect to either $\psi (x)$ or $\psi ^{+}(x)$ 
\begin{eqnarray}
&&F[\psi (x),\psi ^{+}(x)+\delta \psi ^{+}(x)]
\label{Eq.GrassmannFuncDeriv2B} \\
&\approx &F[\psi (x),\psi ^{+}(x)]+\int dx\delta \psi ^{+}(x)\left[ \frac{%
\overrightarrow{\delta }}{\delta \psi ^{+}(x)}F[\psi (x),\psi ^{+}(x)]\right]
_{x}  \nonumber \\
&\approx &F[\psi (x),\psi ^{+}(x)]+\int dx\left[ F[\psi (x),\psi ^{+}(x)]%
\frac{\overleftarrow{\delta }}{\delta \psi ^{+}(x)}\right] _{x}%
\mbox{\rule{-1mm}{0mm}}\delta \psi ^{+}(x).
\label{Eq.GrassmannRightFuncDeriv2B}
\end{eqnarray}%
Finally, higher order functional derivatives can be defined by applying the
basic definitions to lower order functional derivatives.\smallskip

\subsection{Examples of Grassmann Functional Derivatives}

For the case of the functional $F_y[\psi (x)]$ in Eq.(\ref%
{Eq.GrassmannFnalExample1}) that gives the function $\psi (y).$ As 
\begin{eqnarray*}
F_y[\psi (x)+\delta \psi (x)]-F_y[\psi (x)] &=&\psi (y)+\delta \psi (y)-\psi
(y) \\
&=&\int dx\,\delta \psi (x)\,\delta (y-x) \\
&=&\int dx\,\delta \psi (x)\,\left( \frac{\overrightarrow{\delta }}{\delta
\psi (x)}F_y[\psi (x)]\right) _x \\
&=&\int dx\,\delta (y-x)\,\delta \psi (x) \\
&=&\int dx\,\left( F_y[\psi (x)]\frac{\overleftarrow{\delta }}{\delta \psi
(x)}\right) _x\,\delta \psi (x)
\end{eqnarray*}
we have the same result for the left and right Grassmann functional
derivative%
\begin{eqnarray}
\left( \frac{\overrightarrow{\delta }}{\delta \psi (x)}F_y[\psi (x)]\right)
_x &=&\left( \frac{\overrightarrow{\delta }\psi (y)}{\delta \psi (x)}\right)
_x \\
&=&\delta (y-x)  \label{Eq.GrassmannFuncDerivExample1} \\
\left( F_y[\psi (x)]\frac{\overleftarrow{\delta }}{\delta \psi (x)}\right)
_x &=&\left( \frac{\psi (y)\overleftarrow{\delta }}{\delta \psi (x)} \right)
_x \\
&=&\delta (y-x)  \label{Eq.GrassmannRightFuncDerivEx1}
\end{eqnarray}
so here the left and right functional derivatives are a delta function, just
as for c-numbers.

A similar situation applies to the functional $F_{\nabla y}[\psi (x)]$ in
Eq.(\ref{Eq.GrassmannFnalExample2}) that gives the spatial derivative
function $\bigtriangledown _{y}\psi (y)$. We have the same result for the
left and right Grassmann functional derivatives%
\begin{eqnarray}
\left( \frac{\overrightarrow{\delta }}{\delta \psi (x)}F_{\nabla y}[\psi
(x)]\right) _{x} &=&\left( \frac{\overrightarrow{\delta }\,\bigtriangledown
_{y}\psi (y)}{\delta \psi (x)}\right) _{x}  \nonumber \\
&=&\bigtriangledown _{y}\delta (y-x)  \label{Eq.GrassmannFuncDerivExample2}
\\
\left( F_{\nabla y}[\psi (x)]\frac{\overleftarrow{\delta }}{\delta \psi (x)}%
\right) _{x} &=&\left( \frac{\nabla _{y}\psi (y)\overleftarrow{\delta }}{%
\delta \psi (x)}\right) _{x}  \nonumber \\
&=&\nabla _{y}\delta (y-x)  \label{Eq.GrassmannRightFuncDerivEx2}
\end{eqnarray}%
so again the functional derivatives are the spatial derivative of a delta
function. \smallskip

\subsection{Grassmann Functional Derivative and Mode Functions}

We consider functionals $F[\psi (x),\psi ^{+}(x)]$ of both fields $\psi
(x),\psi ^{+}(x)$ and first consider functional derivatives with respect to $%
\psi (x)$. If a mode expansion for the Grassmann field $\psi (x)$ as in Eq.(%
\ref{Eq.GrassmannFieldFn2}) is performed, then we can obtain an expression
for the functional derivative with respect to $\psi (x)$ in terms of mode
functions. With 
\begin{equation}
\delta \psi (x)=\sum\limits_{k}\delta g_{k}\phi _{k}(x)
\label{Eq.GrassmannFieldVariatn}
\end{equation}%
where $\delta g_{1}$, $\delta g_{2}$, ..., $\delta g_{n}$ are Grassmann
variables that determmine the change $\delta \psi (x)$ in the Grassmann
field $\psi (x)$, we see that 
\begin{eqnarray*}
&&F[\psi (x)+\delta \psi (x),\psi ^{+}(x)]-F[\psi (x),\psi ^{+}(x)] \\
&\approx &\int dx\,\delta \psi (x)\,\left( \frac{\overrightarrow{\delta }}{%
\delta \psi (x)}F[\psi (x),\psi ^{+}(x)]\right) _{x} \\
&\approx &\sum\limits_{k}\delta g_{k}\int dx\,\phi _{k}(x)\,\left( \frac{%
\overrightarrow{\delta }}{\delta \psi (x)}F[\psi (x),\psi ^{+}(x)]\right)
_{x}
\end{eqnarray*}%
Suppose we write $F[\psi (x),\psi ^{+}(x)]$ as a Grassmann function, $F[\psi
(x),\psi ^{+}(x)]=f(g_{1},\cdots ,g_{k},\cdots g_{n},g_{1}^{+},\cdots
,g_{k}^{+},\cdots g_{n}^{+})$. Applying a Taylor series expansion (see Eq.
(155) in Ref. \cite{Dalton16b}) we have 
\begin{eqnarray*}
&&f(g_{1}+\delta g_{1},\cdots ,g_{k}+\delta g_{k},\cdots g_{n}+\delta
g_{n},g_{1}^{+},\cdots ,g_{k}^{+},\cdots g_{n}^{+}) \\
&&-f(g_{1},\cdots ,g_{k},\cdots g_{n},g_{1}^{+},\cdots ,g_{k}^{+},\cdots
g_{n}^{+}) \\
&=&\sum\limits_{k}\delta g_{k}\,\left\{ \frac{\overrightarrow{\partial }}{%
\partial g_{k}}f(g_{1},\cdots ,g_{k},\cdots g_{n},g_{1}^{+},\cdots
,g_{k}^{+},\cdots g_{n}^{+})\right\}
\end{eqnarray*}%
correct to first order in the $\delta g_{k}$ so that we can write 
\begin{eqnarray*}
&&F[\psi (x)+\delta \psi (x),\psi ^{+}(x)]-F[\psi (x),\psi ^{+}(x)] \\
&=&\sum\limits_{k}\delta g_{k}\,\frac{\overrightarrow{\partial }}{\partial
g_{k}}f(g_{1},\cdots ,g_{k},\cdots g_{n},g_{1}^{+},\cdots ,g_{k}^{+},\cdots
g_{n}^{+})
\end{eqnarray*}%
We therefore have found that 
\begin{eqnarray*}
&&\sum\limits_{k}\delta g_{k}\int dx\,\phi _{k}(x)\left( \frac{%
\overrightarrow{\delta }}{\delta \psi (x)}F[\psi (x),\psi ^{+}(x)]\right)
_{x} \\
&=&\sum\limits_{k}\delta g_{k}\,\frac{\overrightarrow{\partial }}{\partial
g_{k}}f(g_{1},\cdots ,g_{k},\cdots g_{n},g_{1}^{+},\cdots ,g_{k}^{+},\cdots
g_{n}^{+})
\end{eqnarray*}%
Equating the coefficient of the $\delta g_{k}$ and then using the
completeness relationship analogous to Eq.(\ref{Eq.Completeness}) gives the
key result%
\begin{equation}
\left( \frac{\overrightarrow{\delta }}{\delta \psi (x)}F[\psi (x),\psi
^{+}(x)]\right) _{x}=\sum\limits_{k}\phi _{k}^{\ast }(x)\,\frac{%
\overrightarrow{\partial }}{\partial g_{k}}f(g_{1},\cdots ,g_{k},\cdots
g_{n},g_{1}^{+},\cdots ,g_{k}^{+},\cdots g_{n}^{+}).
\label{Eq.GrassmannLeftFnalDerivResult}
\end{equation}%
This relates the left functional derivative to the mode functions and to the
ordinary left Grassmann derivatives of the function $f(g_{1},\cdots
,g_{k},\cdots g_{n},g_{1}^{+},\cdots ,g_{k}^{+},\cdots g_{n}^{+})$ that was
equivalent to the original Grassmann functional $F[\psi (x),\psi ^{+}(x)]$.
Again, we see that the result is a function of $x$. Note that the left
functional derivative involves an expansion in terms of the conjugate mode
functions $\phi _{k}^{\ast }(x)$ rather than the original modes $\phi
_{k}(x) $. The last result may be put in the form of a useful operational
identity 
\begin{equation}
\left( \frac{\overrightarrow{\delta }}{\delta \psi (x)}\right)
_{x}=\sum\limits_{k}\phi _{k}^{\ast }(x)\,\frac{\overrightarrow{\partial }}{%
\partial g_{k}}  \label{Eq.GrassmannLeftFnalDerivResult2}
\end{equation}%
where the left side is understood to operate on an arbitary functional $%
F[\psi (x),\psi ^{+}(x)]$ and the right side is understood to operate on the
equivalent function

$f(g_{1},\cdots ,g_{k},\cdots g_{n},g_{1}^{+},\cdots ,g_{k}^{+},\cdots
g_{n}^{+})$.

Similar results can be obtained for the right functional derivative with
respect to $\psi (x)$ 
\begin{equation}
\left( F[\psi (x),\psi ^{+}(x)]\frac{\overleftarrow{\delta }}{\delta \psi (x)%
}\right) _{x}=\sum\limits_{k}\phi _{k}^{\ast }(x)\,f(g_{1},\cdots
,g_{k},\cdots g_{n},g_{1}^{+},\cdots ,g_{k}^{+},\cdots g_{n}^{+})\frac{%
\overleftarrow{\partial }}{\partial g_{k}}
\label{Eq.GrassmannRightFnalDerivResult}
\end{equation}%
and 
\begin{equation}
\left( \frac{\overleftarrow{\delta }}{\delta \psi (x)}\right)
_{x}=\sum\limits_{k}\phi _{k}^{\ast }(x)\,\frac{\overleftarrow{\partial }}{%
\partial g_{k}}.  \label{Eq.GrassmannRightFnalDerivResult2}
\end{equation}

The equivalent results for left and right functional derivatives with
respect to $\psi ^{+}(x)$ are%
\begin{eqnarray}
\left( \frac{\overrightarrow{\delta }}{\delta \psi ^{+}(x)}F[\psi (x),\psi
^{+}(x)]\right) _{x} &=&\sum\limits_{k}\phi _{k}(x)\,\frac{\overrightarrow{%
\partial }f(g_{1},\cdots ,g_{k},\cdots g_{n},g_{1}^{+},\cdots
,g_{k}^{+},\cdots g_{n}^{+})}{\partial g_{k}^{+}}  \nonumber \\
&&  \label{Eq.GrassmannLeftFnalDerivResultB} \\
\left( \frac{\overrightarrow{\delta }}{\delta \psi ^{+}(x)}\right) _{x}
&=&\sum\limits_{k}\phi _{k}(x)\,\frac{\overrightarrow{\partial }}{\partial
g_{k}^{+}}  \label{Eq.GrassmannLeftFnalDerivResult2B} \\
\left( F[\psi (x),\psi ^{+}(x)]\frac{\overleftarrow{\delta }}{\delta \psi
^{+}(x)}\right) _{x} &=&\sum\limits_{k}\phi _{k}(x)\,f(g_{1},\cdots
,g_{k},\cdots g_{n},g_{1}^{+},\cdots ,g_{k}^{+},\cdots g_{n}^{+})\frac{%
\overleftarrow{\partial }}{\partial g_{k}^{+}}  \nonumber \\
&&  \label{Eq.GrassmannRightFnalDerivResultB} \\
\left( \frac{\overleftarrow{\delta }}{\delta \psi ^{+}(x)}\right) _{x}
&=&\sum\limits_{k}\phi _{k}(x)\,\frac{\overleftarrow{\partial }}{\partial
g_{k}^{+}}.  \label{Eq.GrassmannRightFnalDerivResult2B}
\end{eqnarray}

The results for left and right functional derivatives can be inverted to
give 
\begin{eqnarray}
\frac{\overrightarrow{\partial }}{\partial g_{k}} &=&\int dx\phi
_{k}(x)\left( \frac{\overrightarrow{\delta }}{\delta \psi (x)}\right) _{x}
\label{Eq.GrassLeftModeDerivC} \\
\frac{\overrightarrow{\partial }}{\partial g_{k}^{+}} &=&\int dx\phi
_{k}^{\ast }(x)\left( \frac{\overrightarrow{\delta }}{\delta \psi ^{+}(x)}%
\right) _{x}  \label{Eq.GrassLeftModeDerivD} \\
\frac{\overleftarrow{\partial }}{\partial g_{k}} &=&\int dx\phi
_{k}(x)\left( \frac{\overleftarrow{\delta }}{\delta \psi (x)}\right) _{x}
\label{Eq.GrassRightModeDerivC} \\
\frac{\overleftarrow{\partial }}{\partial g_{k}^{+}} &=&\int dx\phi
_{k}^{\ast }(x)\left( \frac{\overleftarrow{\delta }}{\delta \psi ^{+}(x)}%
\right) _{x}.  \label{Eq.GrassRightModeDerivD}
\end{eqnarray}%
\smallskip

\subsection{Basic Rules for Grassmann Functional Derivatives}

It is possible to establish useful rules for the functional derivative of
the sum of two Grassmann functionals. It is easily shown that 
\begin{eqnarray}
\left( \frac{\overrightarrow{\delta }}{\delta \psi (x)}\right) _{x}%
\mbox{\rule{-2mm}{0mm}}\{F[\psi (x)]+G[\psi (x)]\} &=&\mbox{\rule{-1mm}{0mm}}%
\left( \frac{\overrightarrow{\delta }}{\delta \psi (x)}\right) _{x}%
\mbox{\rule{-2mm}{0mm}}F[\ ]+\left( \frac{\overrightarrow{\delta }}{\delta
\psi (x)}\right) _{x}\mbox{\rule{-2mm}{0mm}}G[\ ]
\label{Eq.GrassmannLefsumRule} \\
\{F[\ ]+G[\ ]\}\left( \frac{\overleftarrow{\delta }}{\delta \psi (x)}\right)
_{x} &=&F[\ ]\left( \frac{\overleftarrow{\delta }}{\delta \psi (x)}\right)
_{x}\mbox{\rule{-2mm}{0mm}}+G[\ ]\left( \frac{\overleftarrow{\delta }}{%
\delta \psi (x)}\right) _{x}  \label{Eq.GrassmannRighsumRule}
\end{eqnarray}%
with similar results for functionals of $\psi (x),\psi ^{+}(x)$.

Rules can be established for the functional derivative of the product of two
Grassmann functionals that depend on the parity of the functions equivalent
to the functionals, and the proofs are quite different to the c-number case.
For functionals that are neither even nor odd, results can be obtained by
expressing the relevant functional as a sum of even and odd contributions.
We will keep the functionals in order to cover the case where the
functionals are operators.

Correct to first order in $\delta \psi (x)$, we have from the definitions 
\begin{eqnarray*}
&&\int dx\,\delta \psi (x)\,\left( \frac{\overrightarrow{\delta }}{\delta
\psi (x)}\right) _{x}\{F[\psi (x)]G[\psi (x)]\} \\
&\approx &F[\psi (x)+\delta \psi (x)]G[\psi (x)+\delta \psi (x)]-F[\psi
(x)]G[\psi (x)] \\
&\approx &F[\psi (x)+\delta \psi (x)]G[\psi (x)+\delta \psi (x)]-F[\psi
(x)]G[\psi (x)+\delta \psi (x)] \\
&&+F[\psi (x)]G[\psi (x)+\delta \psi (x)]-F[\psi (x)]G[\psi (x)] \\
&\approx &\left\{ \int dx\,\delta \psi (x)\,\left( \frac{\overrightarrow{%
\delta }}{\delta \psi (x)}\right) _{x}F[\psi (x)]\right\} G[\psi (x)+\delta
\psi (x)] \\
&&+F[\psi (x)]\left\{ \int dx\,\delta \psi (x)\,\left( \frac{\overrightarrow{%
\delta }}{\delta \psi (x)}\right) _{x}G[\psi (x)]\right\} \\
&\approx &\int dx\,\delta \psi (x)\,\left\{ \left( \frac{\overrightarrow{%
\delta }}{\delta \psi (x)}\right) _{x}F[\psi (x)]\right\} G[\psi (x)] \\
&&+\int dx\,\delta \psi (x)\,\sigma (F)F[\psi (x)]\left\{ \left( \frac{%
\overrightarrow{\delta }}{\delta \psi (x)}\right) _{x}G[\psi (x)]\right\}
\end{eqnarray*}%
where $\sigma (F,G)=+1,-1$ depending on the parity of the function $f,g$
that is equivalent to the functionals $F,G$. In the first term the factor $%
G[\psi (x)+\delta \psi (x)]$ is replaced by $G[\psi (x)]$ to discard second
order contributions. Hence 
\begin{eqnarray}
\left( \frac{\overrightarrow{\delta }}{\delta \psi (x)}\right) _{x}\{F[\psi
(x)]G[\psi (x)]\} &=&\left\{ \left( \frac{\overrightarrow{\delta }}{\delta
\psi (x)}\right) _{x}F[\psi (x)]\right\} G[\psi (x)]  \nonumber \\
&+&\sigma (F)F[\psi (x)]\left\{ \left( \frac{\overrightarrow{\delta }}{%
\delta \psi (x)}\right) _{x}G[\psi (x)]\right\} \qquad
\label{Eq.GrassmannLefprodRule}
\end{eqnarray}%
A similar derivation covers the right functional derivative. The result is 
\begin{eqnarray}
\{F[\psi (x)]G[\psi (x)]\}\left( \frac{\overleftarrow{\delta }}{\delta \psi
(x)}\right) _{x} &=&F[\psi (x)]\left\{ G[\psi (x)]\left( \frac{%
\overleftarrow{\delta }}{\delta \psi (x)}\right) _{x}\right\}  \nonumber \\
&+&\sigma (G)\left\{ F[\psi (x)]\left( \frac{\overleftarrow{\delta }}{\delta
\psi (x)}\right) _{x}\right\} G[\psi (x)]\qquad
\label{Eq.GrassmannRighprodRule}
\end{eqnarray}%
These two rules are the functional derivative extensions of the previous
left and right product rules (see Eq. (154) in Ref. \cite{Dalton16b}) for
Grassmann derivatives. The extension to functionals of $\psi (x),\psi
^{+}(x) $ are obvious. These results are essentially the same as for
Grassmann ordinary differentiation.\smallskip

\subsection{Other Rules for Grassmann Functional Derivatives}

There are several rules that are needed because of the distinction between
left and right functional differentiation. These are analogous to rules
applying for left and right differentiation of Grassmann functions and may
be established using the mode based expressions for Grassmann functional
derivatives. With $\psi (x)$ a general Grassmann field these include:

(1) Right and left functional derivative relations for even and odd
functionals 
\begin{eqnarray}
\frac{\overrightarrow{\delta }}{\delta \psi (x)}F_{E}[\psi (x)]
&=&-F_{E}[\psi (x)]\frac{\overleftarrow{\delta }}{\delta \psi (x)}  \nonumber
\\
\frac{\overrightarrow{\delta }}{\delta \psi (x)}F_{O}[\psi (x)]
&=&+F_{O}[\psi (x)]\frac{\overleftarrow{\delta }}{\delta \psi (x)}
\label{Eq.LeftRightFnalDerivsResult}
\end{eqnarray}

(2) Altering order of functional derivatives 
\begin{equation}
\frac{\overrightarrow{\delta }}{\delta \psi (x)}\frac{\overrightarrow{\delta 
}}{\delta \psi (y)}F[\psi (x)]=-\frac{\overrightarrow{\delta }}{\delta \psi
(y)}\frac{\overrightarrow{\delta }}{\delta \psi (x)}F[\psi (x)]
\label{Eq.LeftDoubleFnalDerivResult}
\end{equation}
An analogous result applies for right functional derivatives.

(3) Mixed functional derivatives 
\begin{eqnarray}
\left( \frac{\overrightarrow{\delta }}{\delta \psi (x)}F[\psi (x)]\right) 
\frac{\overleftarrow{\delta }}{\delta \psi (y)} &=&\frac{\overrightarrow{%
\delta }}{\delta \psi (x)}\left( F[\psi (x)]\frac{\overleftarrow{\delta }}{%
\delta \psi (y)}\right)  \nonumber \\
&=&\frac{\overrightarrow{\delta }}{\delta \psi (x)}F[\psi (x)]\frac{%
\overleftarrow{\delta }}{\delta \psi (y)}  \label{Eq.MixedFnalDerivResult}
\end{eqnarray}%
\smallskip

\subsection{Functional Integration for Grassmann Fields}

\label{Sec10.8}

If the range over variable $x$ for the Grassmann field $\psi (x)$ is divided
up into $n$ small intervals $\Delta x_{i}=x_{i+1}-x_{i}$ (the $i$th
interval), then we may specify the value $\psi _{i}$ of the function $\psi
(x)$ in the $i$th interval via the spatial average over the interval 
\begin{equation}
\psi _{i}=\frac{1}{\Delta x_{i}}\int\limits_{\Delta x_{i}}dx\,\psi (x).
\label{Eq.GrassmannAverageValue}
\end{equation}%
Averaging a Grassmann field over a position interval still results in a
linear form involving the Grassmann variables $g_{1},\cdots ,g_{k},\cdots
g_{n}$. As previously, for simplicity we will choose the same number $n$ of
intervals as mode functions.

The functional $F[\psi (x)]$ may be regarded as a function $F(\psi
_{1},\cdots ,\psi _{i},\cdots ,\psi _{n})$ of all the $n$ different $\psi
_{i}$, which in the present case are a set of Grassmann variables. As we
will see, these Grassmann variables $\psi _{1},\cdots ,\psi _{i},\cdots
,\psi _{n}$ just involve a linear transformation from the $g_{1},\cdots
,g_{k},\cdots g_{n}$. Introducing a suitable \emph{weight function} $w(\psi
_{1},\cdots ,\psi _{i},\cdots ,\psi _{n})$ we may then define the \emph{%
functional integral} via the multiple Grassmann integral 
\begin{eqnarray}
\int D\psi \,F[\psi (x)] &=&\lim_{n\rightarrow \infty }\lim_{\epsilon
\rightarrow 0}\int \cdots \int d\psi _{n}\cdots d\psi _{i}\cdots d\psi
_{1}\,w(\psi _{1},\cdots ,\psi _{i},\cdots ,\psi _{n})  \nonumber \\
&&\times F(\psi _{1},\cdots ,\psi _{i},\cdots ,\psi _{n})
\label{Eq.GrassmannFuncIntegral1}
\end{eqnarray}%
where $\epsilon >\Delta x_{i}$. As previously, we use left integration and
follow the convention in which the symbol $D\psi $ stands for $d\psi
_{n}\cdots d\psi _{i}\cdots d\psi _{1}\,w(\psi _{1},\cdots ,\psi _{i},\cdots
,\psi _{n})$. A total functional integral of a functional of a Grassmann
field gives a c-number.

If the functional $F[\psi (x),\psi ^{+}(x)]$ involves pairs of Grassmann
fields, then the functional integral will be of the form $\int \int D\psi
^{+}\,D\psi \,F[\psi (x),\psi ^{+}(x)]$, where $D\psi ^{+}D\psi =d\psi
_{n}^{+}\cdots d\psi _{i}^{+}\cdots d\psi _{1}^{+}d\psi _{n}\cdots d\psi
_{i}\cdots d\psi _{1}\,w(\psi _{1},..,\psi _{i}..,\psi _{n},\psi
_{1}^{+},..,\psi _{i}^{+}..,\psi _{n}^{+})$. Similarly to differentiation
and integration in ordinary Grassmann calculus, functional integration and
differentiation are not inverse processes.\smallskip

\subsection{Functional Integrals and Phase Space Integrals}

For a mode expansion such as in Eq.(\ref{Eq.GrassmannFieldFn2}) the value $%
\phi _{ki}$ of the mode function in the $i$th interval is also defined via
the average 
\begin{equation}
\phi _{ki}=\frac{1}{\Delta x_{i}}\int\limits_{\Delta x_{i}}dx\,\phi _{k}(x)
\label{Eq.AverModeFn}
\end{equation}%
Unlike the Grassmann field, this is just a c-number. It is then easy to see
that the Grassmann variables $\psi _{1},\cdots ,\psi _{i},\cdots ,\psi _{n}$
are related to the g-number expansion coefficients $g_{1},\cdots
,g_{k},\cdots g_{n}$ via the linear transformation with c-number
coefficients $\phi _{ki}$ 
\begin{eqnarray}
\psi _{i} &=&\sum\limits_{k}\phi _{ki}\,g_{k}
\label{Eq.GrassmannAverValInter} \\
&=&\sum\limits_{k}g_{k}\,\phi _{ki}.  \label{Eq.GrassmannAverValInter2}
\end{eqnarray}%
This shows that the average values in the $i$th interval of the function $%
\psi _{i}$ and the mode function $\phi _{ki}$ are related via the expansion
coefficients $g_{k}$. This linear relation enables us to transform the
functional Grassmann integral (\ref{Eq.GrassmannFuncIntegral1}) into a
Grassmann phase space integral, thereby establishing the link between these
integrals.

Using the expression Eq.(\ref{Eq.GrassmannExpnCoefts1}) for the expansion
coefficients we then obtain the inverse formula to Eq.(\ref%
{Eq.GrassmannAverValInter}) 
\begin{equation}
g_{k}=\sum\limits_{i}\Delta x_{i}\,\phi _{ki}^{\ast }\,\psi _{i}.
\label{Eq.GrassmannAverValInter3}
\end{equation}%
The relationship in Eq.(\ref{Eq.GrassmannAverValInter}) shows that the
functions $F(\psi _{1},\cdots ,\psi _{i},\cdots ,\psi _{n})$ and $w(\psi
_{1},\cdots ,\psi _{i},\cdots ,\psi _{n})$ of all the interval values $\psi
_{i}$ can also be regarded as functions of the expansion coefficients $g_{k}$
which we may write as%
\begin{eqnarray}  \label{Eq.GrassmannPhaseSpaceWeightFn}
\,f(g_{1},\cdots ,g_{k},\cdots g_{n}) &\equiv &F(\psi _{1}(g_{1},\cdots
,g_{k},\cdots g_{n}),\cdots ,\psi _{i}(g_{1},\cdots ,g_{k},\cdots
g_{n}),\cdots ,\psi _{n})  \nonumber  \label{Eq.GrassmannPhaseSpaceFn} \\
v(g_{1},\cdots ,g_{k},\cdots g_{n}) &\equiv &w(\psi _{1}(g_{1},\cdots
,g_{k},\cdots g_{n}),\cdots ,\psi _{i}(g_{1},\cdots ,g_{k},\cdots
g_{n}),\cdots ,\psi _{n}).  \nonumber \\
&&
\end{eqnarray}%
Thus the various values $\psi _{1},\cdots ,\psi _{1},\cdots ,\psi
_{i},\cdots ,\psi _{n},\cdots ,\psi _{n}$ that the function $\psi (x)$ takes
on in the $n$ intervals - and which are integrated over in the functional
integration process - are all determined by the choice of the expansion
coefficients $g_{1},\cdots ,g_{k},\cdots g_{n}$. Hence Grassmann integration
over all the $\psi _{i}$ is equivalent to Grassmann integration over all the 
$g_{k}$. This enables us to express the functional integral in Eq.(\ref%
{Eq.GrassmannFuncIntegral1}) as a Grassmann phase space integral over the
expansion coefficients $g_{1},\cdots ,g_{k},\cdots g_{n}$. However, the
derivation of the result differs from the c-number case because the
transformation of the product of Grassmann differentials $d\psi _{n}\cdots
d\psi _{i}\cdots d\psi _{1}$ into the new product of Grassmann differentials 
$dg_{n}\cdots dg_{i}\cdots dg_{2}dg_{1}$ requires a similar treatment to
that explained in Appendix A of Ref. \cite{Dalton16b} where the
transformation between Grassmann integration variables is linear. We cannot
just write $d\psi _{i}=\sum_{k}\phi _{ki}\,dg_{k}$ because the differentials
are also Grassmann variables. Hence the usual c-number transformation
involving the Jacobian does not apply. The required result can be obtained
from Appendix A in Paper I \cite{Dalton16b} (see Eqs.(157), (158), (161)
therein) by making the idenifications $g_{i}\rightarrow \psi
_{i},h_{k}\rightarrow g_{k},A_{ik}\rightarrow \phi _{ki}$ so 
\begin{equation}
d\psi _{n}\cdots d\psi _{i}\cdots d\psi _{1}=(\mathrm{Det}%
A)^{-1}dg_{n}\cdots dg_{i}\cdots dg_{1}.
\end{equation}%
Now with $A_{ik}=\phi _{ki}$ we have using the completeness relationship in
Eq.(\ref{Eq.Completeness}) 
\begin{eqnarray}
(AA^{\dag })_{ij} &=&\sum\limits_{k}\phi _{ki}\phi _{kj}^{\ast }  \nonumber
\\
&=&\sum\limits_{k}\frac{1}{\Delta x_{i}}\int\limits_{\Delta x_{i}}dx\,\phi
_{k}(x)\frac{1}{\Delta x_{j}}\int\limits_{\Delta x_{j}}dy\,\phi _{k}^{\ast
}(y)  \nonumber \\
&=&\frac{1}{\Delta x_{i}}\int\limits_{\Delta x_{i}}dx\,\frac{1}{\Delta x_{j}}%
\int\limits_{\Delta x_{j}}dy\,\delta (x-y)  \nonumber \\
&=&\,\delta _{ij}\frac{1}{(\Delta x_{i})^{2}}\int\limits_{\Delta x_{i}}dx 
\nonumber \\
&=&\,\delta _{ij}\,(\Delta x_{i})^{-1}.
\end{eqnarray}%
Thus we have 
\begin{equation}
|(\mathrm{Det}A)|=\prod\limits_{i}(\Delta x_{i})^{-1/2}
\end{equation}

Hence we have using the result in Eq.(165) in Ref. \cite{Dalton16b} for
transforming phase space integrals 
\begin{eqnarray}
\int D\psi \,F[\psi (x)] &=&\lim_{n\rightarrow \infty }\lim_{\epsilon
\rightarrow 0}\int \cdots \int dg_{n}\cdots dg_{k}\cdots
dg_{1}\,\prod\limits_{i}(\Delta x_{i})^{1/2}  \nonumber \\
&&\times v(g_{1},\cdots ,g_{k},\cdots g_{n})\,f(g_{1},\cdots ,g_{k},\cdots
g_{n})  \label{Eq.GrassmannPhaseSpaceIntegral2}
\end{eqnarray}%
This key result expresses the original Grassmann functional integral as a
Grassmann phase space integral over the g-number expansion coefficients $%
g_{k}$ for the Grassmann field $\psi (x)$ in terms of the mode functions $%
\phi _{k}(x)$. Note that this result is different to the previous c-number
case, where the factor is $\prod_{i}(\Delta x_{i})^{-1/2}$instead of $%
\prod_{i}(\Delta x_{i})^{1/2}$. This is because the Grassmann differentials
transform via $(\mathrm{Det}A)^{-1}$ instead of $(\mathrm{Det}A)^{+1}$.

The general result can be simplified with a special choice of the weight
function%
\begin{equation}
w(\psi _{1},\cdots ,\psi _{i},\cdots ,\psi _{n})=\prod\limits_{i}(\Delta
x_{i})^{-1/2}  \label{Eq.GrassmannSimpleWeightFn}
\end{equation}%
and we then get a simple expression for the Grassmann functional integral 
\begin{equation}
\int D\psi \,F[\psi (x)]=\lim_{n\rightarrow \infty }\lim_{\epsilon
\rightarrow 0}\int \cdots \int dg_{1}\cdots dg_{k}\cdots
dg_{n}\,\,f(g_{1},\cdots ,g_{k},\cdots g_{n}).
\label{Eq.GrassmannPhaseSpaceIntegral3}
\end{equation}%
In this form of the Grassmann functional integral the original Grassmann
functional $F[\psi (x)]$ has been replaced by the equivalent function $%
f(g_{1},\cdots ,g_{k},\cdots g_{n})$ of the g-number expansion coefficients $%
g_{k}$, and the functional integration is now replaced by a Grassmann phase
space integration over the expansion coefficients.

For two Grassmann fields $\psi (x),\psi ^{+}(x)$ a straightforward extension
of the last result gives 
\begin{eqnarray}
&&\int D\psi ^{+}\,D\psi \,F[\psi (x),\psi ^{+}(x)]  \nonumber \\
&=&\lim_{n\rightarrow \infty }\lim_{\epsilon \rightarrow 0}\int \cdots \int
dg_{n}^{+}\cdots dg_{k}^{+}\cdots dg_{1}^{+}dg_{n}\cdots dg_{k}\cdots
dg_{1}\,  \nonumber \\
&\times &f(g_{1},\cdots ,g_{k},\cdots g_{n},g_{1}^{+},\cdots
,g_{k}^{+},\cdots g_{n}^{+})  \label{Eq.GrassmannPhaseSpaceIntegral4}
\end{eqnarray}%
where the weight function is now%
\begin{equation}
w(\psi _{1},\cdots ,\psi _{i},\cdots ,\psi _{n},\psi _{1}^{+},\cdots ,\psi
_{i}^{+},\cdots ,\psi _{n}^{+})=\prod\limits_{i}(\Delta x_{i})^{-1}
\label{Eq.GrassmannWeightFn4}
\end{equation}%
and $\psi ^{+}(x)$ is given via (\ref{Eq.GrassmannFunction3}).\smallskip

\subsection{Functional Integration by Parts}

A useful integration by parts rule can often be established from Eq.(\ref%
{Eq.GrassmannLefprodRule}). Consider the Grassmann functional $H[\psi
(x)]=F[\psi (x)]G[\psi (x)]$. Then 
\begin{eqnarray*}
&&F[\psi (x)]\left\{ \left( \frac{\overrightarrow{\delta }}{\delta \psi (x)}%
\right) _{x}G[\psi (x)]\right\} \\
&=&\sigma (F)\left( \frac{\overrightarrow{\delta }}{\delta \psi (x)}\right)
_{x}\{F[\psi (x)]G[\psi (x)]\}-\sigma (F)\left\{ \left( \frac{%
\overrightarrow{\delta }}{\delta \psi (x)}\right) _{x}F[\psi (x)]\right\}
G[\psi (x)]
\end{eqnarray*}%
Then 
\begin{eqnarray*}
\int D\psi \,F[\ ]\left\{ \left( \frac{\overrightarrow{\delta }}{\delta \psi
(x)}\right) _{x}G[\ ]\right\} &=&\sigma (F)\int D\psi \,\left( \frac{%
\overrightarrow{\delta }}{\delta \psi (x)}\right) _{x}H[\ ] \\
&-&\sigma (F)\int D\psi \,\left\{ \left( \frac{\overrightarrow{\delta }}{%
\delta \psi (x)}\right) _{x}F[\ ]\right\} G[\ ]
\end{eqnarray*}%
If we now introduce mode expansions and use Eq.(\ref%
{Eq.GrassmannLeftFnalDerivResult}) for the functional derivative of $H[\psi
(x)]$ and Eq.(\ref{Eq.GrassmannPhaseSpaceIntegral3}) for the first of the
two functional integrals on the right hand side of the last equation then%
\begin{eqnarray*}
&&\int D\psi \left( \frac{\overrightarrow{\delta }}{\delta \psi (x)}\right)
_{x}H[\psi (x)] \\
&=&\mbox{\rule{-1.5mm}{0mm}}\lim_{n\rightarrow \infty }\lim_{\epsilon
\rightarrow 0}\int \mbox{\rule{-1mm}{0mm}}dg_{1}\cdots dg_{k}\cdots dg_{n}%
\mbox{\rule{-1mm}{0mm}}\sum\limits_{k}\phi _{k}^{\ast }(x)\frac{%
\overrightarrow{\partial }}{\partial g_{k}}h(g_{1},\mbox{\rule{-1mm}{0mm}}%
\cdots ,g_{k},\mbox{\rule{-1mm}{0mm}}\cdots ) \\
&=&\lim_{n\rightarrow \infty }\lim_{\epsilon \rightarrow
0}\sum\limits_{k}\phi _{k}^{\ast }(x)\,\int \cdots \int dg_{1}\cdots
dg_{k-1}dg_{k+1}\cdots dg_{n} \\
&\times &(-1)^{n-k}\{\int dg_{k}\,\frac{\overrightarrow{\partial }}{\partial
g_{k}}h(g_{1},\cdots ,g_{k},\cdots )\}
\end{eqnarray*}%
so that the functional integral of this term reduces to the Grassmann
integral of a Grassmann derivative. This is zero, since differentiation
removes the $g_{k}$ dependence. Hence the Grassmann functional integral
involving the functional derivative of $H[\psi (x)]$ vanishes and we have
the integration by parts result%
\begin{equation}
\int D\psi \,F[\psi (x)]\left\{ \left( \frac{\overrightarrow{\delta }}{%
\delta \psi (x)}\right) _{x}\mbox{\rule{-1mm}{0mm}}G[\psi (x)]\right\}
=-\sigma (F)\int D\psi \,\left\{ \left( \frac{\overrightarrow{\delta }}{%
\delta \psi (x)}\right) _{x}\mbox{\rule{-1mm}{0mm}}F[\psi (x)]\right\}
G[\psi (x)]  \label{Eq.GrassmannFnlIntegParts}
\end{equation}%
A similar result involving right functional differentiation can be
established.\smallskip

\subsection{Differentiating a Functional Integral}

Functionals can be defined via functional integration processes and it is
useful to find rules for their functional derivatives. This leads to a rule
for differentiating a functional integral.

Suppose we have a functional $G[\chi (x)]$ determined from another
functional $F[\psi (x)]$ via a functional integral that involves a left
transfer functional $A_{GF}[\chi (x),\psi (x)]$ 
\begin{equation}
G[\chi (x)]=\int D\psi \,A_{GF}[\chi (x),\psi (x)]\,F[\psi (x)].
\label{Eq.GrassmannFuncIntegral2}
\end{equation}%
Applying the definition of the left Grassmann functional derivatives of $%
G[\chi (x)]$ and $A_{GF}[\chi (x)+\delta \chi (x),\psi (x)]$ with respect to 
$\chi (x)$ we have%
\begin{eqnarray*}
&&G[\chi (x)+\delta \chi (x)] \\
&=&\int D\psi \,A_{GF}[\chi (x)+\delta \chi (x),\psi (x)]\,F[\psi (x)] \\
&=&\int D\psi \,\left\{ A_{GF}[\chi (x),\psi (x)]\,\right\} F[\psi (x)] \\
&&+\int D\psi \,\left\{ \int dx\,\delta \chi (x)\,\left\{ \left( \frac{%
\overrightarrow{\delta }}{\delta \psi (x)}\right) _{x}A_{GF}[\chi (x),\psi
(x)]\right\} \right\} F[\psi (x)] \\
&=&G[\chi (x)]+\int dx\,\delta \chi (x)\,\int D\psi \,\left\{ \left( \frac{%
\overrightarrow{\delta }}{\delta \psi (x)}\right) _{x}A_{GF}[\chi (x),\psi
(x)]\right\} F[\psi (x)]
\end{eqnarray*}%
since (for reasonably well-behaved quantities) the functional integration
over $D\psi $ and the ordinary integration over $dx$ can be carried out in
either order, given that both just involve processes that are limits of
sumnations. Hence from the definition of the functional derivative we have 
\begin{eqnarray}
\left( \frac{\overrightarrow{\delta }}{\delta \chi (x)}\right) _{x}G[\chi
(x)] &=&\int D\psi \,\left\{ \left( \frac{\overrightarrow{\delta }}{\delta
\psi (x)}\right) _{x}A_{GF}[\chi (x),\psi (x)]\right\} F[\psi (x)]  \nonumber
\\
&&  \label{Eq.DiffnLeftGrassmannFnalIntegral}
\end{eqnarray}%
which is the required rule for left differentiating a functional defined via
a functional of another function. Clearly the rule is to just differentiate
the transfer functional under the functional integration sign, a rule
similar to that applying in ordinary calculus.

A similar rule can be obtained for a functional $H[\chi (x)]$ determined
from another functional $F[\psi (x)]$ via a functional integral that
involves a right transfer functional $A_{GF}[\chi (x),\psi (x)]$ in the form%
\begin{equation}
H[\chi (x)]=\int D\psi \,F[\psi (x)]\,A_{GF}[\chi (x),\psi (x)].
\label{Eq.GrassmannFuncIntegral3}
\end{equation}%
We find that 
\begin{eqnarray}
H[\chi (x)]\left( \frac{\overleftarrow{\delta }}{\delta \chi (x)}\right)
_{x} &=&\int D\psi \,F[\psi (x)]\left\{ A_{GF}[\chi (x),\psi (x)]\left( 
\frac{\overleftarrow{\delta }}{\delta \psi (x)}\right) _{x}\right\} 
\nonumber \\
&&  \label{Eq.DiffnRightGrassmannFnalIntegral}
\end{eqnarray}%
which is the required rule for right differentiating a functional defined
via a functional of another function. The proof is left as an exercise.
Clearly the rule is to just differentiate the transfer functional under the
functional integration sign, a rule similar to that applying in ordinary
calculus (except that right and left differentiation are different).

As a particular case, consider the Fourier like Grassmann transfer
functional 
\begin{equation}
A_{GF}[\chi (x),\psi (x)]=\exp \left\{ i\int dx\,\chi (x)\,\psi (x)\right\}
\label{Eq.FourierGrassmannFnalIntegral}
\end{equation}%
This Grassmann transfer functional is equivalent to a even Grassmann
function of the expansion coefficients $g_{k}$ for $\psi (x)$ and $h_{k}$
for $\chi (x)$. In this case 
\begin{eqnarray*}
A_{GF}[\chi (x)+\delta \chi (x),\psi (x)] &=&\exp \left\{ i\int dx\,(\chi
(x)+\delta \chi (x))\,\psi (x)\right\} \\
&=&\exp \left\{ i\int dx\,\chi (x)\,\psi (x)\right\} \exp \left\{ i\int
dx\,\delta \chi (x)\,\psi (x)\right\} \\
&\approx &\exp \left\{ i\int dx\,\chi (x)\,\psi (x)\right\} (1+i\int
dx\,\delta \chi (x)\,\psi (x)) \\
&=&A_{GF}[\chi (x),\psi (x)]+A_{GF}[\chi (x),\psi (x)]\,i\int dx\,\delta
\chi (x)\,\psi (x) \\
&=&A_{GF}[\chi (x),\psi (x)]+\int dx\,\delta \chi (x)\,iA_{GF}[\chi (x),\psi
(x)]\,\psi (x)
\end{eqnarray*}%
where we have used the Baker-Haussdorff theorem (see Eq. (130) in Ref. \cite%
{Dalton16b}) with $A=\int dx\,\chi (x)\,\psi (x)$ and $B=\int dx\,\delta
\chi (x)\,\psi (x)$ together with the commutator result based on Grassmann
fields anti-commuting%
\begin{eqnarray*}
\lbrack A,B] &=&\int dx\,\chi (x)\,\psi (x)\int dy\,\delta \chi (y)\,\psi
(y)-\int dy\,\delta \chi (y)\,\psi (y)\int dx\,\chi (x)\,\psi (x) \\
&=&\int dx\int dy\,\chi (x)\,\psi (x)\delta \chi (y)\,\psi
(y)-(-1)^{2+2}\int dx\int dy\,\chi (x)\,\psi (x)\delta \chi (y)\psi (y) \\
&=&0
\end{eqnarray*}%
to establish the third line of the derivation. The last line follows from $%
A_{GF}[\chi (x),\psi (x)]$ being equivalent to an even Grassmann function
and therefore commuting with the Grassmann field $\delta \chi (x)$. Hence
for the left functional derivative%
\begin{equation}
\left( \frac{\overrightarrow{\delta }}{\delta \psi (x)}\right)
_{x}A_{GF}[\chi (x),\psi (x)]=A_{GF}[\chi (x),\psi (x)]\times i\psi (x)
\end{equation}%
and 
\begin{eqnarray}
G[\chi (x)] &=&\int D\psi \,A_{GF}[\chi (x),\psi (x)]\,F[\psi (x)] \\
\left( \frac{\overrightarrow{\delta }}{\delta \chi (x)}\right) _{x}G[\chi
(x)] &=&\int D\psi \,\left\{ A_{GF}[\chi (x),\psi (x)]\,\times (i\psi
(x))\right\} \,F[\psi (x)]  \nonumber \\
&&  \label{Eq.LeftDiffnGrassFourierFnalIntegral}
\end{eqnarray}

A similar result follows for the right transfer functional case. We have 
\begin{eqnarray}
H[\chi (x)] &=&\int D\psi \,F[\psi (x)]\,A_{GF}[\chi (x),\psi (x)] \\
H[\chi (x)]\left( \frac{\overleftarrow{\delta }}{\delta \chi (x)}\right)
_{x} &=&\int D\psi \,F[\psi (x)]\left\{ A_{GF}[\chi (x),\psi (x)]\times
(-i\psi (x))\right\}  \nonumber \\
&&  \label{Eq.RightDiffnGrassFourierFnalIntegral}
\end{eqnarray}

\pagebreak

\section{Appendix C - Functional $P$ Distribution}

\label{Appendix - Functional P Distribution}

\subsection{Characteristic and $P$ Distribution Functional Relation}

The characteristic functional - distribution functional relationship'can be
established in terms of functional integrals.

In the characteristic functional definition (\ref{Eq.FermiCharFnal}) the
Grassmann fields $h,h^{+}$ are expanded in terms of mode functions%
\begin{equation}
h(x)=\sum\limits_{i}h_{i}\phi _{i}(x)\hspace{1cm}h^{+}(x)=\sum%
\limits_{i}h_{i}^{+}\phi _{i}^{\ast }(x)  \label{Eq.ModeExpnHFields}
\end{equation}%
and it is easy to see using orthogonality of the modes that 
\begin{equation}
\int dx\,\hat{\Psi}(x)h^{+}(x)=\sum_{i}\hat{c}_{i}h_{i}^{+}\qquad \int
dx\,h(x)\hat{\Psi}^{\dag }(x)=\sum_{i}h_{i}\hat{c}_{i}^{\dag }
\end{equation}%
Hence from Eqs.(\ref{Eq.FermiCharFnal}) and (\ref{Eq.FermionCharFunction})
we see that 
\begin{equation}
\chi \lbrack h(x),h^{+}(x)]\equiv \chi (h,h^{+})
\label{Eq.EquivFermiCharFnalFn}
\end{equation}%
so the characteristic functional of the fields $h(x),h^{+}(x)$ is entirely
equivalent to the original fermionic characteristic function of the g-number
expansion coefficients $h_{i},h_{i}^{+}$.

The distribution function\ $P(g,g^{+})\equiv $\ $P[\psi (x),\psi ^{+}(x)]$
is related to the characteristic function $\chi (h,h^{+})$ as in Eq. (34) in
Ref. \cite{Dalton16b}. Using%
\begin{equation}
\int dx\,\psi (x)h^{+}(x)=\sum_{i}g_{i}h_{i}^{+}\qquad \int dx\,h(x)\psi
^{+}(x)=\sum_{i}h_{i}g_{i}^{+}
\end{equation}%
we see that 
\begin{eqnarray}
&&\chi \lbrack h(x),h^{+}(x)]  \nonumber \\
&=&\tint \tprod\limits_{i}dg_{i}^{+}dg_{i}\,\exp i\left\{ \int dx\,\psi
(x)h^{+}(x)\right\} P[\psi (x),\psi ^{+}(x)]\,\exp i\left\{ \int
dx\,h(x)\psi ^{+}(x)\right\}  \nonumber \\
&=&\int D\psi ^{+}D\psi \,\exp i\left\{ \int dx\,\psi (x)h^{+}(x)\right\}
P[\psi (x),\psi ^{+}(x)]\,\exp i\left\{ \int dx\,h(x)\psi ^{+}(x)\right\} 
\nonumber \\
&&
\end{eqnarray}%
giving the characteristic functional as a functional integral involving the $%
P$ distribution functional. \smallskip

\subsection{Correspondence Rules}

For the $P$ distribution functional these are

\begin{eqnarray}
\hat{\rho} &\Rightarrow &\hat{\Psi}_{\alpha }(x)\,\hat{\rho}\qquad P[\psi
(x)]\Rightarrow \psi _{\alpha }(x)\,P  \label{Eq.GrassFnalCorr5} \\
\hat{\rho} &\Rightarrow &\hat{\rho}\,\hat{\Psi}_{\alpha }(x)\qquad P[\psi
(x)]\Rightarrow P\,\left( +\frac{\overleftarrow{\delta }}{\delta \psi
_{\alpha }^{+}(x)}-\psi _{\alpha }(x)\right)  \label{Eq.GrassFnalCorr6} \\
\hat{\rho} &\Rightarrow &\hat{\Psi}_{\alpha }^{\dag }(x)\,\hat{\rho}\qquad
P[\psi (x)]\Rightarrow \left( +\frac{\overrightarrow{\delta }}{\delta \psi
_{\alpha }(x)}-\psi _{\alpha }^{+}(x)\right) \,P  \label{Eq.GrassFnalCorr7}
\\
\hat{\rho} &\Rightarrow &\hat{\rho}\,\hat{\Psi}_{\alpha }^{\dag }(x)\qquad
P[\psi (x)]\Rightarrow P\,\psi _{\alpha }^{+}(x).  \label{Eq.GrassFnalCorr8}
\end{eqnarray}%
and can be derived from those for the $B$ distribution functional together
with the relationship 
\begin{equation}
P[\psi (x)]=B[\psi (x)]\,\exp (+\dint dx\,\psi (x)\psi ^{+}(x))
\label{Eq.BandPDistnFnalReln}
\end{equation}%
As will be seen there is a mixture of Grassmann functional derivatives and
Grassman fields.\pagebreak

\section{Appendix D - Ito Stochastic Field Equations}

\label{Appendix - Ito Stochastic Field Equations}

In the notation where $\widetilde{g}_{p}\rightarrow $ $\widetilde{g}_{\alpha
i}^{A}$ the Ito stochastic equations (\ref{Eq.ItoStochEqns2}) for the
stochastic phase variables become

\begin{equation}
\widetilde{g}_{\alpha i}^{A}(t+\delta t)-\widetilde{g}_{\alpha
i}^{A}(t)=C_{\alpha i}^{A}(\widetilde{g}(t))\delta t+\sum_{a}B_{a}^{\alpha
i\,A}(\widetilde{g}(t))\int_{t}^{t+\delta t}dt_{1}\Gamma _{a}(t_{1})
\end{equation}%
so multiplying by the mode functions $\xi _{\alpha i}^{A}(r)$ and summing
over $i$ leads to 
\begin{equation}
\widetilde{\psi }_{\alpha A}(r,t+\delta t)-\widetilde{\psi }_{\alpha
A}(r,t)=C_{\alpha A}[\widetilde{\psi }(r),r]\delta t+\sum_{a}B_{a}^{\alpha
A}[\widetilde{\psi }(r),r]\int_{t}^{t+\delta t}dt_{1}\Gamma _{a}(t_{1})
\label{Eq.ItoStochFldEqnIntegralForm}
\end{equation}%
using the expansion in (\ref{Eq.ItoStochFieldExpn}) for the stochastic
fields and where%
\begin{eqnarray}
C_{\alpha A}[\widetilde{\psi }(r),r] &=&\dsum\limits_{i}C_{\alpha i}^{A}(%
\widetilde{g})\,\xi _{\alpha \,i}^{A}(r)\,  \nonumber \\
B_{a}^{\alpha A}[\widetilde{\psi }(r),r] &=&\dsum\limits_{i}B_{a}^{\alpha
i\,A}(\widetilde{g})\,\xi _{\alpha i}^{A}(r)
\end{eqnarray}

Hence we have 
\begin{equation}
\frac{\partial }{\partial t}\widetilde{\psi }_{\alpha A}(r)=C_{\alpha A}[%
\widetilde{\psi }(r),r]+\tsum\limits_{a}B_{a}^{\alpha A}[\widetilde{\psi }%
(r),r]\;\Gamma _{a}(t_{+})
\end{equation}%
as required. The right side is the sum of a classical field term and a noise
field term. The classical and noise field terms involve Grassmann
functionals $C_{\alpha A}[\widetilde{\psi }(r),r]$ and $B_{a}^{\alpha A}[%
\widetilde{\psi }(r),r]$ defined by\ 
\begin{eqnarray}
C_{\alpha A}[\widetilde{\psi }(r),r] &=&-\dsum\limits_{i}A_{\alpha i}^{A}(%
\widetilde{g})\,\xi _{\alpha \,i}^{A}(r)\, \\
B_{a}^{\alpha A}[\widetilde{\psi }(r),r] &=&\dsum\limits_{i}B_{a}^{\alpha
i\,A}(\widetilde{g})\,\xi _{\alpha i}^{A}(r)
\end{eqnarray}%
Note that the classical field term is also stochastic because it involves a
functional of the stochastic fields $\widetilde{\psi }_{\alpha A}(r)$. The
noise field term is stochastic, not only for the same reason but also
because it involves the Gaussian-Markoff nose terms.

\pagebreak

\section{Appendix E - Fermi Gas Functional Fokker-Planck Equation}

\label{Appendix - Fermi Gas FFPE}

In this Appendix we denote the distribution functional $B[\psi _{u}(\mathbf{r%
}),\psi _{u}^{+}(\mathbf{r}),\psi _{d}(\mathbf{r}),\psi _{d}^{+}(\mathbf{r}%
)] $ as $B[\mathbf{\psi (r)}]$ for short. The Hamiltonian is given in Eq.(%
\ref{Eq.HamiltonianFieldModel}). The correspondence rules are given in Eqs.(%
\ref{Eq.FermionFieldOprsCorresRules}) with a simple extension to give rules
for the spatial derivatives of field operators. We will assume the modes are
restricted to a cut-off $K$.\smallskip

\subsection{Kinetic Energy Terms}

If $\widehat{\rho }\rightarrow \widehat{T}\widehat{\rho }$\ then 
\[
B[\mathbf{\psi }(\mathbf{r})]\rightarrow \frac{\hbar ^{2}}{2m}%
\sum\limits_{\alpha }\sum\limits_{\mu }\int d\mathbf{s}\left\{ \left(
\partial _{\mu }\frac{\overrightarrow{\delta }}{\delta \psi _{\alpha 1}(%
\mathbf{s})}\right) \left( \partial _{\mu }\psi _{\alpha 1}(\mathbf{s}%
)\right) \right\} B[\mathbf{\psi }(\mathbf{s})] 
\]%
and if $\widehat{\rho }\rightarrow \widehat{\rho }\widehat{T}$ then 
\[
B[\mathbf{\psi }(\mathbf{r})]\rightarrow \frac{\hbar ^{2}}{2m}%
\sum\limits_{\alpha }\sum\limits_{\mu }\int d\mathbf{s}B[\mathbf{\psi }(%
\mathbf{s})]\left\{ \left( \partial _{\mu }\psi _{\alpha 2}(\mathbf{s}%
)\right) \left( \partial _{\mu }\frac{\overleftarrow{\delta }}{\delta \psi
_{\alpha 2}(\mathbf{s})}\right) \right\} 
\]%
so for $\widehat{\rho }\rightarrow -i/\hbar \lbrack \widehat{T},\widehat{%
\rho }]$\ then 
\begin{eqnarray}
&&B[\mathbf{\psi }(\mathbf{r})]  \nonumber \\
&&\qquad \rightarrow -\frac{i}{\hbar }\left\{ \frac{\hbar ^{2}}{2m}%
\sum\limits_{\alpha }\sum\limits_{\mu }\int d\mathbf{s}\left\{ \left(
\partial _{\mu }\frac{\overrightarrow{\delta }}{\delta \psi _{\alpha 1}(%
\mathbf{s})}\right) \left( \partial _{\mu }\psi _{\alpha 1}(\mathbf{s}%
)\right) \right\} B[\mathbf{\psi }(\mathbf{s})]\right\}  \nonumber \\
&&\qquad \quad -\frac{i}{\hbar }\left\{ -\frac{\hbar ^{2}}{2m}%
\sum\limits_{\alpha }\sum\limits_{\mu }\int d\mathbf{s}B[\mathbf{\psi }(%
\mathbf{s})]\left\{ \left( \partial _{\mu }\psi _{\alpha 2}(\mathbf{s}%
)\right) \left( \partial _{\mu }\frac{\overleftarrow{\delta }}{\delta \psi
_{\alpha 2}(\mathbf{s})}\right) \right\} \right\} \,.  \nonumber \\
&&
\end{eqnarray}

Using the mode expansions in Eqs.(\ref{Eq.GrassmannLeftFnalDerivResult2})
and (\ref{Eq.GrassmannRightFnalDerivResult2B}) for the functional
derivatives we have for the first term 
\begin{eqnarray}
&&\int d\mathbf{s}\left\{ \left( \partial _{\mu }\frac{\overrightarrow{%
\delta }}{\delta \psi _{\alpha 1}(\mathbf{s})}\right) \left( \partial _{\mu
}\psi _{\alpha 1}(\mathbf{s})\right) \right\} B[\mathbf{\psi }(\mathbf{s})] 
\nonumber \\
&&\qquad \qquad =\sum_{i}\sum\limits_{j}\int d\mathbf{s}\left\{ \left(
\partial _{\mu }\xi _{\alpha i}^{2}(\mathbf{s})\frac{\overrightarrow{%
\partial }}{\partial g_{\alpha i}^{1}}\right) \left( \partial _{\mu }\xi
_{\alpha j}^{1}(\mathbf{s})\,g_{\alpha j}^{1}\right) \right\} B(\mathbf{g}) 
\nonumber \\
&&\qquad \qquad =-\sum_{i}\sum\limits_{j}\int d\mathbf{s}\left\{ \left( \xi
_{\alpha i}^{2}(\mathbf{s})\frac{\overrightarrow{\partial }}{\partial
g_{\alpha i}^{1}}\right) \left( \partial _{\mu }^{2}\xi _{\alpha j}^{1}(%
\mathbf{s})\,g_{\alpha j}^{1}\right) \right\} B(\mathbf{g})  \nonumber \\
&&\qquad \qquad =-\int d\mathbf{s}\left\{ \left( \frac{\overrightarrow{%
\delta }}{\delta \psi _{\alpha 1}(\mathbf{s})}\right) \left( \partial _{\mu
}^{2}\psi _{\alpha 1}(\mathbf{s})\right) \right\} B[\mathbf{\psi }(\mathbf{s}%
)]  \nonumber \\
&&\qquad \qquad =-\int d\mathbf{s}B[\mathbf{\psi }(\mathbf{s})]\left\{
\partial _{\mu }^{2}\psi _{\alpha 1}(\mathbf{s})\left( \frac{\overleftarrow{%
\delta }}{\delta \psi _{\alpha 1}(\mathbf{s})}\right) \right\}
\end{eqnarray}%
where we have used spatial integration by parts and then Eq.(\ref%
{Eq.LeftRightFnalDerivsResult}) for reversing the functional derivative,
noting that $\left( \partial _{\mu }^{2}\psi _{\alpha 1}(\mathbf{s})\right)
B[\mathbf{\psi }(\mathbf{s})]$ is an odd Grassmann function. Similarly 
\begin{eqnarray}
&&\int d\mathbf{s}B[\mathbf{\psi }(\mathbf{s})]\left\{ \left( \partial _{\mu
}\psi _{\alpha 2}(\mathbf{s})\right) \left( \partial _{\mu }\frac{%
\overleftarrow{\delta }}{\delta \psi _{\alpha 2}(\mathbf{s})}\right) \right\}
\nonumber \\
&&\qquad \qquad \qquad \qquad =-\int d\mathbf{s}B[\mathbf{\psi }(\mathbf{s}%
)]\left\{ \left( \partial _{\mu }^{2}\psi _{\alpha 2}(\mathbf{s})\right)
\left( \frac{\overleftarrow{\delta }}{\delta \psi _{\alpha 2}(\mathbf{s})}%
\right) \right\}
\end{eqnarray}%
though here no reversal of the functional derivative is needed. Combining
these results for $\widehat{\rho }\rightarrow -i/\hbar \lbrack \widehat{T},%
\widehat{\rho }]$\ gives the kinetic energy term in the functional
Fokker-Planck equation as 
\begin{eqnarray}
&&\left( \frac{\partial }{\partial t}B[\mathbf{\psi }(\mathbf{s})]\right)
_{K}  \nonumber \\
&=&\frac{-i}{\hbar }\left\{ -\sum\limits_{\alpha }\int d\mathbf{s}\left\{
\left( \sum\limits_{\mu }\frac{\hbar ^{2}}{2m}\partial _{\mu }^{2}\psi
_{\alpha 1}(\mathbf{s})\,B[\mathbf{\psi }(\mathbf{s})]\right) \frac{%
\overleftarrow{\delta }}{\delta \psi _{\alpha 1}(\mathbf{s})}\right\}
\right\}  \nonumber \\
&&\qquad \frac{-i}{\hbar }\left\{ +\sum\limits_{\alpha }\int d\mathbf{s}%
\left\{ \left( \sum\limits_{\mu }\frac{\hbar ^{2}}{2m}\partial _{\mu
}^{2}\psi _{\alpha 2}(\mathbf{s})\,B[\mathbf{\psi }(\mathbf{s})]\right) 
\frac{\overleftarrow{\delta }}{\delta \psi _{\alpha 2}(\mathbf{s})}\right\}
\right\}
\end{eqnarray}%
\smallskip

\subsection{Potential Energy Terms}

If $\widehat{\rho }\rightarrow \widehat{V}\widehat{\rho }$\ then 
\begin{eqnarray}
B[\mathbf{\psi }(\mathbf{r})] &\rightarrow &\sum\limits_{\alpha }\int d%
\mathbf{s}\left\{ \left( \frac{ \overrightarrow{\delta }}{\delta \psi
_{\alpha 1}(\mathbf{s})}\right) V_{\alpha }(\mathbf{s})\left( \psi _{\alpha
1}(\mathbf{s})\right) \right\} B[ \mathbf{\psi }(\mathbf{s})]  \nonumber \\
&=&\sum\limits_{\alpha }\int d\mathbf{s}\left\{ V_{\alpha }(\mathbf{s}
)\left( \psi _{\alpha 1}(\mathbf{s})\right) \right\} B[\mathbf{\psi }( 
\mathbf{s})]\left( \frac{\overleftarrow{\delta }}{\delta \psi _{\alpha 1}( 
\mathbf{s})}\right)
\end{eqnarray}
where the functional derivative in the first term has been reversed, and we
have used the fact that $\left\{ V_{\alpha }(\mathbf{s})\left( \psi _{\alpha
1}(\mathbf{s} )\right) \right\} B[\mathbf{\psi }(\mathbf{s})]$ is an odd
Grassman function.

If $\widehat{\rho }\rightarrow \widehat{\rho }\widehat{V}$\ then 
\[
B[\mathbf{\psi }(\mathbf{r})]\rightarrow \sum\limits_{\alpha }\int d\mathbf{s%
}B[\mathbf{\psi }(\mathbf{\ s})]\left\{ \left( \psi _{\alpha 2}(\mathbf{s}%
)\right) V_{\alpha }(\mathbf{s})\left( \frac{\overleftarrow{\delta }}{\delta
\psi _{\alpha 2}(\mathbf{s})}\right) \right\} 
\]%
so for $\widehat{\rho }\rightarrow -i/\hbar \lbrack \widehat{V},\widehat{%
\rho }]$\ then the potential energy term in the functional Fokker-Planck
equation becomes 
\begin{eqnarray}
&&\left( \frac{\partial }{\partial t}B[\mathbf{\psi }(\mathbf{s})]\right)
_{V}  \nonumber \\
&=&-i/\hbar \left\{ +\sum\limits_{\alpha }\int d\mathbf{s}\left( V_{\alpha }(%
\mathbf{s})\,\psi _{\alpha 1}(\mathbf{s})B[\mathbf{\psi }(\mathbf{s}%
)]\right) \left( \frac{\overleftarrow{\delta }}{\delta \psi _{\alpha 1}(%
\mathbf{s})}\right) \right\}  \nonumber \\
&&\qquad -i/\hbar \left\{ -\sum\limits_{\alpha }\int d\mathbf{s(}V_{\alpha }(%
\mathbf{s})\,\psi _{\alpha 2}(\mathbf{s})B[\mathbf{\psi }(\mathbf{s}%
)])\left( \frac{\overleftarrow{\delta }}{\delta \psi _{\alpha 2}(\mathbf{s})}%
\right) \right\} \,
\end{eqnarray}%
\smallskip

\subsection{Interaction Energy Terms}

If $\widehat{\rho }\rightarrow \widehat{U}\widehat{\rho }$\ then 
\begin{eqnarray}
&&B[\mathbf{\psi }(\mathbf{s})]  \nonumber \\
&\rightarrow &\frac{g}{2}\sum\limits_{\alpha }\int d\mathbf{s}\left\{ \left(
+\frac{\overrightarrow{\delta }}{\delta \psi _{\alpha 1}(\mathbf{s})}\right)
\left( +\frac{\overrightarrow{\delta }}{\delta \psi _{-\alpha 1}(\mathbf{s})}%
\right) \psi _{-\alpha 1}(\mathbf{s})\psi _{\alpha 1}(\mathbf{s})\right\} B[%
\mathbf{\psi }(\mathbf{s})]  \nonumber \\
&=&-\frac{g}{2}\sum\limits_{\alpha }\int d\mathbf{s}\left\{ \left( +\frac{%
\overrightarrow{\delta }}{\delta \psi _{\alpha 1}(\mathbf{s})}\right)
\left\{ \psi _{-\alpha 1}(\mathbf{s})\psi _{\alpha 1}(\mathbf{s})B[\mathbf{\
\psi }(\mathbf{s})]\left( +\frac{\overleftarrow{\delta }}{\delta \psi
_{-\alpha 1}(\mathbf{s})}\right) \right\} \right\}  \nonumber \\
&=&-\frac{g}{2}\sum\limits_{\alpha }\int d\mathbf{s}\left\{ \left\{ \psi
_{-\alpha 1}(\mathbf{s})\psi _{\alpha 1}(\mathbf{s})B[\mathbf{\psi }(\mathbf{%
\ s})]\left( +\frac{\overleftarrow{\delta }}{\delta \psi _{-\alpha 1}(%
\mathbf{s})}\right) \right\} \left( +\frac{\overleftarrow{\delta }}{\delta
\psi _{\alpha 1}(\mathbf{s})}\right) \right\}  \nonumber \\
&&
\end{eqnarray}%
where the second line is obtained using Eq.(\ref%
{Eq.LeftRightFnalDerivsResult}) noting that $\psi _{-\alpha 1}(\mathbf{s}%
)\psi _{\alpha 1}(\mathbf{s})B[\mathbf{\psi }(\mathbf{s})]$ is an even
Grassmann function, and the third line is obtained from the same equation,
but noting that $\left\{ \psi _{-\alpha 1}(\mathbf{s})\psi _{\alpha 1}(%
\mathbf{s})B[\mathbf{\psi }(\mathbf{s})]\left( +\frac{\overleftarrow{\delta }%
}{\delta \psi _{-\alpha 1}(\mathbf{s})}\right) \right\} $ is an odd
Grassmann function.

If $\widehat{\rho }\rightarrow \widehat{\rho }\widehat{U}$\ then 
\begin{eqnarray}
&&B[\mathbf{\psi }(\mathbf{s})]  \nonumber \\
&\rightarrow &\frac{g}{2}\sum\limits_{\alpha }\int d\mathbf{s}B[\mathbf{\
\psi }(\mathbf{s})]\left\{ \psi _{\alpha 2}(\mathbf{s})\psi _{-\alpha 2}(%
\mathbf{s})\left( +\frac{\overleftarrow{\delta }}{\delta \psi _{-\alpha 2}(%
\mathbf{s})}\right) \left( +\frac{\overleftarrow{\delta }}{\delta \psi
_{\alpha 2}(\mathbf{s})}\right) \right\}  \nonumber \\
&=&-\frac{g}{2}\sum\limits_{\alpha }\int d\mathbf{s}B[\mathbf{\psi }(\mathbf{%
s})]\left\{ \psi _{-\alpha 2}(\mathbf{s})\psi _{\alpha 2}(\mathbf{s})\left( +%
\frac{\overleftarrow{\delta }}{\delta \psi _{-\alpha 2}(\mathbf{s})}\right)
\left( +\frac{\overleftarrow{\delta }}{\delta \psi _{\alpha 2}(\mathbf{s})}%
\right) \right\}  \nonumber \\
&&
\end{eqnarray}%
where we have reversed the order of the fields.

Hence if $\widehat{\rho }\rightarrow -i/\hbar \lbrack \widehat{U},\widehat{%
\rho }]$\ then the interaction energy term in the functional Fokker-Planck
equation becomes 
\begin{eqnarray}
&&\left( \frac{\partial }{\partial t}B[\mathbf{\psi }(\mathbf{s})]\right)
_{U}  \nonumber \\
&=&\frac{i}{\hbar }\frac{g}{2}\sum\limits_{\alpha }\int d\mathbf{s}\left\{
\psi _{-\alpha 1}(\mathbf{s})\psi _{\alpha 1}(\mathbf{s})B[\mathbf{\psi }(%
\mathbf{s})]\left( +\frac{\overleftarrow{\delta }}{\delta \psi _{-\alpha 1}(%
\mathbf{s})}\right) \left( +\frac{\overleftarrow{\delta }}{\delta \psi
_{\alpha 1}(\mathbf{s})}\right) \right\}  \nonumber \\
&&-\frac{i}{\hbar }\frac{g}{2}\sum\limits_{\alpha }\int d\mathbf{s}\left\{
\psi _{-\alpha 2}(\mathbf{s})\psi _{\alpha 2}(\mathbf{s})B[\mathbf{\psi }(%
\mathbf{s})]\left( +\frac{\overleftarrow{\delta }}{\delta \psi _{-\alpha 2}(%
\mathbf{s})}\right) \left( +\frac{\overleftarrow{\delta }}{\delta \psi
_{\alpha 2}(\mathbf{s})}\right) \right\} \,.  \nonumber \\
&&
\end{eqnarray}%
\pagebreak

\end{document}